\documentclass[prb,aps,twocolumn]{revtex4}
\usepackage{amsmath}
\usepackage{amssymb}
\usepackage {graphicx}

\newcommand{\dtildeu}{\tilde{\tilde u}}

\newcommand{\beq}{\begin{equation}}
\newcommand{\eeq}{\end{equation}}
\newcommand{\bea}{\begin{eqnarray}}
\newcommand{\eea}{\end{eqnarray}}

\begin{document}

\title{Renormalization group flow, competing phases,
and the structure of superconducting gap in multi-band models of
Iron based superconductors}
\author{Saurabh Maiti, Andrey V. Chubukov}
\affiliation {~Department of Physics, University of Wisconsin,
Madison, Wisconsin 53706, USA}
\date{\today}


\begin{abstract}
We perform an analytical  renormalization group (RG) study to
address the role of Coulomb repulsion, the competition between
extended s-wave  superconducting order ($s\pm$), and the spin
density wave (SDW) order and the angular dependence of the
superconducting gap in multi-pocket models of Iron based
superconductors. Previous analytic RG studies considered a toy
model of one hole and one electron pocket. We consider more
realistic models of two electron pockets and either two or three
hole pockets, and also incorporate angular dependence of
 the interactions. In a toy 2-pocket model, SDW order always wins over
$s\pm$ order at perfect nesting; $s\pm$ order only appears when
doping is finite and RG flow extends long enough to overcome
intra-pocket Coulomb repulsion. For multi-pocket models, there are
two new effects. First, in most cases there  exists an attractive
component of the interaction in $s\pm$ channel no matter how
strong intra-pocket repulsion is, such that the system necessary
becomes a superconductor once it overcomes the competition from
the SDW state. Second, in 4-pocket case (but not in 5-pocket
case),  $s\pm$ order wins over SDW order even for perfect nesting,
if RG flow extends long enough, suggesting that SDW order is not a
necessary pre-condition for the $s\pm$ order. Our analytic results
are in full agreement with recent numerical functional RG studies
by Thomale et al. [arXiv:1002.3599v1]

\end{abstract}

\maketitle

\section{Introduction}

Since $2008$ a lot of efforts in condensed-matter community have
been devoted to  solve the puzzle of high-$T_c$ superconductivity
(SC) in newly discovered Fe-based superconductors. To a large
extent, in two years the community managed to obtain the data for
the pnictides in the amount comparable to that collected  for the
cuprates  over twenty years~\cite{review}.

The family of $Fe$-based superconductors is already large and
keeps growing. It includes  doped $1111$ systems $RFeAsO$
($R=$Rare earth element)\cite{bib:Hosono,bib:X Chen,bib:G
Chen,bib:ZA Ren}, doped $122$ systems $XFe_2As_2$(X=alkaline earth
metals)\cite{bib:Rotter,bib:Sasmal,bib:Ni}, as well as 111 and 11
systems like $LiFeAs$\cite{bib:Wang} and $FeTe/Se$
~\cite{bib:Mizuguchi}. The parent compounds of most of these
materials exhibit  a spin density wave (SDW) order \cite{Cruz},
and superconductivity emerges upon either hole or electron doping,
or upon gradual substitution of one pnictide by the other ($As$ by
$P$). In some systems, like $LiFeAs$\cite{bib:Wang} and
$LaFePO$\cite{bib:Kamihara}, SC was found already without doping,
instead of a magnetically ordered state.

ARPES\cite{Li}, de-Haas van Alphen oscillations
measurements~\cite{bib:Suchitra}, and first-principle calculations
~\cite{bib:first,Boeri,Mazin} all show that low-energy electronic
structure of pnictides in 2D basal plane consists of two  nearly
circular, non-equivalent hole pockets located at the center of the
Brillouin zone(BZ), and two symmetry-related elliptical electron
pockets, located near the corners of the BZ in the folded zone
scheme, or near $(0,\pi)$ and $(\pi,0)$ points, respectively, in
the unfolded zone scheme. [Folded and unfolded zones differ in
treating the pnictides -- folded zone takes into account the fact
that there are two non-equivalent positions of pnictides above and
below $Fe$ plane, and has two $Fe$ atoms in the unit cell, while
unfolded zone incorporates only $Fe$ atoms and has one $Fe$ atom
in the unit cell.] These hole and electron pockets form warped
cylinders in 3D space. In addition, in some
pnictides  there is a fifth cylindrical
hole pocket centered at $(\pi,\pi)$ in the unfolded zone (at
$(0,0)$ in the folded zone, like other two hole pockets), while in
other pnictides this fifth Fermi surface (FS) becomes a 3D sphere
centered near $k_z = \pi$ along $z-$direction.

A lot of work has been done over the last two years regarding the
symmetry of the order parameter and the interplay between SDW and
SC orders.  Most of researchers (but not all, see Ref.
\onlinecite{jap}) believe that the gap symmetry is extended
$s-$wave ($s\pm$), meaning the gap transforms according to
$A_{1g}$ representation of $D_{4h}$ tetragonal symmetry group, but
the average gap values along hole and electron Fermi surfaces have
different sign. However, the structure of the gap is still a
puzzle. Early works based on either spin-fluctuation scenario~
\cite{Mazin,Kuroki,Kuroki_2,Seo} or on renormalization group (RG)
study of a toy model of one hole and one electron
FS~\cite{Chubukov,Stanev} found a simple angle-independent $s\pm$
gap. Subsequent more sophisticated numerical studies, which take
into account multi-orbital nature of low-energy excitations in the
pnictides, have reported angular dependence of the $s\pm$ gap,
with $cos2\phi$ variations on the electron FSs and $cos4\phi$
variations along the hole
FSs~\cite{Kuroki_2,Graser,FWang,Rthomale,Thomale}. The $cos2\phi$
modulations of the $s\pm$ gap on the electron FSs has also been
obtained in the analytical study~\cite{bib:Chu_Vav_vor}.  If $\cos
2\phi$ variation is strong enough, the gap has ``accidental''
nodes along electron FSs, still preserving $s\pm$ symmetry.

Other recent theory proposals include $s^{++}$ state~\cite{jap},
$s\pm$ state with nodes on hole FSs due to strong $\cos 4 \phi$
modulations~\cite{peter}, and $s^{+-}$ state with
nodes at particular $k_z$  along
$z-$direction~\cite{peter_2}.

Experiments are generally consistent with $s\pm$ gap symmetry, but
whether or not the gap has nodes in particular materials is still
subject of debate
\cite{bib:Nakai,bib:Grafe,bib:YWang,bib:Millo,bib:Fletcher,bib:Hicks,
bib:Hashimoto1,reid,nakai_2,gordon,hashimoto_3,bib:Hashimoto2,bib:Malone}.
In addition, there is no information yet from the experiments
where the gap nodes are located, if present. ARPES measurements of
the gap along hole FSs (taken at fixed $k_z$) on various
$Fe-$pnictides~\cite{bib:Ding,borisenko,kaminski} indicate that
the gap is almost angle independent , but the detailed
measurements of the gap separately along each of the two electron
FSs  are still lacking, with only few exceptions~\cite{ding_2}.

From  theoretical perspective, the most relevant issue is the
nature of the pairing interaction. Conventional electron-phonon
coupling is always a candidate, particularly when the gap has
$s-$wave symmetry, but has been shown to be rather
weak~\cite{Boeri}, and is incapable to account for $T_c \sim 50K$
even if one neglects destructive effect of Coulomb interaction.
This leaves electron-electron interaction (i.e., dressed Coulomb
repulsion at a finite momentum transfer) as the dominant pairing
interaction. Such interaction cannot give rise to a constant
$s-$wave gap, but it can give rise to either momentum-dependent,
sign reversing gap along a given FS, like $d_{x^2-y^2}$ gap in the
cuprates, or to gaps of different signs along different FSs.
Coulomb interaction at large momentum transfer contributes to the
pairing both directly  and by creating effective pairing
interactions mediated by collective excitations in either spin or
charge channel.  The close proximity to magnetism makes spin
fluctuations a preferable candidate~\cite{Mazin,Kuroki,Graser},
although orbital fluctuations were also considered
recently~\cite{kont}. Both direct Coulomb interaction (the pair
hopping) and magnetically-mediated interaction are attractive for
$s\pm$ gap, and the total attractive pairing interaction is a
combination of the two.  Still, to give rise to a pairing, this
combined attractive pairing interaction has to overcome repulsion
coming from Coulomb interaction at small momentum transfers. A
conventional   McMillan-Tolmachev  renormalization~\cite{mcmillan}
does not help here because both repulsive and attractive parts of
the interaction renormalize in the same way, and if repulsive part
is initially stronger, the renormalization just reduces the
strength of the total repulsive interaction, but cannot change its
sign.

How to overcome Coulomb interaction became the major issue for
$s-$wave pairing in the pnictides. The situation is further
complicated by the fact that low-energy excitations are composed
of all 5 hybridized  $Fe$ 3d-orbitals, and Coulomb interactions
between fermions belonging to a given orbital and belonging to
different orbitals are equally important.  Only if all of these
interactions are approximated by the same momentum-independent
Hubbard $U$, this $U$ cancels out in the pairing problem and one
is left with a pure spin-mediated interaction (this was termed
``Coulomb avoidance''\cite{IMazin}). In reality, Coulomb
interaction is momentum-dependent and is larger at a small
momentum transfer than at a large momentum transfer,\ and
intra-orbital and inter-orbital interactions are also different.
As a result, direct Coulomb pairing interaction  is repulsive for
superconductivity with angle-independent plus-minus gap.
Spin-mediated interaction is attractive and can potentially
compete with direct Coulomb repulsion.  However, at least at
weak/moderate coupling   direct Coulomb repulsion is the largest
term. This holds even when magnetic correlation length $\xi$
diverges because for the pairing one generally needs the
interaction at non-zero frequencies, where it remains finite.

There are two possibilities to obtain $s\pm$ superconductivity
despite strong Coulomb repulsion. First, multi-orbital character
of excitations in the pnictides implies that the attractive
pairing interaction (either pair hopping or spin-fluctuation
exchange) is angle-dependent. The angle-dependence comes from
coherent factors which dress up the interactions when one
transforms from the orbital picture -in which different parts of
the Fermi surface are made of different orbitals- to the band
picture, in which free-fermion part is simply $\epsilon_k
c^\dagger_k c_k$, and all information about multi-orbital
character is passed onto interactions.  Once interaction is
angle-dependent, the gap  also becomes angle-dependent, and the
system adjusts the angle-dependence of the gap to minimize the
effect of Coulomb repulsion (we discuss this in more detail
below). The most natural is the case when the gap acquires $\pm
\cos 2\phi$ components along the two electron FSs, and the
magnitude of this component is adjusted to balance the interplay
between small $q$ Coulomb repulsion and the combined attractive
interaction in $s\pm$ channel. When Coulomb repulsion dominates,
the angle-dependent part is large, and the gap has four nodes
along each of the electron FSs.

Second, one can analyze how the interactions evolve as the system
flows towards smaller energies, relevant to superconductivity.
This flow involves renormalizations of interactions in both
particle-hole(p-h) and particle-particle(p-p) channel, and goes
beyond RPA. This flow has been studied numerically, using
functional renormalization group (fRG)
technique\cite{FWang,Rthomale, Thomale} and
analytically\cite{Chubukov,bib:Chu_physica,bib:Chu_Vav_vor},
within parquet RG. Both are weak-coupling studies, based on the
Hamiltonian which contains screened Coulomb interaction, but no
additional spin-fluctuation interaction. The results of both types
of studies are quite similar: it turns out that Coulomb repulsion
at small momentum transfers decreases upon system flow to smaller
energies, but the pairing interaction at large momentum transfers
(the pair-hopping from hole to electron FSs), which is attractive
for $s^{+-}$ superconductivity increases. The increase of the pair
hopping is the result of the ``push'' from the inter-pocket
density-density interaction which by itself leads to SDW
instability. If RG flow of the couplings persists over a wide
enough range of energies, pair-hopping interaction exceeds Coulomb
repulsion, and the system develops an attraction in the $s\pm$
channel. In this situation, the $\pm cos 2\phi$ variations of the
gap on electron FSs induced by angle dependence of the
interaction, are not crucial for the pairing, and the system can
develop an $s\pm$ gap without nodes.

While this scenario is quite generic, the earlier parquet RG
study\cite{Chubukov,bib:Chu_physica} was more limited in scope. It
was done for a toy model of one hole and one electron FS centered
at $(0,0)$ and $(\pi,\pi)$, respectively.  For such a model, angle
dependencies of the interactions must be symmetric with respect to
interchanging $x$ and $y$ momentum components both near $k=(0,0)$
and $k = (\pi,\pi)$  and can only be in the form $\cos 4 \phi$,
$\cos 8 \phi$, etc which are subleading terms in the expansion of
$A_{1g}$ gap for a single FS (no $\cos 2 \phi$ terms!). Such $\cos
4 n \phi$ terms are generally irrelevant and were neglected in toy
model analysis, i,e. the gap along each of the FSs was
approximated by a constant.  Like we said, for
momentum-independent gaps, the bare interaction in the $s^{+-}$
channel is repulsive if intra-pocket Coulomb repulsion is the
largest.  The interaction flows under RG and changes sign at some
value of RG parameter. Still, all along parquet RG flow, the
pairing interaction remains secondary to the interaction in the
SDW channel. As a result, for perfect nesting the system develops
an SDW order. Only when the system is doped and the logarithmical
flow of the SDW vertex is cut at low energies, SC channel takes
over and the system develops an $s^{+-}$ superconductivity.

In the present paper we extend earlier parquet RG analysis to
multi-pocket models of Fe-pnictides. We consider two models.  The
first one has  two electron FSs, located at $(0,\pi)$ and
$(\pi,0)$ in the unfolded zone, and two hole FSs located near
$(0,0)$. The second model has an additional hole pocket centered
at $(\pi,\pi)$ in the folded zone. The presence or absence of this
additional FS in different Fe-pnictides is
attributed~\cite{Kuroki,Thomale} to the difference in the distance
between the pnictide (e.g., $As$ or $P$) and the Fe-plane.  To
avoid overly complicated analysis of RG equations we only consider
two limiting cases: one when the  two hole pockets centered at
$(0,0)$ are completely equivalent, and the other when one pocket
is  coupled to electron FSs much weaker than the other one  and
can be neglected [this effectively reduces 4 pocket model to 3
pockets and 5-pocket model to 4 pockets]. The system behavior is
identical in the two limits which gives us confidence that it
remains the same also in between the limits.  Throughout the paper
we assume that all FSs are cylindrical and neglect their
variations along $k_z$.

We argue that new physics:  nodal $s^{\pm}$ SC,  appearance of SC
even at perfect nesting, emerges once one extends the model from 2
to 4 pocket (or to 3-pocket when only one hole FS is relevant).
This result is consistent with early
assertions\cite{Chubukov,bib:Chu_Vav_vor} that the 3-pocket model
is the minimum model needed to understand  the key physics of
$Fe-$pnictides. The behavior of 5 pocket model, on the contrary,
is in many respects similar to that for 2-pocket model (except
that nodal $s^{+-}$ SC is still possible).  That 4-pocket and
5-pocket models behave differently under RG  has been the
conclusion of fRG study by Thomale {\it et al}
(Ref.\onlinecite{Thomale}), and our results fully agree with their
analysis.

We extend previous 2-pocket study in two directions. First, we
incorporate $\cos 2 \phi$ angular dependence of the interactions
which give rise to $\pm \cos 2\phi$ modulations of the gaps on
electron FSs. In this situation, there are at least three
different effective vertices for $s\pm$ gap symmetry, and we argue
that, if the dominant angle dependence comes from electron-hole
interaction,  one of them remains positive (i.e., attractive) over
the entire RG flow even when bare intra-pocket repulsion is the
largest interaction.  We show that the stronger is the angular
dependence of the interaction, the stronger is the tendency to
develop an nodal $s\pm$ order.

Second, we re-analyze the interplay between SC and SDW channels.
For angular-dependent interactions, are also several SDW vertices
of which at least one is attractive along the whole RG trajectory.
We compare the flow of the leading vertices in SDW and SC channel.
We show that in 4-pocket model, the trajectory of the leading SC
vertex are steeper than that of the leading SDW vertex, and the
latter remains larger  only down to some RG scale. At smaller
scales (i.e., when RG flow extends further to lower energies) the
SC vertex overshadows SDW vertex even at perfect nesting.  This
agrees with  fRG study by Thomale \emph{et .al.}\cite{Thomale}. We
argue, based on our analytic consideration, that the crossing
between SC and SDW vertices under RG flow is to a large extent a
combinatoric effect -- compared to 2-pocket case (where SDW and SC
vertices flow to the same value under RG), the presence of the two
electron FSs adds the factor of 2 to the renormalization of the SC
vertex as the pair hopping can, e.g., hop a pair of $k,-k$
fermions from the hole FS to each of the two electron FSs, but
momentum conservation does not allow such factor of 2 to appear in
the renormalization of the SDW vertex.

We further find that for 5-pocket model, SC vertex always remains
secondary compared to SDW vertex, just like in 2-pocket model.
Furthermore, like in 2-pocket model,  SC and SDW vertices flow to
the same value at the fixed point. This again agrees
 with fRG result by  Thomale \emph{et. al.}.

The rest of the paper is organized as follows: Section
\ref{section:Model} outlines our approach, explains the subtleties
in the RG flow, and discusses the technique by which we
incorporate the momentum dependence into our analysis. In Sec.
\ref{section:2Bmodel} we briefly review the results for the
2-pocket case. Section \ref{section:3Bmodel} is the central
section in the paper -- here we consider in detail the 4-pocket
model. We discuss SC vertices in the presence of angular
dependence of the interaction, the RG flow,  and   the competition
between SC and SDW instabilities. Most of our treatment in this
section is for the limit when one of the hole FS can be
neglected.Later in the section we analyze another limit when  the
two hole FS are equivalent and show that the system behavior is
identical in the two limits. In Sec. \ref{section:4Bmodel} we
discuss 5-pocket model and argue that its behavior to a large
extent is similar to that in 2-pocket model. We summarize our
results in the Conclusion.  \\

\section{Discussions}\label{section:Model}

Before starting the detailed description of each model, we present
a brief discussion on the outline of our approach. We first
describe the central idea behind our analysis, then describe the
technique through which we incorporate the angular dependence of
the vertices and finally discuss how we incorporate these into RG
equations.

\subsection{Approach}

We consider four-fermion interactions between fermions located
close to the FSs of two or more pockets.  We consider  Hamiltonian
with all possible quartic interactions allowed by symmetry and ask
what can be said about the onset of SC, SDW, and, possibly, CDW
instabilities. The usual approach is to write down equations for
effective vertices $\Gamma_i$ in SDW, SC, CDW channels and check
for the existence of critical temperatures $T^{(i)}_{ins}$ at
which $\Gamma_i$ diverges. In case of competing instabilities, the
equations for the effective vertices are  coupled and, once the
coupled system is solved, the instability with the highest
$T_{ins}(T_c)$ takes over.

As we said in the Introduction, when one does such an analysis,
one deals with interactions taken at the scale of $T_c$, which are
not the same as the terms in the Hamiltonian. To account for the
flow of the couplings from the scale of the bandwidth down to
$T_c$, we need RG analysis. This analysis  assumes
renormalizability of the theory and can be rigorously justified
only when the  RG flow is logarithmical (i.e., interactions vary
as functions of the logarithm of the running scale $E$). One
well-known example of logarithmical RG flow is the renormalization
in the particle-particle channel (Cooper renormalization).
Another, specific to our case, is the renormalization in the
particle-hole channel, involving intermediate fermions from hole
and electron pockets. Because hole and electron dispersions are of
opposite sign, such a renormalization also generates logarithmical
dependence of the running energy and/or  momentum as long as the
running energy exceeds the sum of energies of the top of the hole
band and the bottom of the electron band (hole and electron masses
do not have to be equal).\

The logarithmical renormalizations in the particle-particle and
particle-hole channels are characterized by corresponding
polarization bubbles. Let $c$ describe a hole band centered at
$k=0$ and $f$ describe an electron band centered at $Q$. Assume
for simplicity that hole and electron masses are equal. For a
perfect nesting, hole and electron dispersions obey $\varepsilon_c
(k) = \epsilon_0 - k^2/(2m_h) =- \varepsilon_f (k+Q)$. The two
logarithmically singular polarization bubbles are

\bea \label{eq:bubbles}
\Pi_{pp}^{cc}(q,\Omega)  &=& \Pi_{pp}^{ff}(q,\Omega) \nonumber\\
&=& \int \frac{d^2 k\;d\omega}{(2\pi)^3}
G^c(k,\omega)G^c(q-k,\Omega-\omega) \nonumber\\
&=& \frac{m}{2\pi}L  + ...\nonumber\\
\Pi_{ph}^{cf}(q+Q,\Omega) &=&\int \frac{d^2 k\;d\omega}{(2\pi)^3}
G^c(k,\omega)G^f(q+Q+k,\Omega+\omega)
\nonumber\\
&=& -\frac{m}{2\pi}L  + ...\nonumber\\
L&=& \frac{1}{2}ln\left( \frac{\Lambda}{E} \right) \eea

where the dots stand for non-logarithmic terms, $E = max\{\Omega,
v_F q\}$ and $E>E_F$,  $\Lambda$ is the upper cutoff of order
bandwidth, and the propagators are given by $G^x =
\frac{1}{i\omega - \epsilon^x_k}$, $x$ being $c$ or $f$.

\begin{figure}[htp]
\includegraphics[width=3.5in]{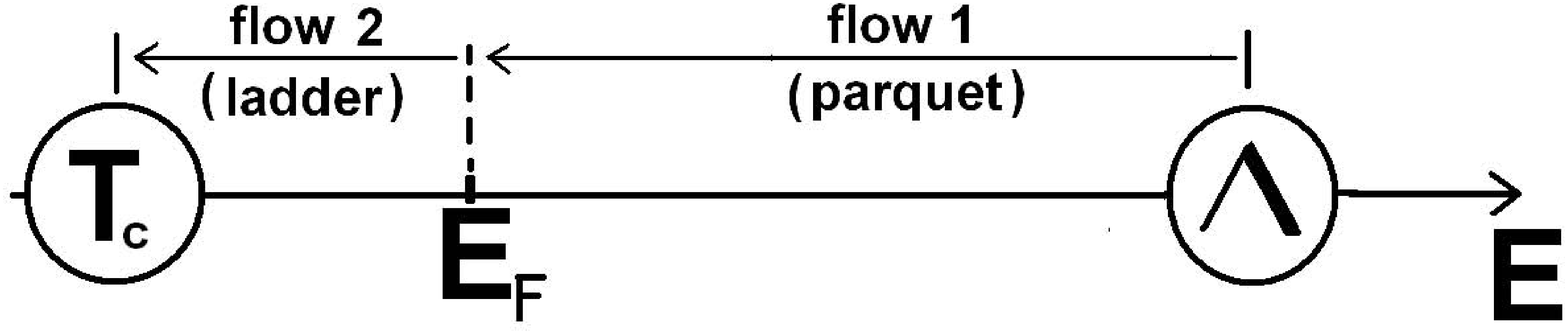}
\caption{\label{fig:flow_schematic} Illustration showing how the
couplings evolve under RG flow. The bare couplings (the parameters
of the Hamiltonian) are defined at energies comparable to the
bandwidth $\Lambda$.  In pnictides, this scale is $2-3 eV$, much
larger than the Fermi energy $E_F \sim 0.1 eV$ SC and SDW
instabilities likely  come from even smaller energies because
instability temperatures are at least order of magnitude smaller
than $E_F$. The couplings vary as one integrates out higher
energies. This variation (i.e., the flow of the couplings from
$\Lambda$ down to the running scale $E$) can be generally captured
by the RG technique. In pnictides, the flow is different above and
below $E_F$. Above $E_F$, each of the couplings changes because of
integrating out higher energy fermions in both particle-hole and
particle-particle channels. The RG equations in this region are
called parquet RG, because renormalizations extend in the two
directions. Below $E_F$, each vertex continue to flow due to
renormalizations in only one channel, either particle-hole or
particle-particle, depending on the external momenta. The RG
equations in this region is called ladder RG,  because
renormalizations extend only in one direction. }
\end{figure}

The RG study requires caution as the couplings flow differently
for energy scales above $E_F$ and below $E_F$. The reasoning is
simple: logarithmical RG analysis requires that internal momenta
in each diagram for vertex renormalization be larger than external
momenta, which are of order $k_F$. When typical internal energies
are larger than $E_F$, internal momenta are larger than $k_F$, and
vertex corrections in both particle-particle and particle-hole
channel are logarithmic. This gives rise to parquet RG. When
typical energies are smaller than $E_F$,  the strength of the
renormalization in the particular channel depends on the interplay
between external momenta.  When total incoming momenta is zero,
renormalization in the particle-particle channel is still
logarithmically singular, but in the renormalization of the
particle-hole channel, the logarithm is cut by external $E_F$.
Conversely, for the vertex with transferred  momentum equal to the
distance between hole and electron FSs, the renormalization in the
particle-hole channel is still logarithmical, but in the
renormalization in the particle-particle channel, the logarithm is
now cut by external $E_F$. As a result, the renormalizations in
the particle-particle and particle-hole channels are coupled at
energies above $E_F$, but become decoupled at energies below
$E_F$. At $E < E_F$ parquet RG equations are replaced by
conventional ladder RG equations $d{\Gamma}_i/dl =\Gamma^2_i$,
where $l = \log{E_F/E}$. Thus the flow of the couplings splits
into the flow from the bandwidth down to $E_F$, where different
vertices are all coupled, and the flow below $E_F$, where
different vertices are decoupled (see Fig.
\ref{fig:flow_schematic}). This reasoning is particularly
important in our case, as in pnictides $E_F \sim 100 meV$ is much
smaller than the bandwidth, which is a few electron volts.

\begin{figure*}[t]
$\begin{array}{cccc}
\includegraphics[width=1.8in]{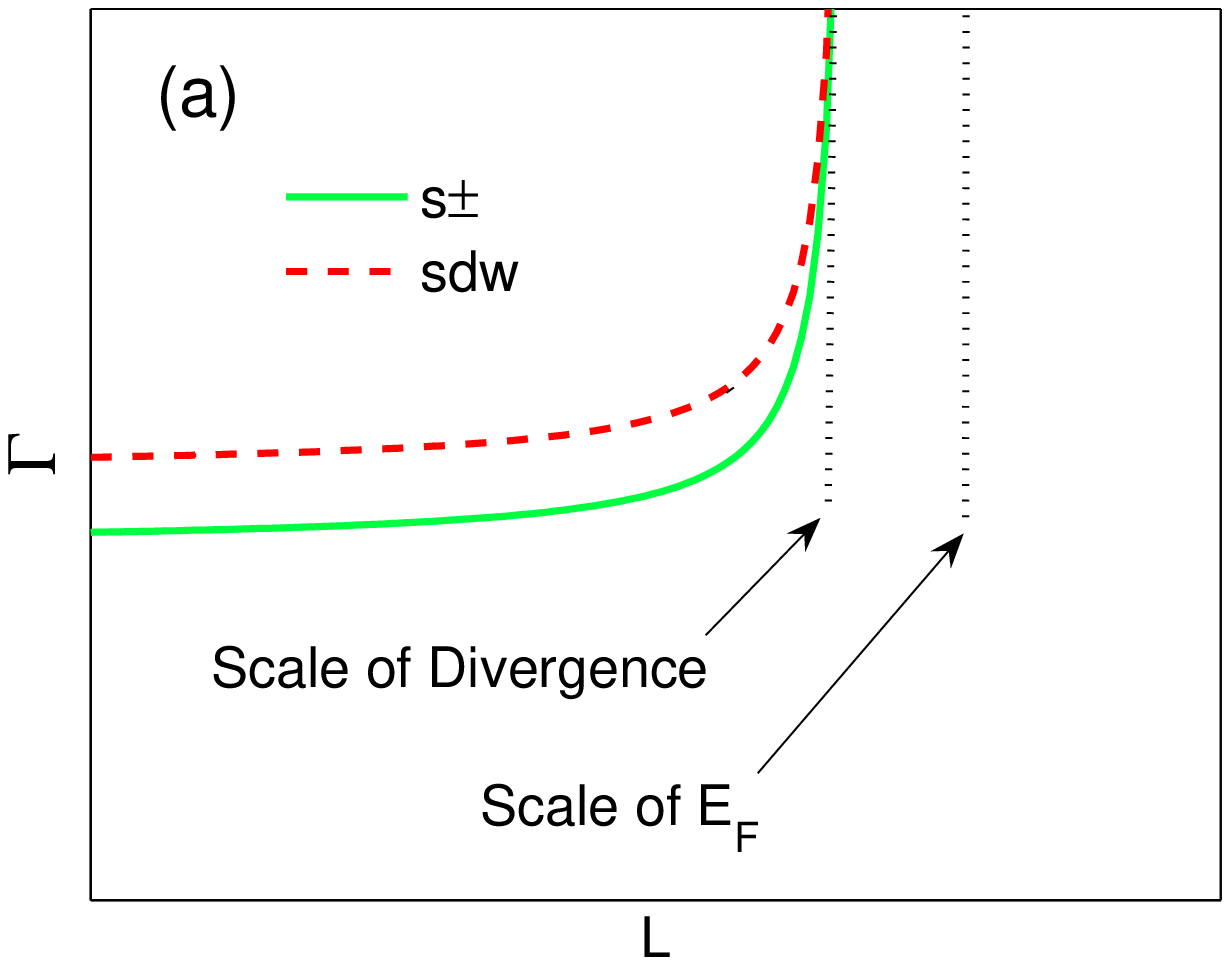}&
\includegraphics[width=1.8in]{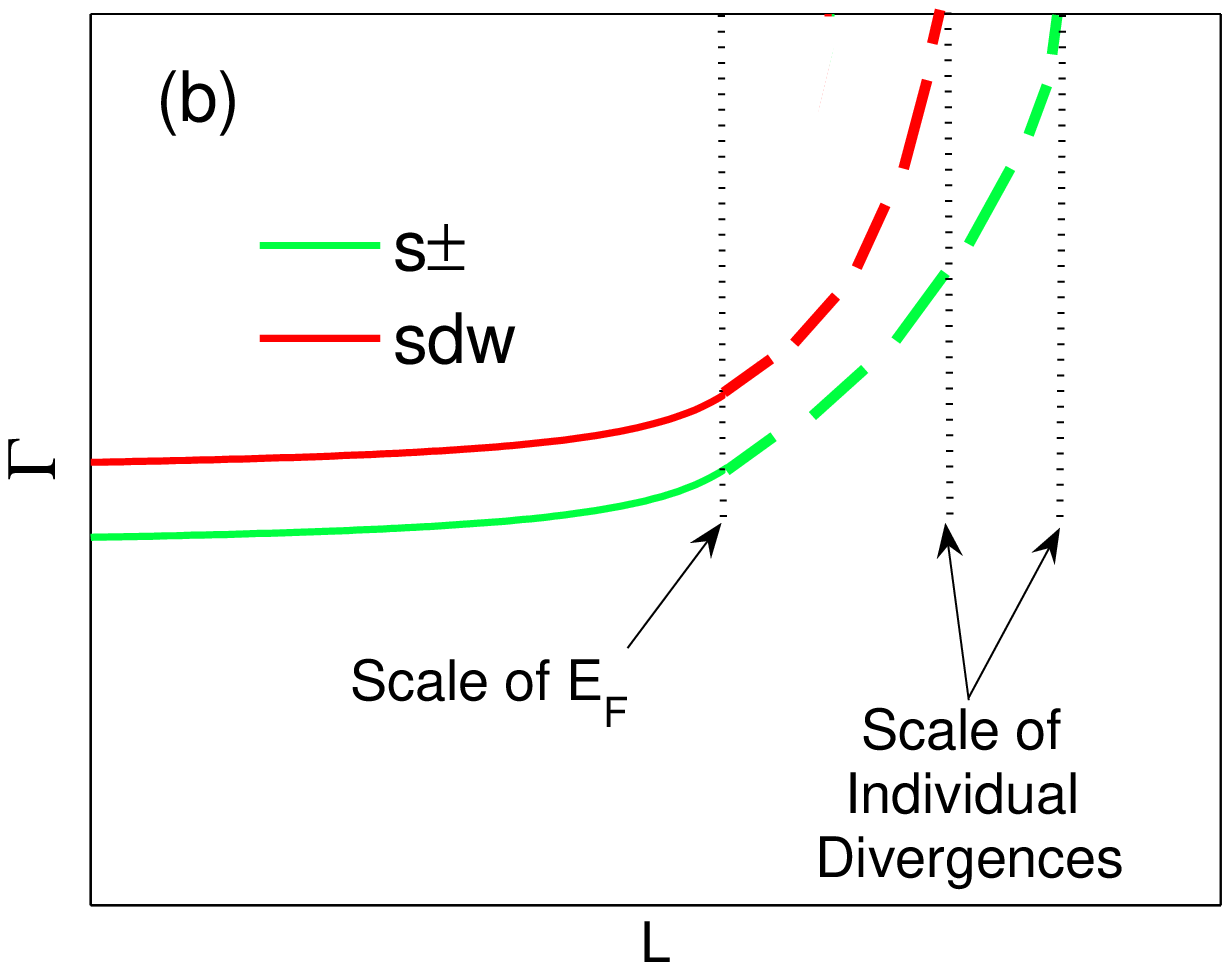}&
\includegraphics[width=1.8in]{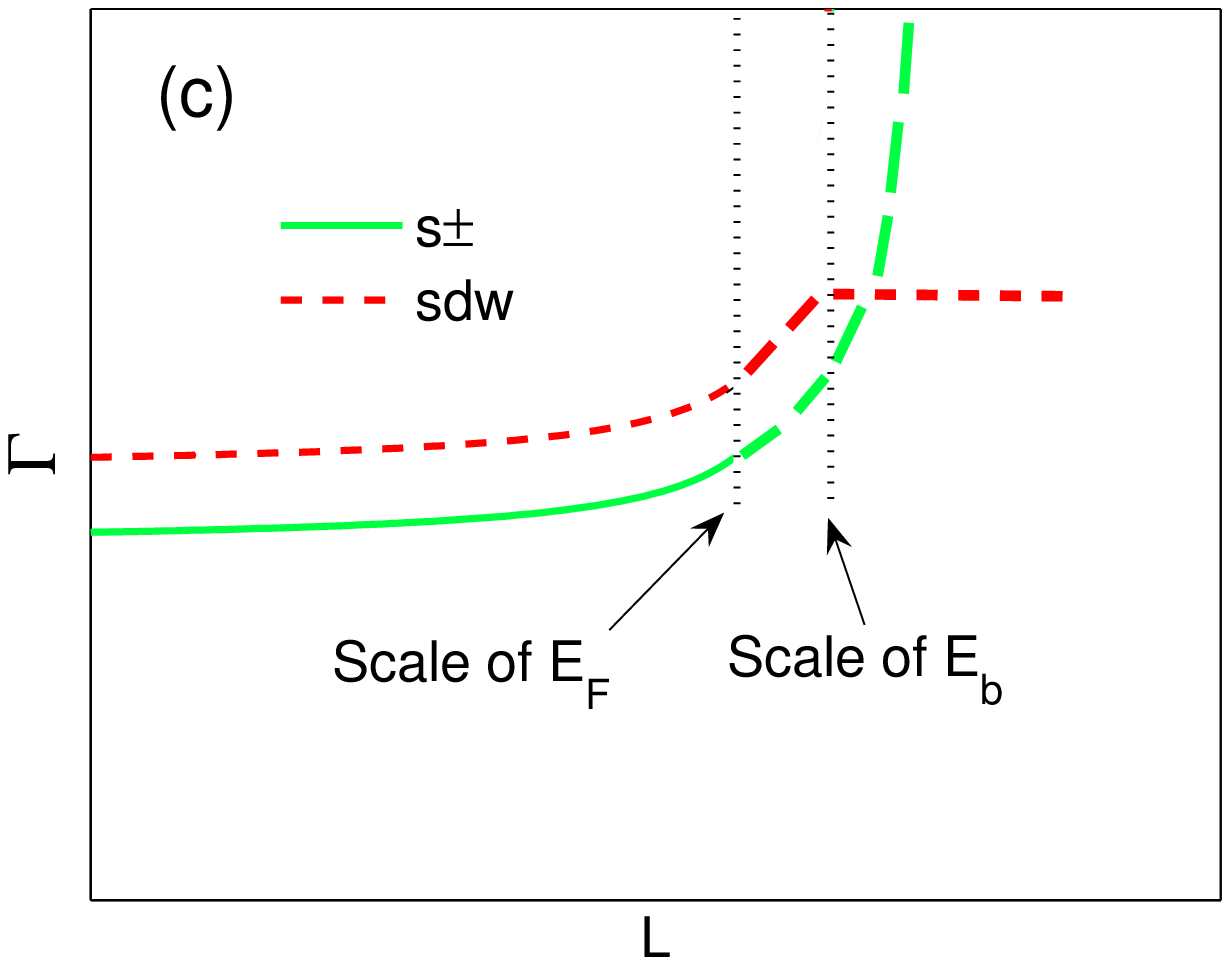}&
\includegraphics[width=1.8in]{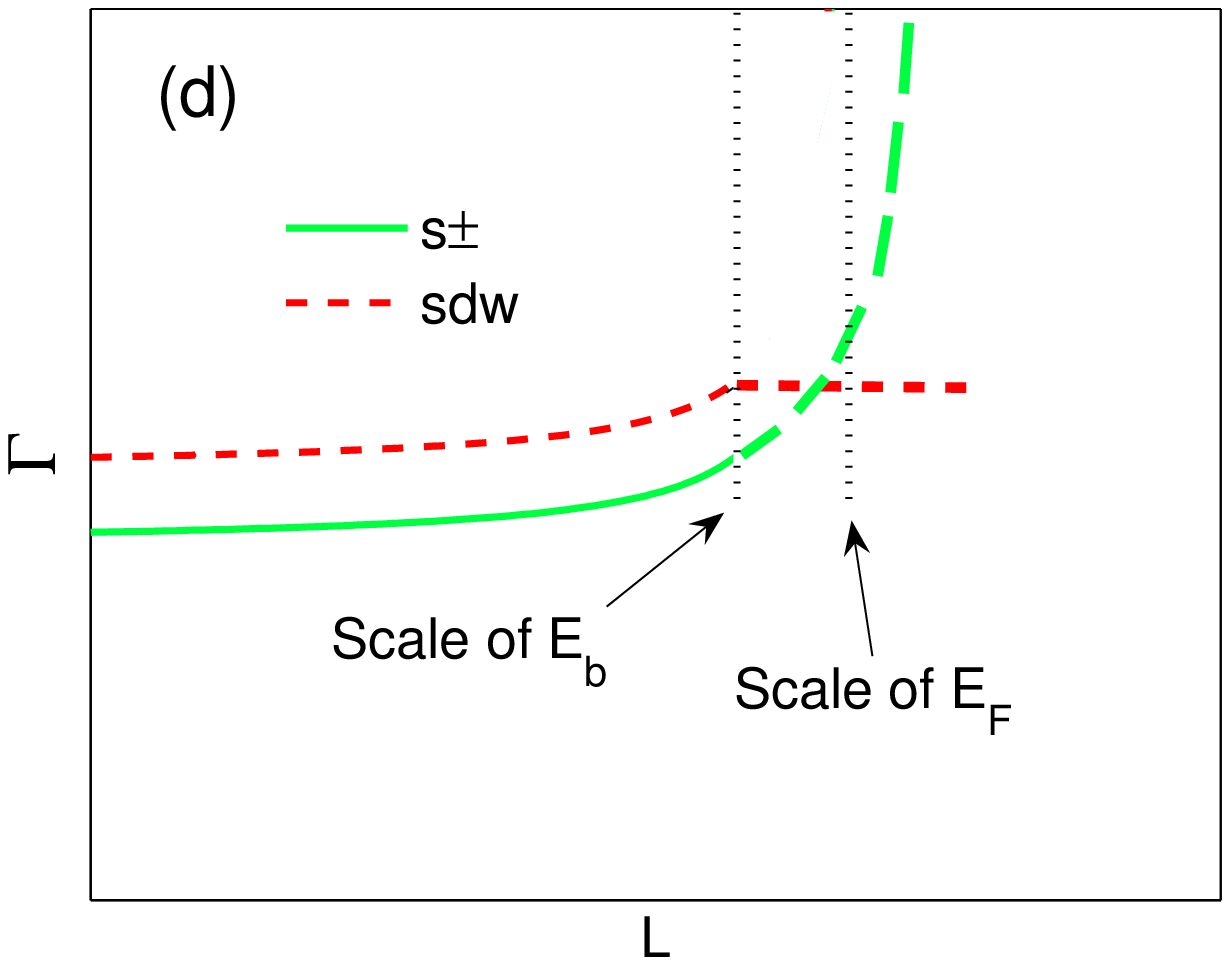}
\end{array}$
\caption{\label{fig:gamma_schem} Schematics of the RG flow of SC
and SDW vertices $\Gamma^{SDW}$ and $\Gamma^{s\pm}$. The
horizontal scale is $L=\frac{1}{2}ln\frac{\Lambda}{E}$ for $E>E_F$
and $\frac{1}{2}ln\frac{\Lambda}{E_F}+ln\frac{E_F}{E}$ for
$E<E_F$. SC and SDW vertices remain coupled at energies larger
than $E_F$ but decouple below $E_F$. Depending on the bare values
of the vertices, the ratio $\Lambda/E_F$, and the doping, four
different scenarios are possible (panels (a)-(d)). For perfect
nesting, there are two possibilities: (a) the vertices diverge at
the same scale before the scale of $E_F$ is reached. The ratio of
the vertices not necessarily tends to one, though. (b) $E_F$ is
reached before the vertices diverge. Then, below $E_F$, the
vertices decouple, flow and diverge independently, each on its own
scale. The vertex that had larger value at $E_F$ diverges first
and sets the instability. For non-perfect nesting (e.g., at a
finite doping), SDW vertex eventually does not diverge. SC vertex
still diverges, and the system becomes a SC even if SC instability
was subleading at perfect nesting. The flow of the SDW vertex
levels off either at $E_b < E_F$ (panel c), or at $E_b> E_F$
(panel d).}
\end{figure*}

Depending on the character of the flow, the bare values of the
couplings, and the ratio of the bandwidth and the Fermi energy,
several situations are possible and shown in  Fig.
\ref{fig:gamma_schem}:

\begin{itemize}
\item
The RG flow diverges and the normal state  becomes unstable before
the scale of $E_F$ is reached (Fig. \ref{fig:gamma_schem}a). In
this situation, the instability is reached already within  parquet
RG. This situation is the most interesting one from physics
perspective particularly because in several cases  different
vertices diverge simultaneously, and the fixed point has an
enhanced symmetry  (e.g., in the 2-band model, the vertices in
SDW, SC, and orbital CDW channels all diverge at the fixed point
which then has $O(6)$ symmetry\cite{bib:Chu_physica,podolsky}).
The instability at energies above $E_F$ is, however, unlikely
scenario for the pnictides  because the largest instability
temperature is only a fraction of $E_F$.
\item
The RG flow  reaches $E_F$ before couplings diverge (Fig.
\ref{fig:gamma_schem}b). In this situation, parquet RG creates a
hierarchy of the couplings at $E_F$: $\Gamma_i (E_F)$. Below
$E_F$, different $\Gamma_i$ decouple and, for a perfect nesting,
each continue evolving according to a ladder RG, i.e., like
$\Gamma_i (E) = \Gamma_i (E_F)/(1 - \Gamma_i (E_F) \log
\frac{E_F}{E})$. The instability occurs at the energy
(temperature) at which  $\Gamma_i (E_F) \log \frac{E_F}{E} =1$.
Obviously, the  winning channel is the one in which the coupling
is the largest at $E_F$.
\item
When nesting is not perfect (i.e., energies of the top of the hole
band and of the bottom of the electron band are not exactly
opposite), the coupling in the SC channel continues to follow
$\Gamma^{SC} (E) = \Gamma^{SC} (E_F)/(1 - \Gamma^{SC} (E_F) \log
\frac{E_F}{E})$ simply because SC instability involves pairs of
fermions with ${\bf k}$ and $-{\bf k}$ from the same FS and is
non-sensitive to a deviation from perfect nesting.  However the
logarithmic flow of SDW and CDW vertices is now cut at some scale
$E_b$. Suppose $E_b < E_F$ (Fig. \ref{fig:gamma_schem}c). In this
situation, SC eventually wins over SDW and CDW instabilities, even
if superconducting $\Gamma$ is subleading at $E_F$ (but it needs
to be attractive at $E_F$).
\item
When $E_b$ exceeds $E_F$ (Fig. \ref{fig:gamma_schem}d)
particle-hole and particle-particle channels decouple already
within the applicability range of parquet RG. At $E < E_b$, SC
vertex continue to grow as $\Gamma^{SC} (E) = \Gamma^{SC} (E_b)/(1
- \Gamma^{SC} (E_b) \log {E_b/E})$, while vertices in density-wave
channels get frozen at their values at $E \sim E_b$. In this
situation, SC instability again wins, even if it was subleading at
$E_b$, provided that  the superconducting vertex is attractive at
$E_b$.
\end{itemize}

A subtle point: below we will be presenting the RG flows of
vertices and couplings at energies both above and below $E_F$ in
terms of the logarithmic variable $L \sim \log{E}$. One has to
bear in mind, however, that the  prefactor for the logarithm
actually  changes between $E > E_F$ and $E < E_F$ because for $E >
E_F$ the integration over intermediate energies involves only
positive excitations for electron states (and negative for hole
state), i.e., $\int \frac{d^2k}{(2\pi)^2} = (m/2\pi)
\int_E^\Lambda d \epsilon_k$, while for $E < E_F$, one has to
linearize the dispersions of holes and electrons near the FS and
integrate on both sides of $E_F$, i.e.,  $\int
\frac{d^2k}{(2\pi)^2} = (m/\pi) \int_E^{E_F} d \epsilon_k$. To
simplify the  presentation we just define
\[ L=\left\{
\begin{array}{cc}
\frac{1}{2}ln\frac{\Lambda}{E} &,E>E_F \\
\frac{1}{2}ln\frac{\Lambda}{E_F}+ln\frac{E_F}{E}&,E<E_F
\end{array}\right. \label{eq:Ldef}
\]
and use  the same symbol $L$ for all energies. This is valid as
long as we describe the system behavior at $E \gg E_F$ and $E \ll
E_F$. The behavior in the crossover regime $E \sim E_F$ is more
complex,  but this is beyond the scope of the present paper. We
will also use $L_{E_F}\equiv\frac{1}{2}ln\frac{\Lambda}{E_F}$.

\subsection{Incorporating angular dependence}

As we said in the Introduction, multi-orbital character of
low-energy excitations in Fe-pnictides implies that interactions
between fermions located near hole or electron FSs depend on the
angles along the FSs. To obtain angular dependence of various
couplings from first principles, one has to transform from
five-orbital to five-band picture and dress up Coulomb
interactions by coherence factors.  This has been done in several
studies (see, e.g., Refs. \onlinecite{Singh,Graser,Rthomale}),
under the assumption that the Coulomb interaction is so strongly
screened that it can be replaced by  Hubbard $U$. This assumption
is generally valid for systems with large FSs, because of many
available particle-hole pairs for screening, but in systems with
small FSs, like pnictides, much fewer number of particle-hole
pairs are available, and the screening of Coulomb interaction is
much weaker, and is actually a rather non-trivial phenomenon at
small $k_F$~\cite{new} Because of this complication, the
``first-principle'' analysis of the angular dependence of the
interaction is a rather difficult task.

One can, however, attempt to extract these angular dependence from
symmetry considerations, like it has been done for the
cuprates~\cite{golden} This is what we will do. Consider first the
pairing vertex between fermions with incoming momenta $k$ and $-k$
and outgoing momenta $p$ and $-p$. Quite generally, for tetragonal
$D_{4h}$ symmetry group, this interaction can be divided into
one-and two-dimensional representation, and one-dimensional
representation can be further divided into $A_{1g}$, $B_{1g}$,
$B_{2g}$, and $A_{2g}$ harmonics, depending on the symmetry under
the transformations under $k_{x,y} \to -k_{x,y}$ and $k_x \to
k_y$. Basic functions from different representations do not mix,
but each contains infinite number of components.    $s-$wave
pairing corresponds to  fully symmetric $A_{1g}$ representation.
The $s-$wave pairing interaction can be quite generally expressed
as

\beq u(k,p)=\sum_{m,n}A_{mn}\Psi_m(k)\Psi_n(p) \label{s_1} \eeq

where $\Psi_m (k)$ are the basis functions of the A$_{1g}$ symmetry
group: $1$, $cosk_x cosk_y$, $cosk_x + cosk_y$, etc, and $A_{mn}$
are coefficients. Suppose for definiteness that $k$ belongs to
hole FS and is close to $k=0$. Expanding {\it any} wave function
with A$_{1g}$ symmetry near $k=0$, one obtains along $|{\bf k}| =
k_F$,

\beq \Psi_m(k) = a_m + b_m \cos 4 \phi_k + c_m \cos 8 \phi_k +...
\label{s_2} \eeq

If ${\bf p}$ is near the same hole FS, the expansion of
$\Psi_n(p)$ also involves $\cos 4 \phi_p,~\cos 8\phi_p$, etc.
There are no fundamental reasons  to expect that $b_m,~c_m$, etc
are much smaller than $a_m$, but sub-leading terms are often small
numerically. Two known examples are the numerical smallness of
$\cos 6 \phi$, etc components of the $d_{x^2-y^2}-$wave gap for
spin-fluctuation mediated pairing in the cuprates~\cite{pines} and
the numerical smallness of $\cos 4 \phi$, etc components of the
gap along the hole FSs in fRG\cite{Thomale} and
RPA\cite{Graser,bib:unp2} calculations for 5-band Hubbard-type
model for the pnictides. Taking these examples as circumstantial
evidence, we assume  that $\cos 4\phi$, etc terms are small.  If
so,  the interaction between fermions belonging to the hole FS can
be approximated by angle-independent term.

The situation changes, however, when we consider pairing
interaction between fermions belonging to different FSs. Suppose
that $k$ are still near the center of the Brillouin zone, but $p$
are near one of the electron FSs, say the one centered at
$(0,\pi)$.  Consider all possible $\Psi_n(p)$ with $A_{1g}$
symmetry. A simple experimentation with trigonometry shows that
there are two different subsets of basic functions:

\bea
&&A: 1,~ \cos{p_x} \cos{p_y},~ \cos{2 p_x} + \cos{2p_y} ... \nonumber \\
&&{\bar A}: \cos{p_x} + \cos{p_y},~\cos{3 p_x} + \cos{3 p_y}...
\label{s_3} \eea

Functions from class A have the same properties as before -- they
can be expanded in series of $\cos 4 l \phi_p$ ($l$ is integer).
Functions from class ${\bar A}$ are different -- they all vanish
at $(0,\pi)$ and are expanded in series of $\cos (2\phi_p + 4 l
\phi_p)$ (i.e., the first term is $\cos 2 \phi_p$, the second is
$~\cos 6 \phi_p$, etc).  Let's make the same approximation as
before and neglect all terms with $l >0$. The functions from class
A can then be approximated by a constant, but the functions from
class ${\bar A}$ are approximated by $\cos 2 \phi_p$. As a result,
$s-$wave pairing interaction involving fermions from hole and one
of the two electron FSs (labeled as $e_1$) has a generic form of

\bea u_{e_1,h}(k,p) &=& u_{e_1,h} + {\bar u}_{e_1,h}\cos 2
\phi_{p_{e1}} +
.... \nonumber\\
&=& u_{e_1,h} \left(1 + 2\alpha \cos 2 \phi_{p_{e1}}\right) + ...
\label{s_4} \eea

where dots stand for $\cos{4 \phi_k}, \cos{4\phi_p},
\cos{6\phi_p}$, etc terms. We emphasize that the constant term
and the $\cos 2 \phi_p$ term in (\ref{s_4}) are the leading terms
of the two subsets of interaction terms, each form series in $\cos
4 \phi_{k,p}$. By the same reasoning, the interaction between
fermions near two electron FSs centered at $(0,\pi)$ and $(\pi,0)$
is expressed as

\bea
u_{e_1,e_2}(k,p) &\sim& u_{e_1,e_2} \left(1+2\alpha'
\left(\cos2\phi_{k_{e1}}+
\cos2\phi_{p_{e2}}\right)\right. \nonumber \\
&&+ 4 \alpha^{''} \cos2\phi_{k_{e1}} \cos2\phi_{p_{e2}} +...
\label{s_5} \eea

Observe also that the $\cos 2 \phi$ terms in (\ref{s_4}) and
(\ref{s_5}) change sign under the  transformation $x\to y$ (like
$d_{x^2-y^2}$ interaction in the cuprates), hence the prefactor
for $\cos {2 \phi_p}$ term in (\ref{s_4})  changes sign between
the two electron FSs [$\cos {2 \phi_{p_{e1}}} \to - \cos{2
\phi_{p_{e2}}}$].

The pairing interaction in the form of Eq. (\ref{s_4}) has been
introduced in Ref. \onlinecite{bib:Chu_Vav_vor}.  The authors of
Ref. \onlinecite{bib:Chu_Vav_vor}, however, didn't include into
consideration the fact that band description is obtained from
multi-orbital description and argued that $\alpha$ must generally
scale as $k^2_F$ and should be small when $k_F$ is small.  In
fact,  the angular dependence produced by the coherent factors
associated with the hybridization of 5 $Fe$ bands are not small in
$k_F$ (Refs. \onlinecite{Singh,Graser}), hence $\alpha$, $\alpha',
\alpha^{''}$  \emph{do not have to be small}. Accordingly, we will
keep $\alpha$'s as just parameters.

Once the pairing interaction has the form of Eqs. (\ref{s_4}) and
(\ref{s_5}), the gap along hole FS is still angle-independent, but
the gaps along the two electron FSs are in the form  $\Delta_e \pm
\bar{\Delta}_e \cos2\phi$. When $\bar{\Delta}_e$ is small compared
to $\Delta_e$, the gaps on electron FSs are nearly angle-independent, but when
$|\bar{\Delta}_e| > |\Delta_e|$, they
have nodes at ``accidental'' values of $\phi$.

\subsection{The RG analysis with angle-dependent interactions}

The angular dependence of the interaction is the key element of fRG
approach, and this approach uses the full momentum dependence of
the interactions, i.e., the full series of $\cos {4l\phi}$ and
$\cos (2\phi + 4 l \phi)$ terms.  At the same time, fRG approach
assumes renormalizability (i.e., that the right-hand side of each fRG
equation contains only renormalized couplings, not some combinations of
bare and renormalized couplings).  In our analytical approach, we
do calculations within logarithmic approximation in which case we explicitly
preserve renormalizability.

We found, after explicitly evaluating the renormalizations of
angle-dependent vertices,  that the only way to justify RG
procedure in this situation is to keep the angular dependence away
from RG flow, i.e.  allow $u_{e_1,e_2}$, $u_{e1,h}$, and
$u_{e2,h}$  to flow under RG, while $\alpha$, $\alpha'$ and
$\alpha^{''}$  are kept unchanged. This can be rigorously
justified when $\alpha$'s are small and all terms of order
$\alpha^2$ are neglected. We will  assume without proof that the
results of RG analysis  are valid even when $\alpha \leq 1$.
There are no new physical effects at $\alpha \leq 1$ compared to
$\alpha <<1$, so the results at $\alpha \leq 1$ should be at least
qualitatively correct by continuity.

As we said in the Introduction, the main goal of our analysis is
to understand whether there are qualitative differences between RG
flow and the pairing in 2, 4, and 5 pocket models for
$Fe-$pnictides. The next three sections deal with the comparison
of 2,4, and 5-pocket models.\\


\section{The 2 pocket model}\label{section:2Bmodel}

This has been studied before ( Refs.
\onlinecite{Chubukov,bib:Chu_physica}) and we briefly review it
here to set notations and for further comparisons with multi-
pocket cases.

\begin{figure}[htp]
\includegraphics[width=2in]{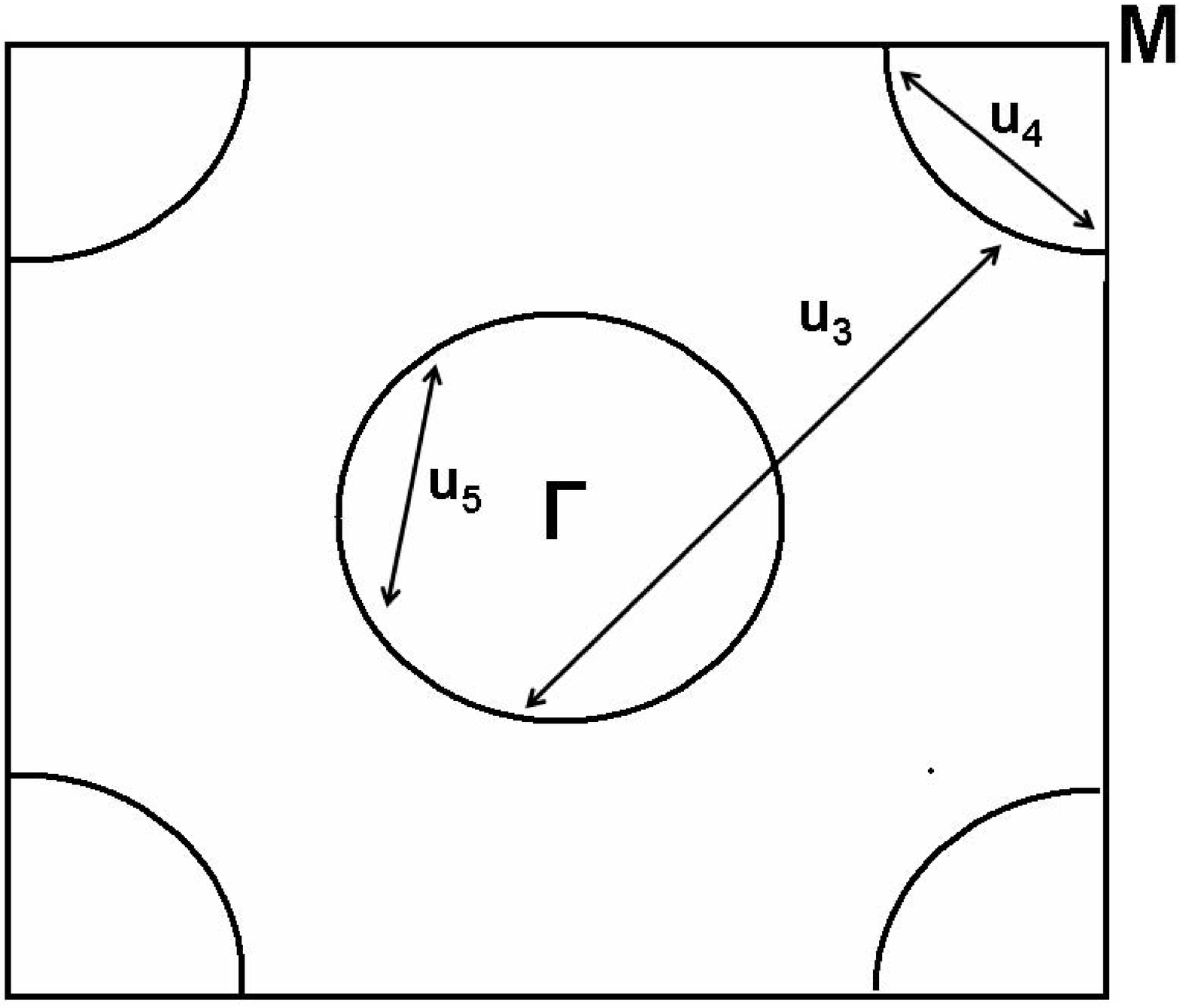}
\caption{\label{fig:UBZ_1H1E} Hole (center) and electron (corners)
FSs in the folded BZ for 2-pocket model. The arrows with symbols
indicate intra-pocket and inter-pocket  pairing interactions
($u_4$ and $u_5$ are intra-pocket interactions, and $u_3$ is
inter-pocket  interaction).  There also exist density-density and
exchange interactions between hole and electron pockets ( $u_1$
and $u_2$ terms, respectively, not shown). }
\end{figure}

The 2 pocket model is a toy model consisting of  one hole pocket
in the center  the folded Brillouin Zone (BZ), and four electron
pockets at the corners as shown in Fig \ref{fig:UBZ_1H1E}. To
obtain parquet RG equations, we consider energies larger than
$E_F$ and $E_b$. At such energies deviations from perfect nesting become
irrelevant, and we can set $E_F, E_b \to 0$ and take hole and electron dispersions to be just opposite in sign.

The interaction Hamiltonian for the 2-pocket model is (following
earlier notations \cite{Chubukov})

\begin{widetext}
\bea \label{eq:H_int 1H1E} \frac{m}{2\pi} H_{int}&=&
        \sum u_1\;c^\dag_{p_1 s}f^\dag_{p_2 s'} f_{p_4 s'}c_{p_3
        s} + \sum u_2\;c^\dag_{p_1 s}f^\dag_{p_2 s'} c_{p_4 s'}f_{p_3
        s} +  \sum \frac{u_3}{2}
        \left(
        c^\dag_{p_1 s}c^\dag_{p_2 s'} f_{p_4 s'}f_{p_3
        s}\;+\;h.c. \right)\\
        &+&\sum \frac{u_4}{2}   f^\dag_{p_1 s}f^\dag_{p_2 s'} f_{p_4 s'}f_{p_3
        s}+\sum \frac{u_5}{2}
        c^\dag_{p_1 s}c^\dag_{p_2 s'}c_{p_4 s'}c_{p_3 s}\nonumber
\eea
\end{widetext}
where $u_i$ are dimensionless couplings, and $m$ is quasiparticle
mass ($\varepsilon_f (k) = k^2/2m -\mu$). The sum is over the spin
indices $s$ and $s'$ and the vector momenta $p_1,p_2,p_3,p_4$ with
momentum conservation assumed. The $c$ and $f$ fermions reside at
the hole and the electron bands respectively. We remind the reader
that there is no angular dependence of the interactions here
because the first angular term that comes in is $\cos{4\phi}$ which we
ignore in our approximation.

\subsection{The Vertices}

\begin{figure}[htp]
\includegraphics[width=3in]{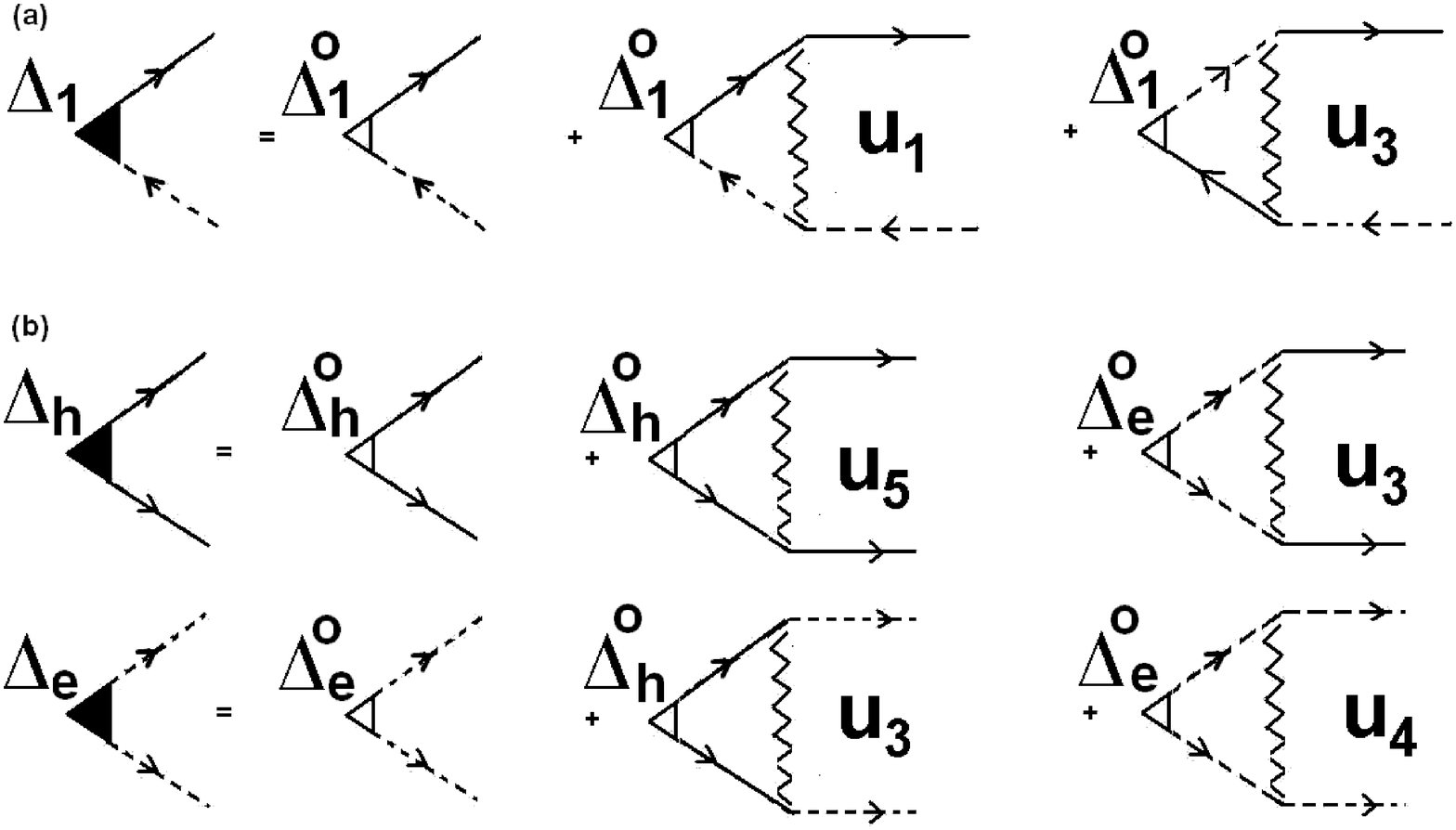}
\caption{\label{fig:sc_sdw_1H1E} Diagrams for the renormalization
of infinitesimally small SDW and SC vertices, added to the
Hamiltonian in order to calculate response functions. Unshaded
triangles - bare vertices, shaded triangles - full vertices, wave
lines -- fully renormalized interactions. Solid lines correspond
to $c$ fermions and the dashed line to $f$ fermions. $\Delta_1$ is
SDW vertex and $\Delta_h$ and $\Delta_e$ are SC vertices on hole
and electron FSs. }
\end{figure}

We begin by looking into the vertices in SC and SDW channels. For
this, we introduce infinitesimally small SC and SDW order
parameters $\Delta^o_{SC}$ and $\Delta^o_{SDW}$, dress them up by
including multiple interactions as shown diagrammatically in Fig.
\ref{fig:sc_sdw_1H1E}, and write the renormalized order parameters
in the form

\beq \Delta_i = \Delta^o_i \left(1 + \Gamma_i L\right) \label{w_1}
\eeq

where $\Gamma_i$ satisfies $\frac{d\Gamma_i}{dL}=\Gamma_i^2$

For a given $i$, $\Delta_i$ becomes nonzero even for vanishing
$\Delta^o_i$ when the corresponding $\Gamma_i$ diverges.

The computations for the 2-pocket model is straightforward. For
the SDW order parameter, we immediately obtain from Fig.
\ref{fig:sc_sdw_1H1E}, $\Gamma^{SDW} = u_1 + u_3$.  For the SC
channel, we obtain from Fig. \ref{fig:sc_sdw_1H1E}

\bea \label{eq:SC 1H1E} \left(
\begin {array}{cc}
 1-u_5 L & -u_3 L \\
 -u_3 L& 1-u_4 L  \\
\end{array}
\right) \left(
\begin {array}{c}
  \Delta^o_h\\
  \Delta^o_e\\
\end{array}
\right) =
 \left(
\begin {array}{c}
  \Delta_h\\
  \Delta_e\\
\end{array}
\right)
 \eea
where $\Delta_{e,h}$ are order parameters on hole and electron
FSs. Diagonalizing this set and casting the result in the form of
Eq. (\ref{w_1}), we obtain two SC $\Gamma's$. One corresponds to a
conventional $s-$wave pairing, is repulsive for all positive $u_i$
and is of no interest to us, another corresponds to $s\pm$ pairing
and is given by

\beq \label{eq:gamma} \Gamma^{s\pm} = \frac{-(u_4+u_5) +
\sqrt{(u_4-u_5)^2+4 u^2_3}}{2} \eeq For $u_4$ = $u_5$,
$\Gamma^{s\pm}$ reduces to $\Gamma^{s\pm}=-u_4 + u_3$.

The SDW vertex is attractive for positive $u_1$ and $u_3$, while
$\Gamma^{s\pm}$ is attractive only when inter-band pair hopping
term exceeds intra-band repulsive interaction.  Like we said, this
is very unlikely because both interactions originate from Coulomb
interaction, and screened Coulomb interaction at small momentum
transfer (i.e., $u_4$ and $u_5$) is larger than that at large
momentum transfer (i.e., $u_3$).

To understand whether the negative sign of $\Gamma^{s\pm}$ can be
reversed, we need to consider RG flow of the couplings. This what
we do next.

\subsection{RG flow between $\Lambda$ and $E_F$}

The RG equations for the couplings have been obtained in
\cite{Chubukov}, and we just quote the result:

\bea \label{eq:1H1E above E_f}
\dot{u}_4 &=& -[u_4^2 + u_3^2]\nonumber\\
\dot{u}_5 &=& -[u_5^2 + u_3^2]\nonumber\\
\dot{u}_1 &=& [u_1^2 + u_3^2]\nonumber\\
\dot{u}_2 &=& [2u_1u_2-2u_2^2]\nonumber\\
\dot{u}_3 &=& [4u_1 u_3 - 2u_2 u_3-(u_4+u_5)u_3]\eea The
derivatives are with respect to $L$. These equations have a single
non-trivial fixed point at which all couplings diverge and tend to
$u_3 = \sqrt{5} u_1, u_4 = u_5 = -u_1$, $u_2 \propto (u_1)^{1/3}$.
The flow of SDW and SC vertices is shown in Fig.
\ref{fig:H1E1_eff_vrtx}a.  In the process of RG flow, the SC
vertex $\Gamma^{s\pm}$ changes sign and become attractive. The
ratio $\Gamma^{s\pm}/\Gamma^{SDW}$ remains smaller than one during
the flow, but tends to one upon approaching the fixed point, i.e.,
if this fixed point is reached within parquet RG, superconducting
and SDW instabilities occur simultaneously, and the system
actually cannot distinguish between the two.  There is another
vertex which tends to the same value as SDW and SC vertices - it
corresponds to an CDW instability with imaginary order parameter
(an instability towards orbital currents),  The combination of
3-component SDW, 2-component SC and 1-component CDW instabilities
makes the fixed point $O(6)$ symmetric~\cite{podolsky}.

The sign change of the superconducting $\Gamma^{s\pm}$ is the most
notable effect within the parquet RG flow. Its physics originates
in the effective ``attraction'' between SC and SDW fluctuations
(not the order parameters!), namely from the fact that $u_3$,
which is the attractive component of $\Gamma^{s\pm}$, gets the
boost from $u_1$, which  contributes to $\Gamma^{SDW}$. The boost
is $4u_1 u_3$ term in the r.h.s. of the RG equation for ${\dot
u}_3$. This term overshadows the negative effect from $u_2$,
$u_4$, and $u_5$, and as a result  $u_3$ increases under RG. At
the same time, intra-pocket repulsions $u_4$ and $u_5$ decrease
under RG. At some point, $u^2_3$ becomes lager compared to $u_4
u_5$, and $\Gamma^{s\pm}$ becomes positive.

However, as we said earlier, parquet RG is only valid at energies
above $E_F$. It is unlikely that the fixed point is reached above
$E_F$, otherwise there would be at least pseudogap effects present
above $E_F$, but there is no strong evidence for pseudogap in the pnictides.
More likely, the couplings   evolve under parquet RG (and
$\Gamma^{s\pm}$ possibly changes sign at some scale), but $u_i$
remain finite at $E_F$.  To continue below this scale we need to
derive a different set of equations, for which $u_i (E_F)$ serve
as initial conditions.

\begin{figure}[htp]
\includegraphics[width = 3in]{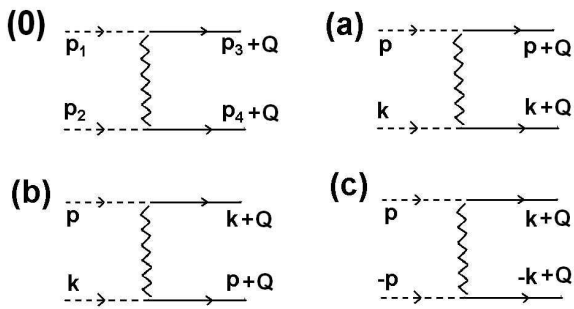}
\caption{\label{fig:low_energy_RG} (0) The $u_3$ vertex with
general momenta $p_1$, $p_2$, $p_3+Q$, $p_4 + Q$ (all $p_i$ are
small and $p_1 + p_2 = p_3 +p_4$). During calculations, three
kinds of $u_3$ vertices arise-(a) the one with $p_1$ = $p_3$, (b)
the one with $p_2$ = $p_3$, and (c) the one with $p_1 + p_2 = 0$.
The vertex `b' contributes to the renormalization in p-h channel,
and the vertex `c' contributes to the renormalization in the p-p
channel.}
\end{figure}

\subsection{RG flow below the scale of $E_F$}

The RG equations below $E_F$ for 2-pocket model have been derived
in Ref. \onlinecite{bib:Chu_physica} and we just quote the result.
The most essential difference with the previous subsection
concerns $u_3$ vertex, which contributes to both SC and SDW
channels. Below $E_F$ the structure of the external momenta
becomes relevant, and one has to distinguish between  $u^{(a)}_3$
with zero transferred momentum, $u^{(b)}_3$ with momentum transfer
$Q$, and $u^{(c)}_3$ with zero total momentum (see Fig. Fig.
\ref{fig:low_energy_RG}). Each of the vertices now undergoes
logarithmic renormalization in its own channel, crossed
renormalizations no longer contribute because internal $E =
O(E_F)$, and the arguments of the corresponding logarithms become
$O(1)$. The new equations are

\bea \label{eq:1H1E below E_f}
\dot{u}_3^{(a)} &=& 2u_1 u_3^{(a)} - 2u_2 u_3^{(a)}\nonumber\\
\dot{u}_3^{(b)} &=& 2u_1 u_3^{(b)}\nonumber\\
\dot{u}_3^{(c)} &=& -[u_4 + u_5]u_3^{(c)}\nonumber\\
\dot{u}_4 &=& -[u_4^2 + (u_3^{(c)})^2]\nonumber\\
\dot{u}_5 &=& -[u_5^2 + (u_3^{(c)})^2]\nonumber\\
\dot{u}_1 &=& [u_1^2 + (u_3^{(b)})^2]\nonumber\\
\dot{u}_2 &=& [2u_1u_2-2u_2^2]\eea where the derivatives are with
respect to $L$. The effective vertices in the SDW and SC channels
also get modified and become

\bea
\Gamma^{SDW} &=& u_1 +u_3^{(b)} \nonumber \\
\Gamma^{\pm} &=& \frac{-(u_4+u_5) +
\sqrt{(u_4-u_5)^2+4(u_3^{(c)})^2}}{2} \label{eq:gamma_1} \eea

One can easily verify that new vertices satisfy

\beq \label{eq:decoupling} \frac{d\Gamma_{i}}{d L} = \Gamma^2_i
\eeq

as it should be because SC and SDW channels are now decoupled (no
cross-terms in RG equations).

\begin{figure*}[t]
$\begin{array}{ccc}
\includegraphics[width=2.2in]{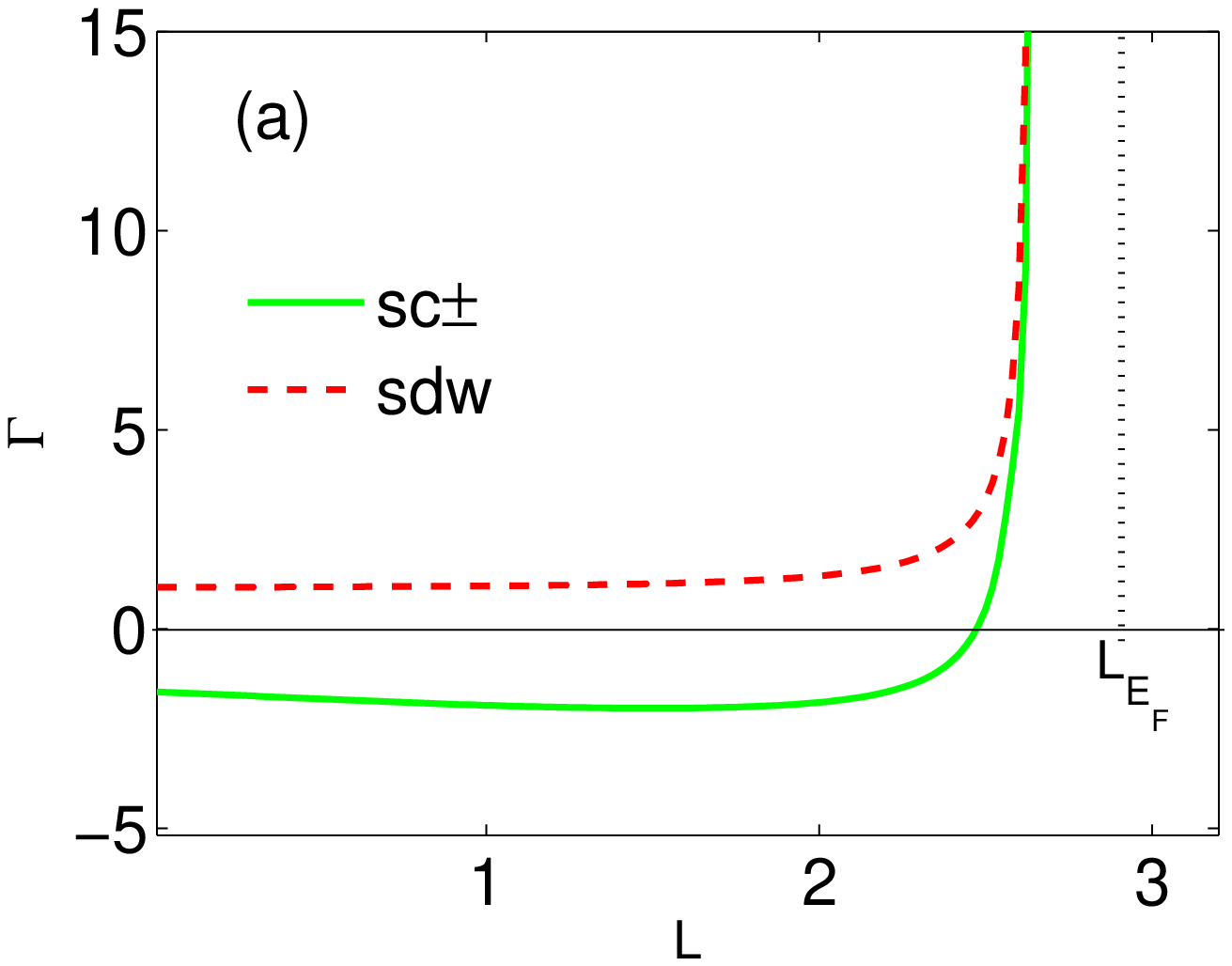}&
\includegraphics[width=2.2in]{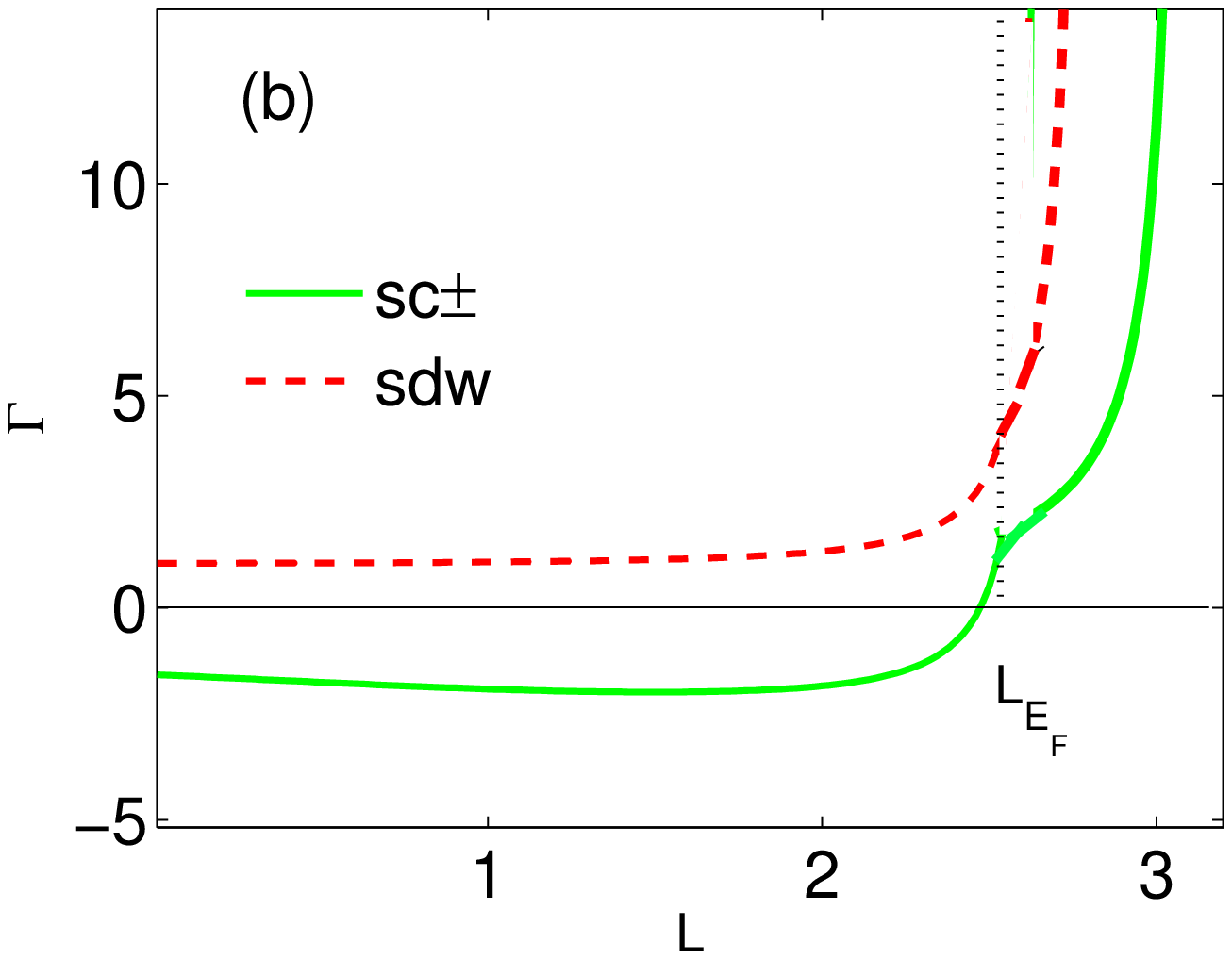}&
\includegraphics[width=2.2in]{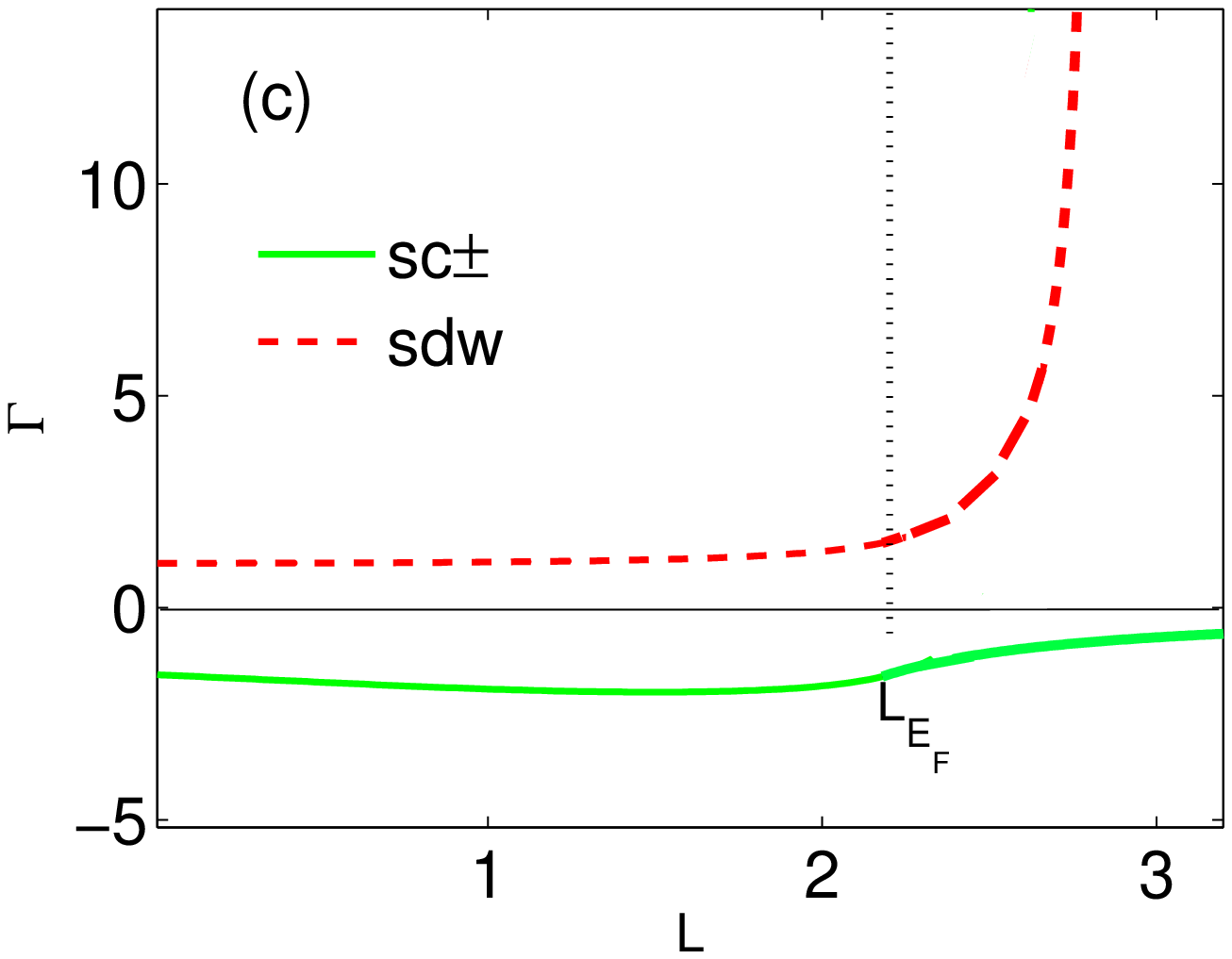}
\end{array}$
\caption{\label{fig:H1E1_eff_vrtx} Running vertices in SDW and SC
$s\pm$ channel for the 2-pocket case at perfect nesting as
functions of $L$. Three qualitatively different scenario are
possible: (a) vertices diverge before $L_{E_F}$ is reached. Both
vertices flow to the same fixed point, their ratio tends to one,
and the fixed point has an extended symmetry. (b) $L_{E_F}$ is
reached before the fixed point, but in the range of $L$ where the
SC vertex is already attractive (positive). The vertices flow
independent of each other beyond $L_{E_F}$(below $E_F$), the SDW
vertex diverges first. On doping SDW yields to SC (see Fig.
\ref{fig:gamma_schem}). (c) The value $L_{E_F}$ is reached when SC
vertex is still repulsive (negative). In this case SC instability
does not occur even after SDW instability is eliminated by
doping.}
\end{figure*}

The behavior of the vertices below $E_F$ is illustrated in Fig.
\ref{fig:H1E1_eff_vrtx} b,c.   If SC vertex is already  positive
(attractive) at $E_F$ (Fig. \ref{fig:H1E1_eff_vrtx}b),  it
diverges at some scale below $E_F$, but for perfect nesting SDW
vertex diverges first. Upon doping, SDW vertex levels off, and the
first instability eventually becomes the SC one. If SC vertex
remains negative at $E_F$ (Fig. \ref{fig:H1E1_eff_vrtx}c), it
decreases below $E_F$ but still remains negative. In this
situation, $s^{\pm}$ SC does not appear even when SDW order gets
killed by non-nesting.

To summarize:  in a 2-pocket model three scenarios are possible:
(i) the instability occurs simultaneously in SDW, SC, and CDW
channels, and a fixed point has $O(6)$ symmetry, (ii) SDW
instability wins at  prefect nesting, but yields to $s\pm$
superconductivity upon doping, and (iii) SDW instability exists
near perfect nesting, but no SC instability emerges when SDW order
is suppressed by doping.  For the cases (i) and (ii), the SC gap
has a simple plus-minus form, i.e., the gaps along hole and
electron FSs are angle independent (up to $\cos 4 \phi$ terms
which we neglected) and are of the opposite signs.

The 2-pocket model is indeed only a toy model for the pnictides.
The actual band structure of the pnictides includes two electron
FSs and at least two hole FSs. The question we now address is
whether qualitatively new behavior emerges when we increase
the number of pockets. We argue below that
there are new features not present in a 2-pocket model.

\section{4 Pocket Model}\label{section:3Bmodel}

We now consider the case of 2 hole and 2 electron FSs. We neglect
$k_z$ variation of the FSs and consider a cross section in XY
plane. The two electron FSs are generally ellipses, centered at
$(0,\pi)$ and $(\pi,0)$ in the unfolded zone.  The two hole FSs
are circles centered at $(0,0)$. They generally are  of non-equal
sizes, and one is less nested with electron FSs than the other.

The RG analysis of a generic 4-pocket model is straightforward but
rather cumbersome. We show the results in the two more easily
trackable limits: one when the two hole FSs are completely
equivalent, and the other when one of the two hole FSs is much
weakly coupled with electron than the other one and can be
neglected. In the second limit, 4-pocket model reduces to 3-pocket
model~\cite{bib:Chu_Vav_vor,Eremin}. We show that the system
behavior is identical in the two limits, and make a conjecture
that it doesn't evolve between the limits. We continue with our
earlier assumption of circular electron FSs that nests with the
hole FS.

We begin with the limit when one hole FS can be neglected and 4
pocket model reduces to 3 pocket model.

\subsection{Effective model with one hole at $(0,0)$.}

The interaction Hamiltonian for the 4-pocket model is
\begin{widetext}
\bea \label{eq:H_int 1H2E} \frac{m}{2\pi} H_{int}&=& \sum
u_1^{(1)}\;c^\dag_{p_1 s}f^\dag_{1 p_2 s'} f_{1 p_4 s'}c_{p_3
        s} +  \sum u_2^{(1)}\;c^\dag_{p_1 s}f^\dag_{1 p_2 s'} c_{p_4 s'}f_{1 p_3
        s} +  \sum \frac{u_3^{(1)}}{2}
        \left(
        c^\dag_{p_1 s}c^\dag_{p_2 s'} f_{1 p_4 s'}f_{1 p_3
        s}\;+\;h.c. \right)\nonumber\\
        &+& f_1 \leftrightarrow f_2\;and\;u_i^{(1)} \leftrightarrow u_i^{(2)}\nonumber\\
        &+& \sum \frac{u_5}{2} c^\dag_{p_1 s}c^\dag_{p_2 s'}c_{p_4 s'}c_{p_3 s}
        + \sum \frac{u_4^{(1)}}{2}   f^\dag_{1 p_1 s}f^\dag_{1 p_2 s'} f_{1 p_4 s'}f_{1 p_3
        s}+ \sum \frac{u_4^{(2)}}{2}  f^\dag_{2 p_1 s}f^\dag_{2 p_2 s'} f_{2 p_4 s'}f_{2 p_3
        s}\nonumber\\
        &+& \sum u_6 \; f^\dag_{1 p_1 s}f^\dag_{2 p_2 s'} f_{2 p_4 s'}f_{1 p_3
        s} +\sum u_7\;  f^\dag_{1 p_1 s}f^\dag_{2 p_2 s'} f_{1 p_4 s'}f_{2 p_3
        s} +  \sum \frac{u_8}{2}
        \left(
        f^\dag_{1 p_1 s}f^\dag_{1 p_2 s'} f_{2 p_4 s'}f_{2 p_3
        s}\;+\;h.c. \right)\nonumber\\
\eea
\end{widetext}

\begin{figure}[htp]
\includegraphics[width=2in]{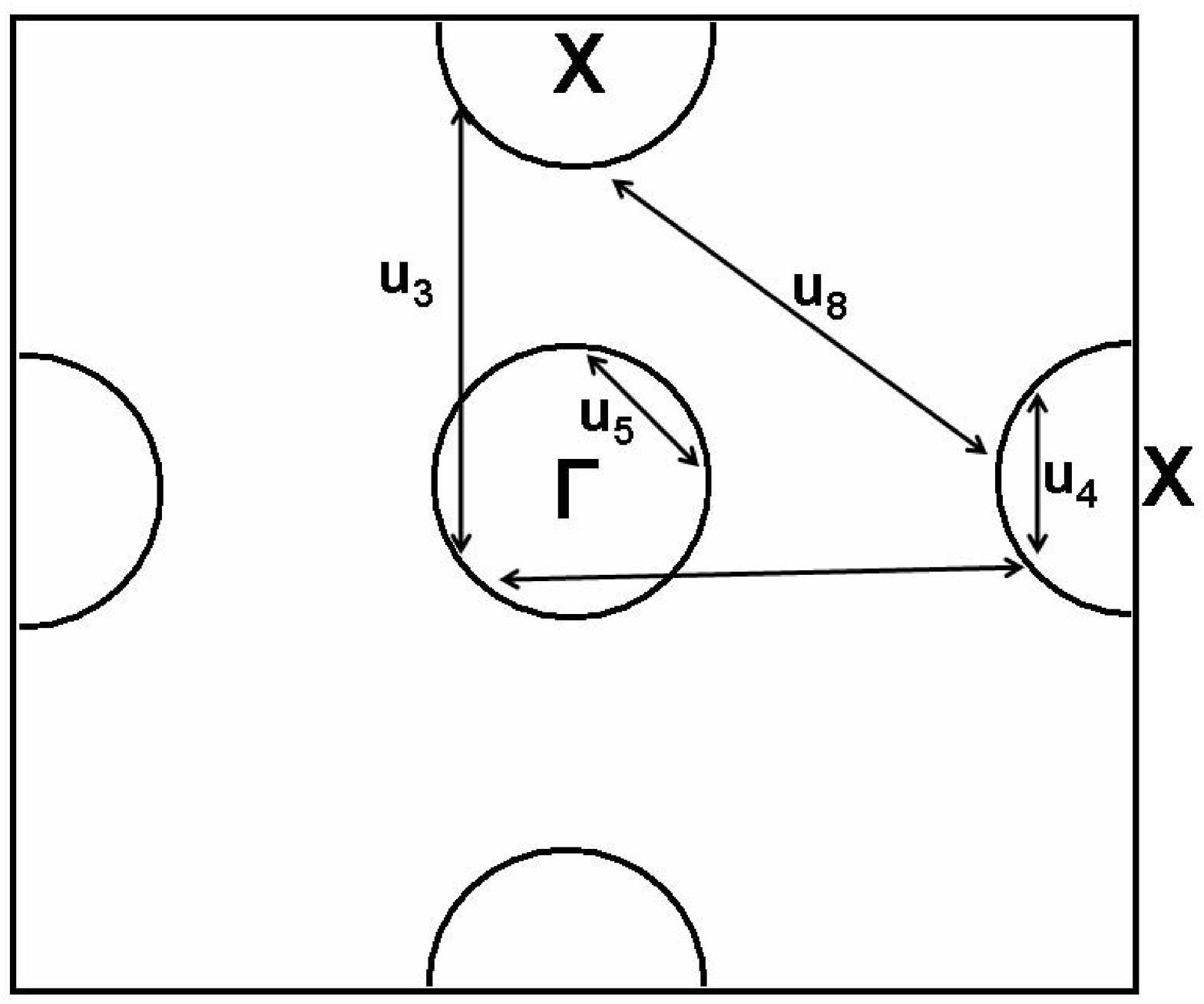}
\caption{\label{fig:UBZ} Hole (center) and electron(edges) FSs in
the unfolded BZ for 4-pocket model. The arrows with symbols
indicate various intra-pocket and inter-pocket interactions which
contribute to the SC vertex. There exist other density-density and
exchange inter-pocket interactions ($u_1$, $u_2$, $u_6$, and $u_7$
terms, not shown).}
\end{figure}

This is a straightforward generalization of the 2-pocket case (see
also Ref. \onlinecite{Eremin}). The notations are the same as for
the 2-pocket model, but now $f_1$ and $f_2$ refer to fermions from
the two different electron bands. The new terms $u_6, u_7$, and
$u_8$ are different inter-pocket interactions between
$f-$fermions. Because we now have two different sets of electron
state, it is convenient to work in the unfolded BZ, and in  Fig
\ref{fig:UBZ} we show the interactions that contribute to the
pairing vertex. There are, however, subtle effects related to the
actual, $As$-induced differences between folded and unfolded
zones,  and we will discuss them below.

The two electron bands are related by symmetry $k_x
\leftrightarrow k_y$ (i.e., $\varepsilon_{f_1} (k_x,k_y) =
\varepsilon_{f_2} (k_y,k_x)$), and it is natural to set  $u_i
^{(1)} = u_x ^{(i)} = u_i$ ($i$ runs between $1$ and $4$). We
verified that no new terms are generated under RG flow, however
the interactions between electron pockets must be included as they
anyway are generated by RG.

The angular dependence of the vertices is incorporated in the same
manner as described in Sec. \ref{section:Model}, by including
$\alpha \cos 2 \phi$ terms into the vertices which involve
fermions near electron pockets.  To simplify calculations, we
first neglect the angular dependence of the intra-pocket
electron-electron interaction $u_4$. Later we show that including
angular dependence of $u_4$ will not change the results
qualitatively.

\subsubsection{The Vertices}

The computational procedure is the same as before. We introduce
infinitesimally  small SC and SDW vertices, dress them up by the
interactions, and express the renormalized vertices in terms of
the running couplings.

\begin{figure}[htp]
\includegraphics[width = 3in]{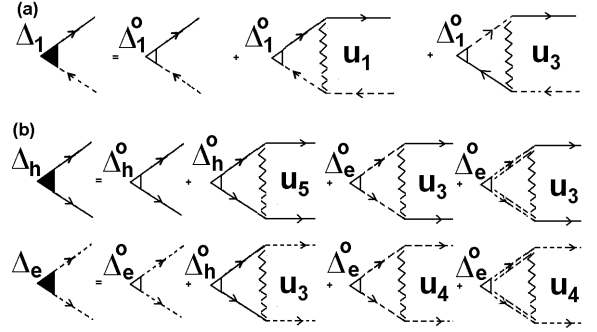}
\caption{\label{fig:sc_sdw_1H2E} The diagrams for the  SDW and
SC vertices for the 4-pocket model (panels (a) and (b)).  The notations
are the same as in Fig.\protect\ref{fig:sc_sdw_1H1E}. Single and double dashed
lines describe fermions from the two electron bands.}
\end{figure}

The diagrammatic expressions for the renormalized vertices are
presented in Fig. \ref{fig:sc_sdw_1H2E}. For the SDW vertex we
have $\Delta^{SDW} = \Delta^{SDW}_1 + \Delta^{SDW}_2 cos2\phi$,
$\phi$ being the angle along the electron FS. For SC vertex on the
hole FS $\Delta^{SC} = \Delta_h$ and on the two electron FSs
$\Delta^{SC} = \Delta_e \pm \bar{\Delta}_{e} cos2\phi$, as
required by symmetry for an $s-$wave gap. If
$|\bar{\Delta}_{e}|>|\Delta_e|$, the SC gap has nodes along the
FS. The nodes are `accidental' in the sense that they are not
protected by any symmetry.

For the SC vertex we then obtain the coupled set of equations \bea
\label{eq:SC 1H2Ea} \left(
\begin {array}{ccc}
 1-u_5 L & -2u_3 L &  -2\alpha u_3 L\\
 -u_3 L& 1-\tilde{u}_4 L  & 0\\
 -2\alpha u_3 L& 0 & 1\\
\end{array}
\right) \left(
\begin {array}{c}
  \Delta^o_h\\
  \Delta^o_{e}\\
  \bar{\Delta}^o_{e}\\
\end{array}
\right) = \left(\begin {array}{c}
  \Delta_h\\
  \Delta_e\\
  \bar{\Delta}_{e}\\
\end{array}\right)\eea
where $\tilde{u}_4=u_4+u_8$. For the  SDW vertex we obtain
 (assuming that $\Delta_{1,2}$ are real)
\bea \label{eq:SDW 1H2E} \left(
\begin {array}{cc}
 1+(u_1+u_3) L & \frac{\alpha}{2}( u_1+u_3) L\\
 \alpha( u_1+u_3) L  & 1\\
\end{array}
\right) \left(
\begin {array}{c}
 \Delta_1^o\\
 \Delta_2^o\\
\end{array}
\right) = \left(\begin {array}{c}
  \Delta_1\\
  \Delta_2\\
\end{array}\right) \eea

We included angular dependence of both $u_3$ and $u_1$ terms and
for simplicity set $\alpha$ to be the same for both. Observe that
$\alpha-$dependent terms appear in SDW vertex with extra $1/2$
compared to SC vertex. This is because  internal and external part
of the SDW vertex each contains one fermion from the electron FS,
while in the SC vertex there are two such fermions, either in the
internal or in the external part. One can easily check that in the
SC vertex all $\alpha-$terms then appear with extra factor of 2
compared to SDW vertex.

The effective vertices are found by diagonalizing these matrices
and casting the results into the forms $\Delta_i = \Delta_i^o (1 +
\Gamma_i L)$. For $\alpha=0$, the formulas simplify  and we have

\bea \label{eq:gap_eq_sp2}
 \Gamma^{SC}_1 = \Gamma^{s\pm} &=&
\frac{-(\tilde{u}_4+u_5)
+ \sqrt{(\tilde{u}_4-u_5)^2+8(u_3)^2}}{2} \nonumber \\
 \Gamma^{SC}_2 = \Gamma^{s++} &=&
\frac{-(\tilde{u}_4+u_5)
- \sqrt{(\tilde{u}_4-u_5)^2+8(u_3)^2}}{2} \nonumber \\
\Gamma^{SDW}&=& \Gamma^{SDW}_1 = u_1+u_3 \eea

The solutions corresponding to $\Gamma_1$ and $\Gamma_2$ are
$\Delta_h/\Delta_e <0$ and $\Delta_h/\Delta_e >0$, accordingly,
hence the notations $\Gamma^{s\pm}$ and $\Gamma^{s++}$. The vertex
$\Gamma^{s++}$ is repulsive for all couplings, while
$\Gamma^{s\pm}$ is repulsive for $2u_3^2 < {\tilde u}_4 u_5$ and
is attractive for $2u^2_3 > {\tilde u}_4 u_5$. The SDW vertex is
attractive. We recall that $u_4$ and $u_5$ are Coulomb
interactions at small momentum transfer, while $u_3$ and $u_8$ are
Coulomb interactions at large momentum transfer and  are supposed
to be smaller than $u_5$, $u_4$. At the bare level then there is
no attractive component of $\Gamma^{SC}$ for $\alpha =0$.

At a finite $\alpha$ we obtain two SDW vertices
$\Gamma^{SDW}_1 = (u_1 + u_3) (1 + \sqrt{1+2\alpha^2})/2$ and
$\Gamma^{SDW}_2 = -(u_1 + u_3) (\sqrt{1+2\alpha^2}-1)/2$.  The
second one is repulsive, which the first one is
attractive and only increases with $\alpha$.
For the SC vertex, the three $\Gamma^{SC}_i$ are obtained by
 diagonalizing Eq. (\ref{eq:SC 1H2Ea}) what requires
solving the cubic equation. The analytical expressions for
$\Gamma^{SC}_i$ are long and we refrain from presenting them (we
will show all three $\Gamma^{SC}_i$ in the figures). It is
essential, however, that for {\it any} $\alpha \neq 0$, one of
three $\Gamma^{SC}_i$ is attractive even when ${\tilde u}_4u_5$ is
larger than $2u^2_3$ and at $\alpha =0$ SC vertices are repulsive.
At small $\alpha$, we have for such induced solution

\beq
\Gamma^{SC}_{1} =
\alpha^2\left[\frac{4u_3^2u_5}{\tilde{u}_4u_5-2u_3^2}\right],~\tilde{u}_4u_5
> 2u_3^2 \label{g_1} \eeq

Other two solutions $\Gamma^{SC}_{2,3}$ are negative, i.e.
repulsive.

For $\tilde{u}_4u_5 <2u_3^2$, the attractive solution exists
already at $\alpha =0$, and is only weakly affected by $\alpha$.
For $\tilde{u}_4u_5 < 2u_3^2$ we have: \beq \Gamma^{SC}_{1}=
\Gamma^{s\pm}\left[1+\alpha^2\left(\frac{2 u_3^2 (-\Delta u +
\sqrt{(\Delta u)^2+8u_3^2})}{(\Gamma^{s\pm})^2 \sqrt{(\Delta
u)^2+8u_3^2}}\right)\right] \label{g_2} \eeq

where $\Delta u = \tilde{u}_4-u_5$ and $\Gamma^{s\pm}$ is given by
Eq. \ref{eq:gap_eq_sp2}. The other two $\Gamma^{SC}_{2,3}$ are
again negative.

We now proceed with the RG flow.

\subsubsection{RG flow between $\Lambda$ and $E_F$} \label{subsection:1H2E RG}

\begin{figure}[htp]
$
\begin{array}{c}
\includegraphics[width = 3in]{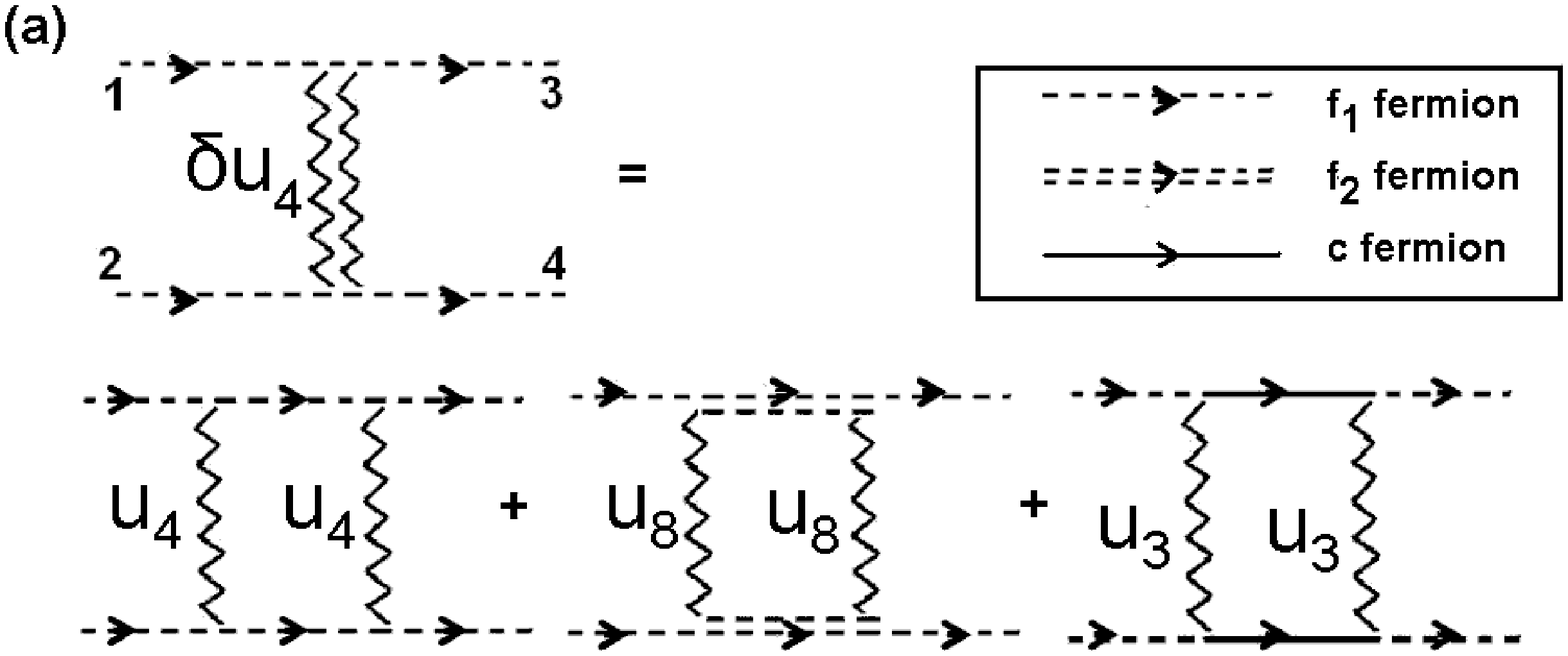}\\
\includegraphics[width = 3in]{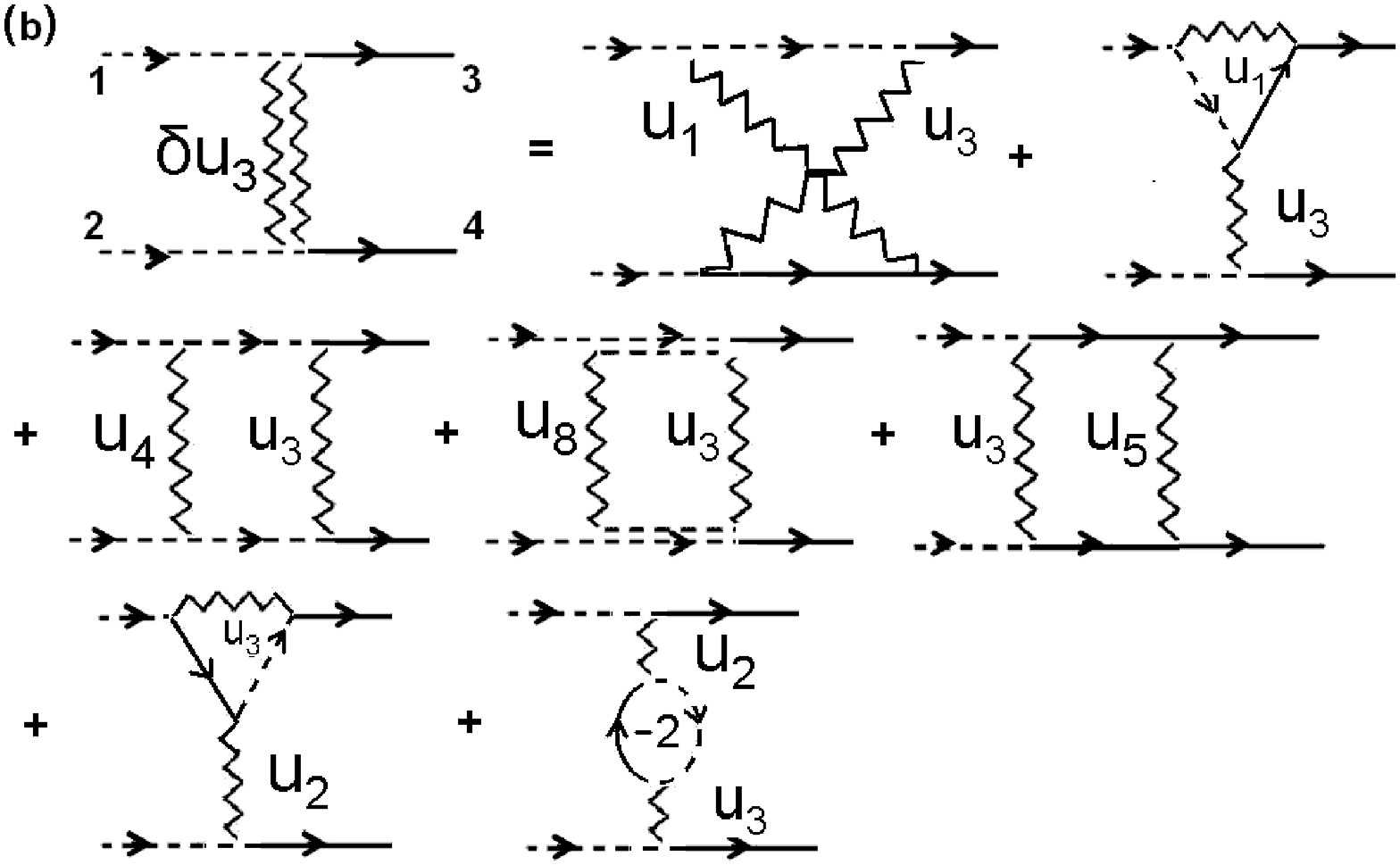}
\end{array}$
\caption{\label{fig:u3v3_vrtx} Second order diagrams for the
renormalizations of $u_4$ and $u_3$ vertices (panels a and b,
respectively). The combinatorial factors are not shown but must
indeed be included.}
\end{figure}

Like in 2-pocket case, at $\Lambda > E > E_F$, renormalizations in
p-h and p-p channels are logarithmical and independent of the
location of external momenta. The  derivation of parquet RG
equations is straightforward but requires more efforts as there
are new terms in the Hamiltonian. For illustration, we show  in
Fig \ref{fig:u3v3_vrtx} the diagrams contributing to the
renormalization of the vertices $u_4$ and $u_3$. The diagrams for
 the renormalization of other vertices are similar.

Collecting the diagrams for the renormalization of all couplings,
we find that the terms $u_6 \pm u_7$ and $u_4-u_8$  are decoupled from the rest
of the terms and are renormalized as $\dot{u}_j =-(u_j)^2$,
Because all these $u_j$ are the differences between Coulomb interactions at small and large momentum transfers, their bare values are positive in which case the these interactions flow to zero under RG and are therefore irrelevant.

\begin{figure}[htp]
\includegraphics[width = 3in]{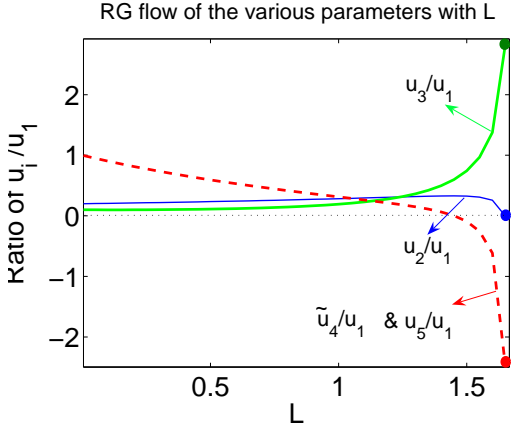}
\caption{\label{fig:1H2E_RG_flow} The solution of Eq. (\ref{eq:RG
1H2E_1}) --
 the RG flow of $u_5$, $\tilde{u}_4$, $u_2$, and $u_3$
(all relative to $u_1$) with $L$.  For simplicity, we set bare
values of ${\tilde u}_4$ and $u_5$ equal.  Note how, as fixed
point is approached, the renormalized Coulomb repulsion at small
momenta ($u_5$ and ${\tilde u}_4$ terms) is suppressed and
eventually changes sign,  while the pair hopping term $u_3$ is
strengthened.}
\end{figure}

The other five vertices are all coupled and flow according to \bea
\label{eq:RG 1H2E_1}
\dot{u}_5 &=& -\left[u_5^2 + 2 u_3^2\right] \nonumber\\
\dot{\tilde{u}}_4 &=& -\left[\tilde{u}_4^2 + 2u_3^2 \right]\nonumber\\
\dot{u}_1 &=& +\left[u_1^2 + u_3^2\right]\nonumber\\
\dot{u}_2 &=& +\left[2 u_1u_2 - 2 u_2^2\right]\nonumber\\
\dot{u}_3 &=& +\left[4u_1u_3-2u_2u_3-u_5u_3-\tilde{u}_4 u_3\right]
\eea This set of equations can be easily solved numerically. Fig
\ref{fig:1H2E_RG_flow} shows the plot of $u_5/u_1$, ${\tilde
u}_4/u_1$, $u_2/u_1$, and $u_3/u_1$ with $L$. For simplicity, we
set bare values of ${\tilde u}_4$ and $u_5$ to be equal -- the two
then remain equal in the process of RG flow.  The values of the
ratios at the fixed point are indicated by the dots. These can be
easily found analytically by requesting that all 5 equations in
(\ref{eq:RG 1H2E_1}) be identical. Imposing this condition we
obtain $u_5/u_1={\tilde u}_4/u_1 =-\sqrt{6} \approx -2.45,~u_3/u_1
= (3+2\sqrt{6})^{1/2} \approx 2.81, u_2/u_1 =0$.  Like in 2-pocket
model, intra-pocket repulsions ${\tilde u}_4$ and $u_5$ decrease
under RG, change sign at some value of $L$, and  become negative
at larger $L$. This sign change (overscreening) goes beyond a
conventional McMillan-Tolmachev screening of the Coulomb
interaction, and  is the result of the ``push'' from $u_3$, which
in turn increases under RG due to the ``push'' from $u_1$ which
contributes to the  SDW vertex. So, eventually,  overscreening is
the result of the ``attraction'' between SDW and SC fluctuations.
The difference ${\tilde u}_4 u_5 - 2 u^2_3$ also changes sign at
some $L$ and becomes negative at larger $L$.

\begin{figure*}[t]
$\begin{array}{cc}
\includegraphics[width = 3in]{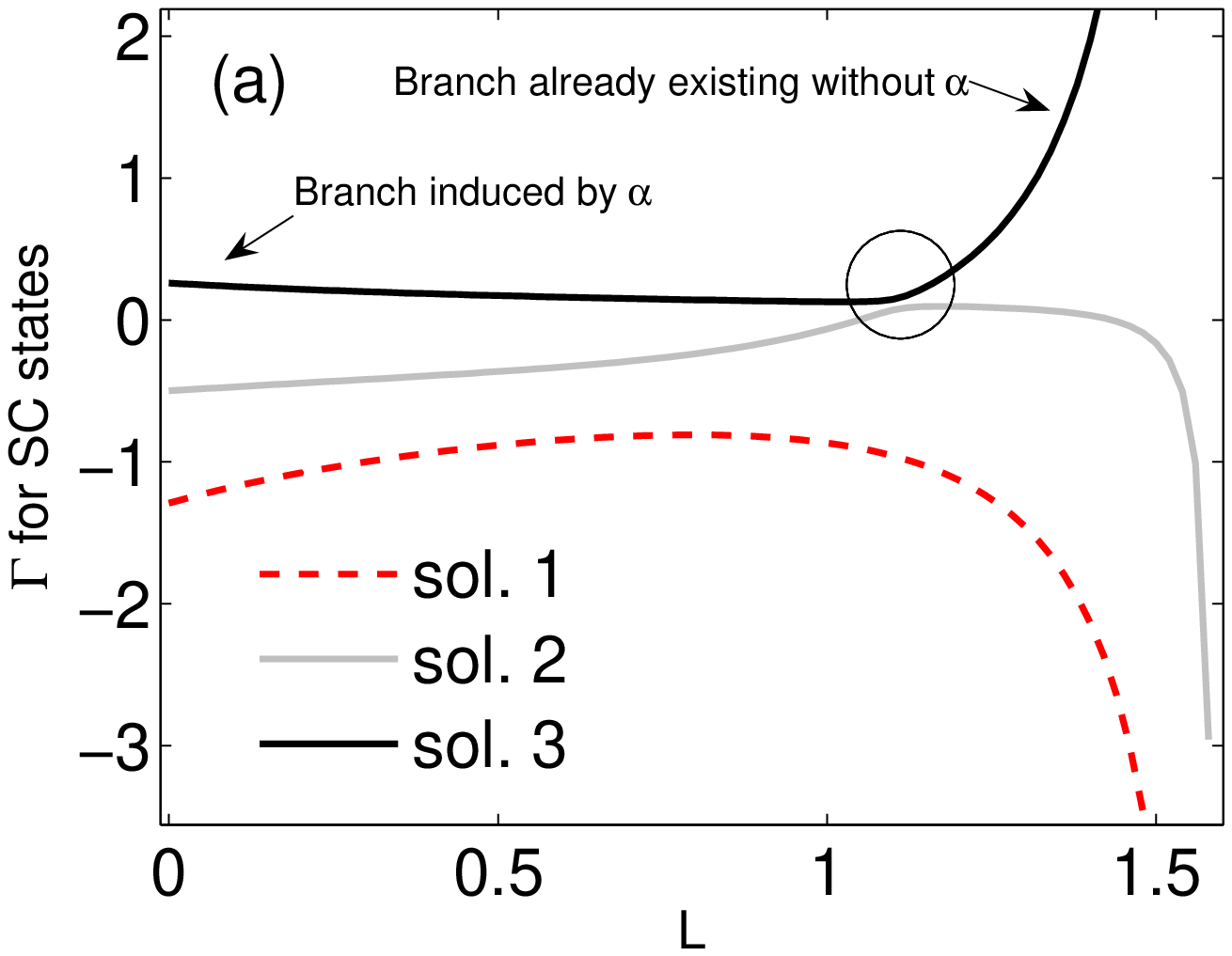}&
\includegraphics[width = 3in]{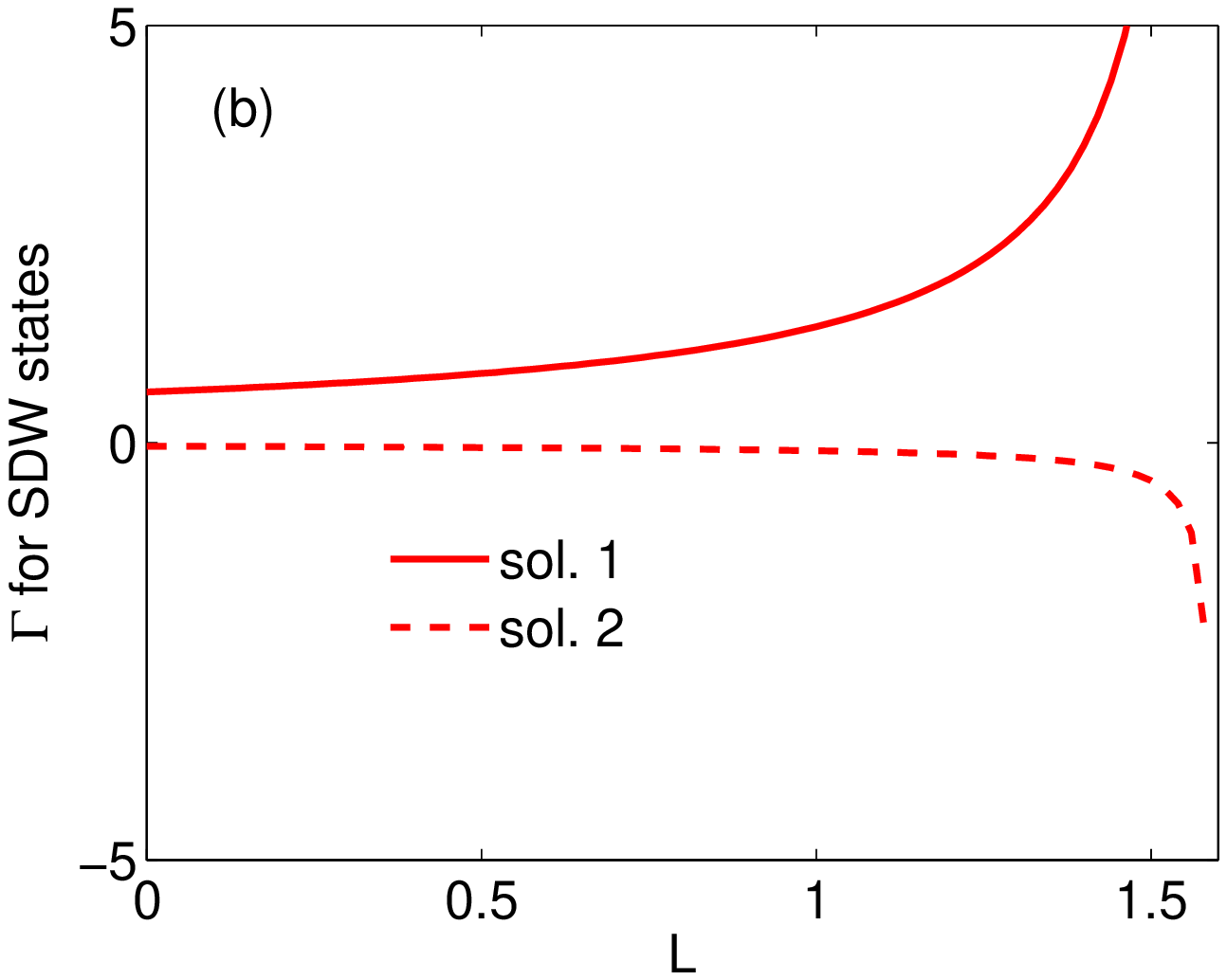}
\end{array}$
\caption{\label{fig:SC_vrtcs} The flow of SC and SDW vertices
under RG in the effective 3-pocket model for $\alpha =0.4$.  Panel (a) 3 SC
vertices. One solution is attractive for all $L$ (and corresponds
to $s\pm$ pairing), the other two are repulsive. One of the
repulsive solutions is of $s\pm$ character, another of $s++$
character. At small $L$, the positive solution is the one induced
by $\alpha$, at large $L$ it almost coincides with the solution
which becomes positive for these $L$ already at $\alpha =0$. The
circle marks the area where positive and negative solutions come
close to each other. The splitting between the two increases with
$\alpha$. (b) The two SDW vertices. One vertex is always
attractive (positive) and the other is repulsive (negative). }
\end{figure*}

We now substitute the running couplings into the expressions for
SC and SDW vertices and check how they flow. The results are
presented in Fig. \ref{fig:SC_vrtcs} for $\alpha =0.4$. For the
SDW vertices, the positive one increases with L, like in the
2-pocket model, while the negative one becomes more negative,
i.e., even less relevant. For the SC vertices,  $\Gamma^{SC}_1$ is
positive for all $L$, the other two $\Gamma^{SC}_{2,3}$ are
negative and hence irrelevant. The positive $\Gamma^{SC}_1$
interpolates between Eq. (\ref{g_1}) at small $L$, and Eq.
(\ref{g_2}) at larger $L$. We emphasize that for all values of $L$
this is the same solution, i.e., there is no level crossing (see
Fig. \ref{fig:SC_vrtcs}).

\begin{figure*}[t]
$\begin{array}{ccc}
\includegraphics[width = 2.2in]{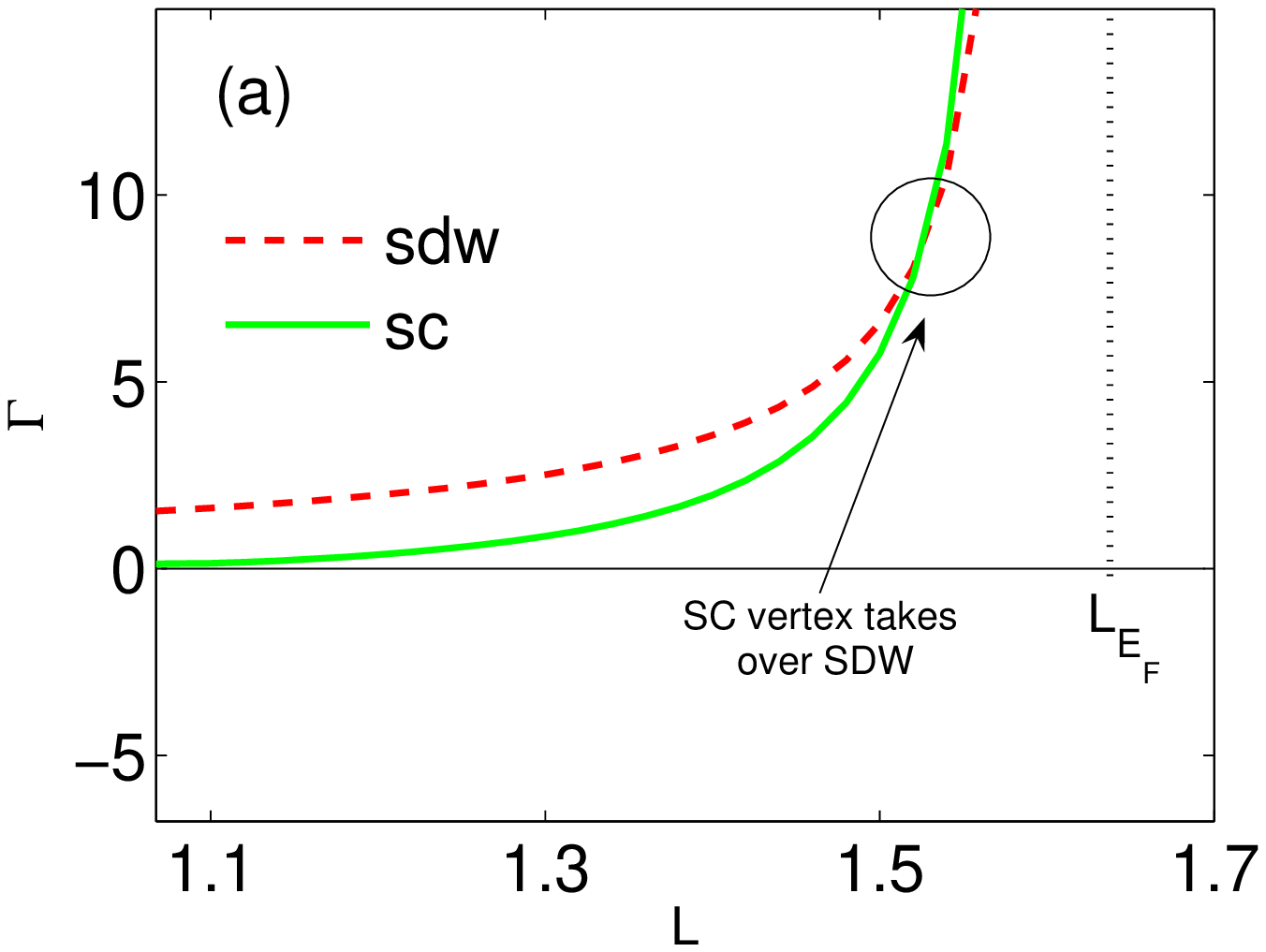}&
\includegraphics[width = 2.2in]{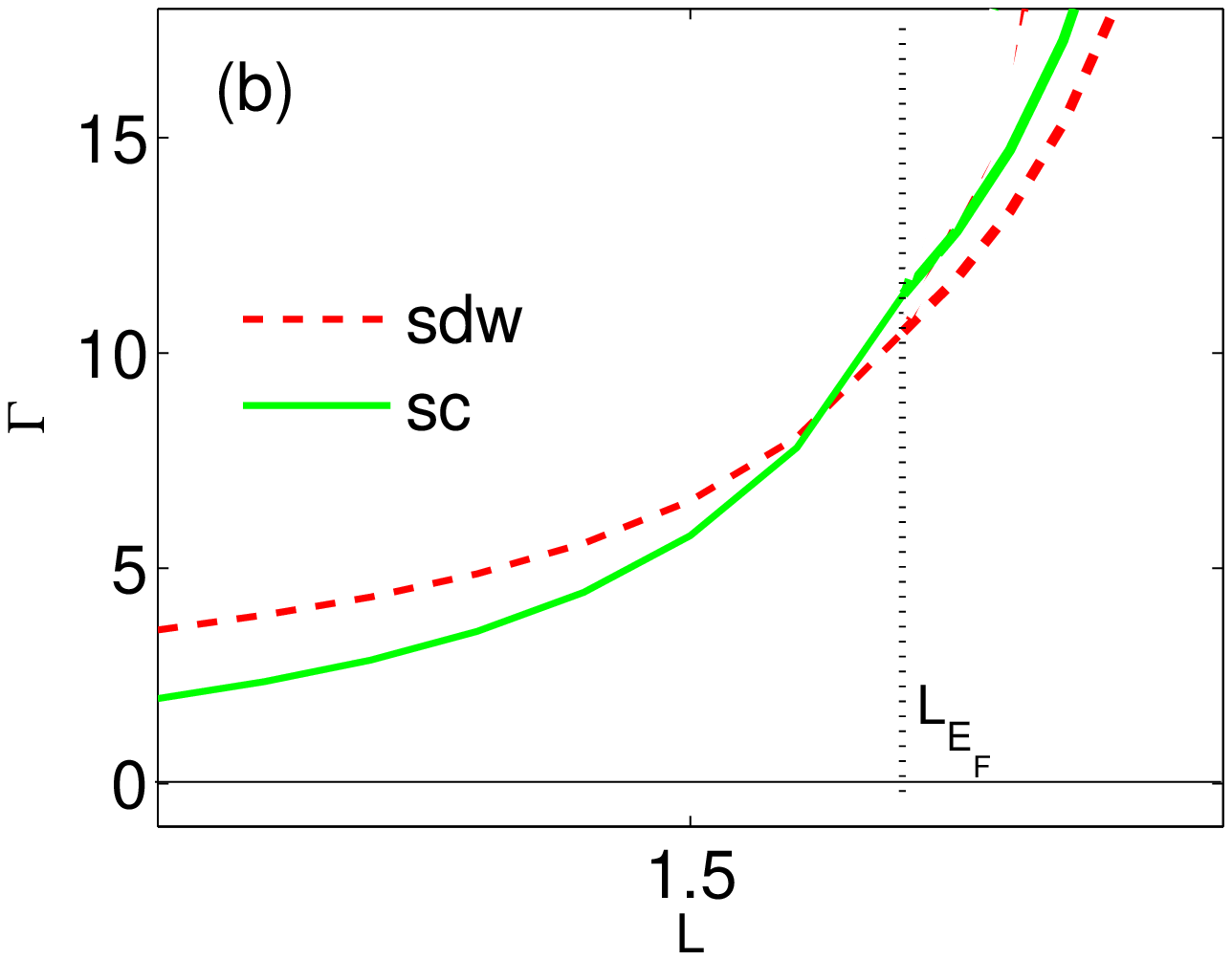}&
\includegraphics[width = 2.2in]{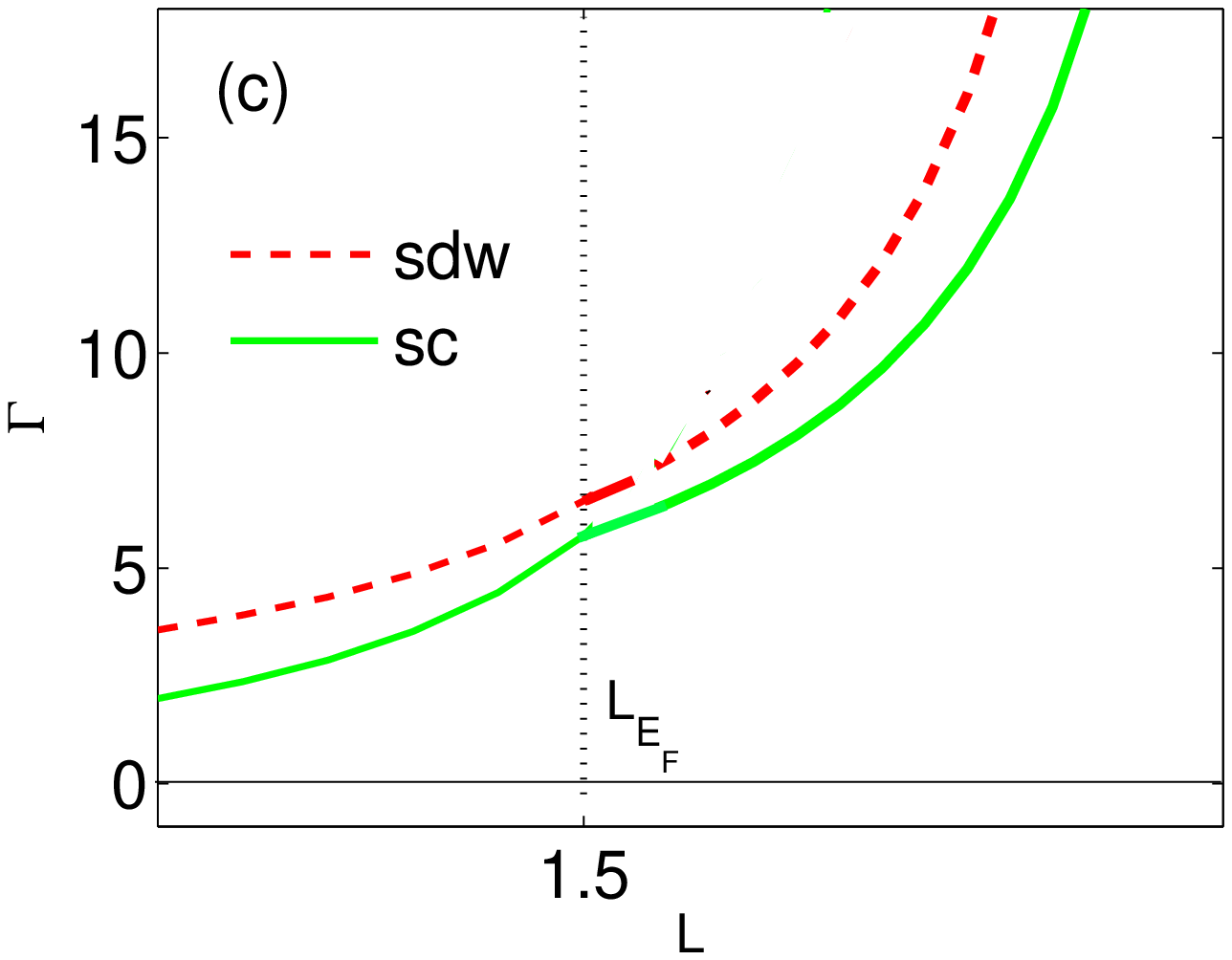}
\end{array}$
\caption{\label{fig:H1E2_eff_vrtx_angle} The flow of vertices for
different values of $L_{E_F}$ for  $\alpha =0.4$. (a) RG flow
reaches fixed point before $L_{E_F}$.  SDW vertex is larger at
small $L$, but SC vertex `crosses' the SDW vertex  at some
distance from the fixed point and becomes the strongest vertex at
the fixed point. As a result, the system develops a SC order. (b)
$L_{E_F}$ is reached after the `crossing' but before reaching
fixed point. The system still develops  a SC order, and SDW order
does not emerge. (c) $L_{E_F}$ is reached before the `crossing'.
In this case the SDW vertex still develops at small dopings, and
SC order  emerges at larger dopings, when SDW order gets
suppressed.}
\end{figure*}

\begin{figure}[htp]
\includegraphics[width = 3in]{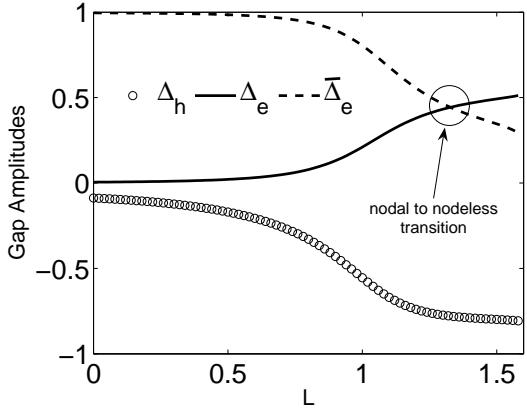}
\caption{\label{fig:deltas_Unoangle} The evolution  of the three
components of the  superconducting gap with $L$ for  $\alpha=0.4$.
$\Delta_h$ is the gap on the hole FS, and $\Delta_e \pm
\bar{\Delta}_{e} \cos 2\phi$ are the gaps along electron FSs. A
circle marks the point where $\Delta_e$ and $\bar{\Delta}_{e}$
cross, and the gap along each of electron FSs changes from nodal
to nodeless. Note that the solution is always of s$\pm$ character;
meaning $\Delta_h$ and $\Delta^e_1$ are of opposite signs.}
\end{figure}

\begin{figure}[htp]
\includegraphics[width = 3in]{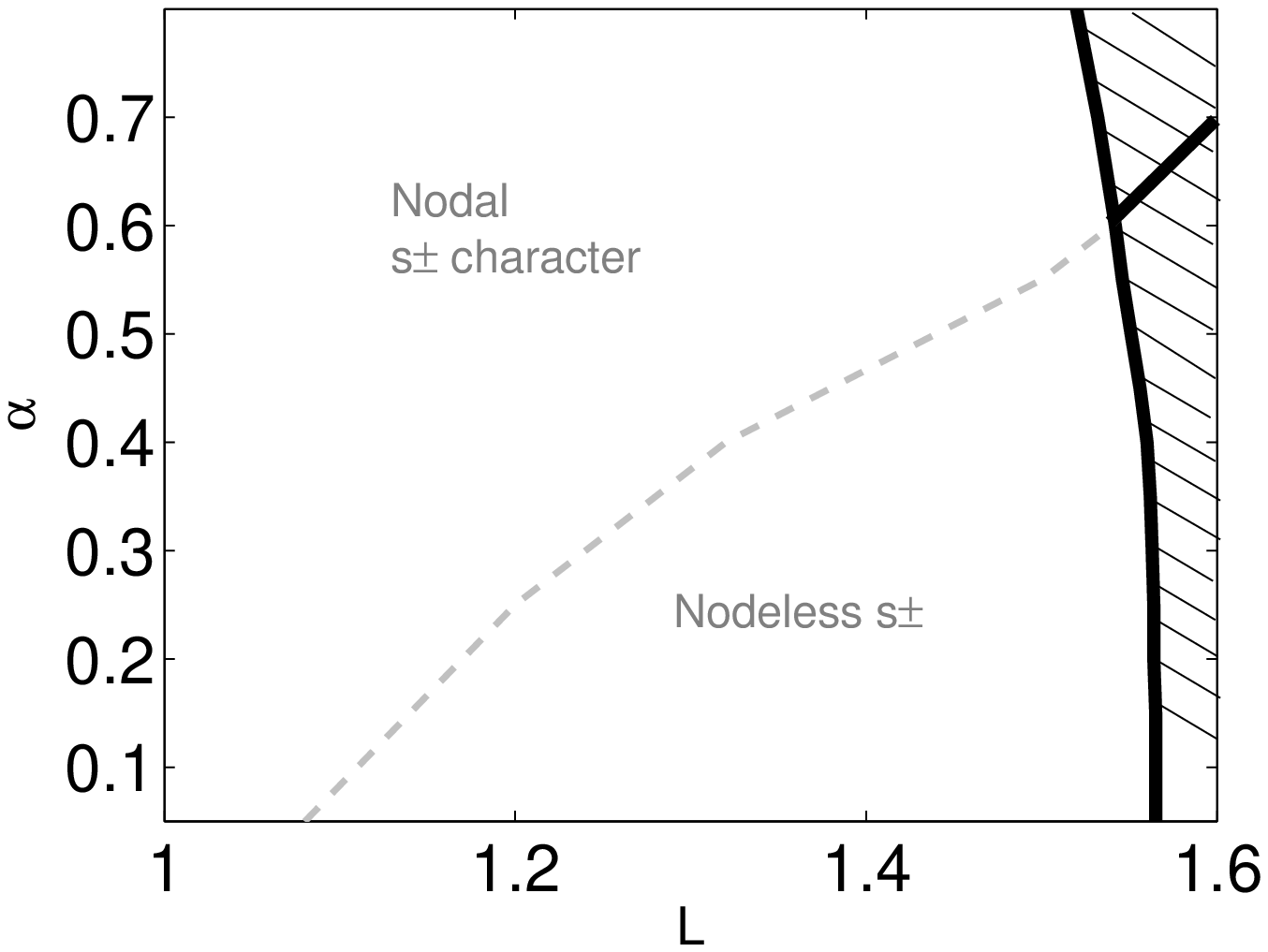}
\caption{\label{fig:new_nodal_pd} The RG flow of the 4-pocket
model in variables $\alpha$ and $L$ over the range of energies
above $E_F$. The white zone is where SDW vertex is the largest,
and the shaded zone is where SC $s\pm$ vertex is the largest. The
thick solid line in the shaded zone marks the transition form
nodal to nodeless s$\pm$ gap in the region where SC vertex wins.
The dashed line is the continuation of this transition line into
the region where SDW vertex takes over. }
\end{figure}

In Fig. \ref{fig:H1E2_eff_vrtx_angle}a we compare the behavior of
$\Gamma^{SDW}$ and $\Gamma^{SC}$ as functions of $L$ assuming that
the fixed point is reached at  energies  above $E_F$. At small
$L$, we have the same situation as in 2-pocket model: SDW vertex
is larger than SC vertex. However, the rate of increase of the SC
vertex exceeds that of the SDW vertex, and  at some $L$ before the
fixed point is reached, $\Gamma^{SC}$  {\it crosses over} $\Gamma^{SDW}$
implying that superconductivity becomes the leading
instability  even at perfect nesting.  Such crossing has been
reported in fRG calculations\cite{Thomale} for the same model. We
view the agreement as a good indication that numerical fRG and
analytical parquet RG approaches describe the same physics. In our
analytical RG, the reason for the crossing is combinatoric:
compared to 2-pocket case (where SDW and SC vertices flow to the
same value under RG), the presence of the second electron FS adds
the factor of 2 to the renormalization of the SC vertex as a pair
of  fermions from the hole FS can hope to each of the two electron
FSs. However, there is no such factor of 2  in the renormalization
of the SDW vertex due to momentum conservation.

We emphasize that the crossing of $\Gamma^{SDW}_1$ and
$\Gamma^{SC}_1$ is not related to the angular dependence of the
interaction. Even when $\alpha =0$, SC vertex exceeds SDW vertex
near the fixed point of parquet RG. At the fixed point, the ratio
of the two is $\Gamma^{SC}_1/\Gamma^{SDW}_1 \approx 1.69$.

The SC order parameter by itself has an interesting character. We
recall that we approximate the gap along the hole FS by a constant
$\Delta_h$ and approximate the gap along the two electron FSs by
$\Delta_e \pm \bar{\Delta}_{e}\cos 2 \phi$. At small $L$, the
attractive  $\Gamma^{SC}$ exists only because of a non-zero
$\alpha$, and $\bar{\Delta}_{e}$ is larger that $\Delta_e$ (Ref.
\onlinecite{bib:Chu_Vav_vor}), hence the gap along the two
electron FSs has ``accidental'' nodes. As $L$ increases, the SC
vertex $\Gamma$ evolves, according to Fig. \ref{fig:SC_vrtcs}, and
eventually gets close to the would-be solution for $\alpha =0$.
For the latter, $\bar{\Delta}_{e} =0$, and the gap obviously has
no nodes.  The crossover from one limit to the other is displayed
in Fig. \ref{fig:deltas_Unoangle}, where we show the flow of the
gaps $\Delta_h, \Delta_e$, and $\bar{\Delta}_{e}$,  corresponding
to the leading SC vertex. For the value of $\alpha$ which we used
in this figure ($\alpha =0.4$) the transition from nodal  to
nodeless gap occurs at $L$ smaller than the one at which SC vertex
crosses the SDW vertex, i.e., when SC becomes the leading
instability, the superconducting gap is already nodeless. But for
other values of $\alpha$ we can get either nodeless or nodal SC in
this regime. In Fig. \ref{fig:new_nodal_pd} we plot the ``phase
diagram'' coming out of parquet RG for different $\alpha$ [the
bare values of $u_i$ are the same for all figures in this
section]. In white region, SDW vertex is the largest and SC vertex
is subleading, implying that superconductivity can be revealed
only after  SDW order is suppressed by doping. In the shaded
region the  SC vertex is the largest. We see that superconducting
gap in this region can actually be either nodal or nodeless,
depending on the value of $\alpha$. At $\alpha =0$, the ratio of
$\Delta_e$ and $\Delta_h$ at the fixed point is $\Delta_h =
-\sqrt{2} \Delta_e$.

\subsubsection{RG flow below the scale of $E_F$}

We now consider the situation below $E_F$. As in the 2-pocket
model, we have to introduce three different $u_3$ couplings
($u^{(a)}_3$, $u^{(b)}_3$, and ($u^{(c)}_3$) of which $u^{(b)}_3$
is the part of SDW vertex, and $u^{(c)}_3$ is the part of the SC
vertex (the corresponding $u^{(i)}_3$ replace $u_3$ in Eqs.
(\ref{eq:SC 1H2Ea}) and (\ref{eq:SDW 1H2E})). The flow of the
couplings is now governed by

\bea \label{eq:RG 1H2E_2}
\dot{u_5} &=& -\left[u_5^2 + 2 (u_3^{(c)})^2\right] \nonumber\\
\dot{\tilde{u}}_4 &=& -\left[\tilde{u}_4^2 + 2(u_3^{(c)})^2 \right]\nonumber\\
\dot{u}_1 &=& +\left[u_1^2 + (u_3^{(b)})^2\right]\nonumber\\
\dot{u}_2 &=& +\left[2 u_1u_2 - 2 u_2^2\right]\nonumber\\
\dot{u}_3^{(a)} &=& 2u_1u_3^{(a)}-2u_2u_3^{(a)}\nonumber\\
\dot{u}_3^{(b)} &=& 2u_1u_3^{(b)}\nonumber\\
\dot{u}_3^{(c)} &=& \left[u_5+\tilde{u}_4\right]u_3^{(c)} \eea

Note the $\tilde{u}_4$ and $u_5$ have identical equations and
hance treated identically. One can then straightforwardly verify
using (\ref{eq:RG 1H2E_2}) that SC and SDW vertices decouple, as
they should, and each satisfies $\dot{\Gamma_i}=\Gamma^2_i$.
Hence, as before, whichever vertex is larger at $E_F$ gives rise
to the first instability as $T$ decreases.
  If SC vertex prevails, the system becomes SC
at perfect nesting and remains a SC at finite dopings
(Fig.  \ref{fig:H1E2_eff_vrtx_angle}b). If SDW
vertex prevails, the system becomes an SDW antiferromagnet at
perfect nesting and then eventually becomes a SC upon doping
 (Fig.  \ref{fig:H1E2_eff_vrtx_angle}c).  In
distinction to the 2-pocket case, we don't need to worry about the
sign of the SC vertex once SDW instability is reduced by doping
because one of $\Gamma^{SC}$ is always attractive (see Fig.
\ref{fig:SC_vrtcs}). We emphasize again that this attractive
$\Gamma^{SC}$ leads to either nodeless $s\pm$ gap, or to $s\pm$
gap with nodes along the electron FSs, depending on $\alpha$ and
on the interplay between $E_F$ and the scale at which parquet RG
flow reaches the fixed point.

\begin{figure}[htp]
\includegraphics[width = 3in]{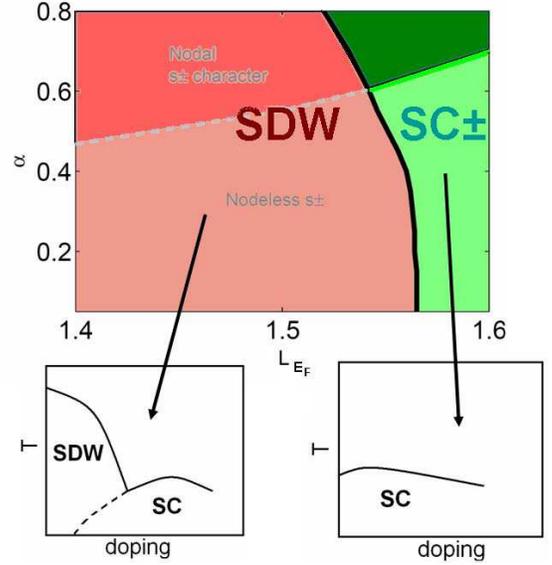}
\caption{\label{fig:phase_schem} The phase diagram of the 4-pocket
model at perfect nesting in variables $\alpha$ and $L_{E_F} =
\log{\Lambda/E_F}$. In red is the region where SDW order develops,
and in green is the region of the SC order. Dark and light green
regions correspond to nodal (dark) and nodeless (light) SC gap
along the electron FSs. The two sub-figures show the behavior at
finite doping. Superconducting state which  is brought out upon
doping in the left sub-figure is either nodeless or nodal
depending on the location with respect a dashed line inside the
red (SDW) region. The transition between SDW and SC states can be
either first order or involve intermediate co-existence
phase~\protect\cite{vor_vav_chub,joerg}. }
\end{figure}

\subsubsection{Effect of the angular dependence of electron-electron interaction}

For completeness, we present the results for the evolution of the
SC gap structure under RG flow for the case when we preserve  the
angular dependence in the electron-electron interactions -- $u_4$
and $u_8$ terms. These interactions only contribute to the pairing
channel, so it will be sufficient co consider $u_4$ and $u_8$
interactions between fermions with momenta $k, -k; p, -p$. The
generic structure of the angular dependence of such interactions
is given by  Eq. (\ref{s_5}). We found earlier that $u_4$ and
$u_8$ terms contribute to the $s-$wave  pairing in the combination
${\tilde u}_4 = u_4 + u_8$ , so we need to consider only this
term. We have

\beq {\tilde u}_4 (k,p) = {\tilde u}_4 \left(1+2\alpha'
\left(\cos2\phi_{k} \pm \cos2\phi_{p}\right) \pm 4 \alpha^{''}
\cos2\phi_{k} \cos2\phi_{p}\right) \label{s_51} \eeq where plus
sign is for intra-pocket interaction and minus sign is for
inter-pocket interaction.

There are two new effects associated with the angular dependence
of ${\tilde u}_4 (k,p)$. First, when ${\tilde u}_4 u_5 > 2 u^2_3$,
the pairing vertex $\Gamma^{SC}$ not necessarily has an attractive
component. It was always the case for angle-independent ${\tilde
u}_4$. Now the existence of the attractive interaction is subject
to condition $K >0$, where

\beq \label{g_3} K = 2{\tilde u}_4 u_5 ((\alpha')^2 -\alpha'') +
u^2_3 (\alpha^2 + 2 \alpha'' -3 \alpha \alpha'). \eeq

If  we set the angular dependence of $u_3$ and ${\tilde u}_4$
terms to be equal, i.e., set $\alpha'=\alpha, \alpha''=0$, this
condition reduces to ${\tilde u}_4 u_5 > u^2_3$, which is well
satisfied. However, for a generic $\alpha'$ and $\alpha''$, Eq.
(\ref{g_3}) is not necessarily satisfied, and if $K <0$,
attractive $\Gamma^{SC}_1$ appears only above some RG scale $L$,
like in 2-pocket model.

\begin{figure}[t]
\includegraphics[width = 3in]{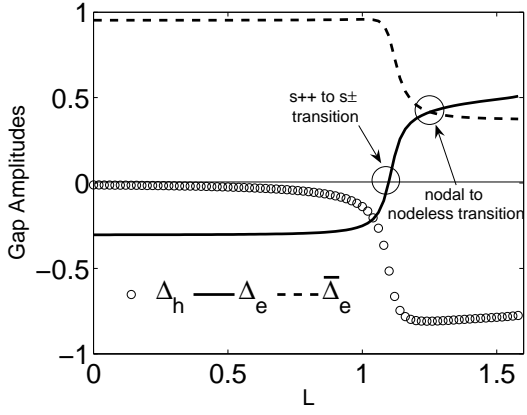}
\caption{\label{fig:deltas_Uangle}Behavior of the different
$\Delta$'s when the angular dependence of the electron
intra--pocket coupling $u_4$ is included. As before, we set
$\alpha=0.4$. The only difference compared to Fig.
\protect\ref{fig:deltas_Unoangle} is the appearance of the region,
at small $L$, where SC order parameter has nodal $s++$ character
meaning that $\Delta_h$ and $\Delta_e$ are of the same sign. In
this range of $L$ the SC vertex is, however, smaller than the SDW
vertex. The character of the SC gap changes to nodal $s\pm$ and then to no-nodal $s\pm$  before SC
vertex takes over SDW vertex.}
\end{figure}

\begin{figure}[t]
\includegraphics[width = 3in]{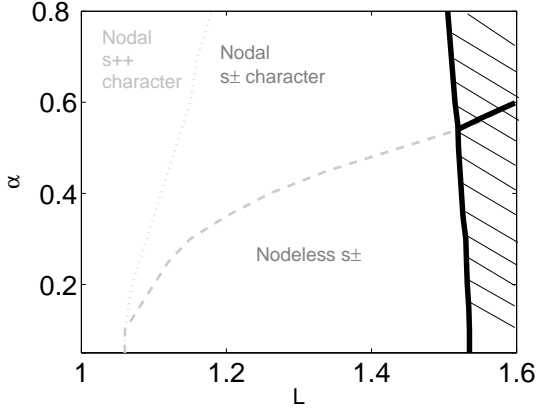}
\caption{\label{fig:special_nodal_pd} Same as in Fig.
\protect\ref{fig:new_nodal_pd}, but with the angular dependence of
the electron intra--pocket  coupling $u_4$ included. The only new
feature is the existence of a range where SC vertex (secondary in
this range to SDW vertex) has  nodal s++ character meaning that
the gaps along electron FSs have nodes, but the average value of
the gap along the electron FS is of the same sign as the gap along
the hole FS.}
\end{figure}

Second, the gap structure may change in some range of $L$. To
demonstrate this, make angular dependence of $u_3$ and ${\tilde
u}_4$ equal, i.e., set
 $\alpha'=\alpha, \alpha''=0$.
The set of equations for the SC vertices then becomes
 \bea \label{eq:SC 1H2E} \left(
\begin {array}{ccc}
 1-u_5 L & -2u_3 L &  -2\alpha u_3 L\\
 -u_3 L& 1-\tilde{u}_4 L  & -\alpha \tilde{u}_4 L\\
 -2\alpha u_3 L& -2\alpha \tilde{u}_4 L & 1\\
\end{array}
\right) \left(
\begin {array}{c}
  \Delta^o_h\\
  \Delta^o_{e}\\
  \bar{\Delta}^o_{e}\\
\end{array}
\right) = \left(
\begin {array}{c}
  \Delta_h\\
  \Delta_e\\
  \bar{\Delta}_{e}\\
\end{array}
\right)\eea As before, we need to diagonalize this set, cast the
result in the form $\Delta_i = \Delta^o_i (1 + \Gamma_i L)$ and
consider the largest $\Gamma_i$. The evolution with $L$ of
$\Delta_h$, $\Delta_e$, and $\bar{\Delta}_{e}$ for such $\Gamma_i$
is shown  in Fig \ref{fig:deltas_Uangle}, and the phase diagram is
shown in Fig. \ref{fig:special_nodal_pd}.  We see that, over some
range when $\bar{\Delta}_{e}$ is the largest and the gap has nodes
along the electron FSs,  the gap actually has $''nodal ~s++''$
character in the sense that $\Delta_h$ and $\Delta_e$ are of the
same sign, although the dominant term is still the oscillating
component $\bar {\Delta}_e$. Note, however, that the character of
the gap changes back to $s\pm$  before it becomes nodeless.

This appearance of the nodal $s++$ like gap might seem unusual,
but it should be kept in mind that  this gap is present in the
parameter range where without the angular dependence there
wouldn't have been a solution. The firm requirement then is that
in the solution induced by $\alpha$  the oscillating ${\bar
\Delta}_e$ component is the largest, because this is the way to
minimize the effect of intra-pocket Coulomb repulsion.  The
relative sign  between the subleading $\Delta_h$ and $\Delta_e$
terms is not uniquely determined by this requirement and be either
minus or plus, depending on the interplay between electron-hole
and electron-electron interactions.

These two potential changes introduced by the angular dependence
of $\tilde {u}_4$,  however, affect only the behavior at small
$L$. At large $L$ (i.e., at low energies), the system behavior
remains unchanged: SDW vertex is the largest at small/intermediate
$L$, but  SC vertex still crosses over SDW vertex at some $L$, and
beyond this scale SC instability comes first.

\subsection{The 4-pocket model in the limit when two hole FSs are identical}

We now consider the opposite limit where the two hole FSs centered
at $(0,0)$ are equivalent. We show  that the system behavior in
this second limit is the same as in the first. The equivalence of
the two limits hints that the system behavior in the intermediate
case is very likely the same as in the two limits.

The computations in the case of two identical hole FSs proceed in
the same way as before, but there are more vertices. The new terms
are  $u_1$, $u_2$, and $u_3$ interactions with hole fermions from
the second hole bands, the analogs of these three interactions for
fermions from the two hole bands, $u_5$ interaction for the second
hole band, and the interactions of the kind $u^*_5\sum
c_1^{\dag}c_1^{\dag}c_1c_2$.  Note that there are no $
f_1^{\dag}f_1^{\dag}f_1f_2$ terms for fermions from the two
electron bands because they would violate momentum conservation.

The full set of RG equations is rather cumbersome, but we verified
that (i) RG flow indeed preserved the invariance between the two
hole bands, and (ii)  all intra-pocket and inter-pocket
interactions involving fermions from the hole bands flow to the
same value $u_5$.  The analysis based on 5-orbital Hubbard model
also yields near-equivalence of all $u_i$ involving fermions from
hole pockets~\cite{bib:unp2}.

To simplify the presentation we set all interactions involving
fermions near hole FSs to be equal to $u_5$ from the start and
also neglect the angular dependence of the interactions. We also
set $u_4 =u_8$ because $u_4-u_8 >0$ again flows to zero under RG
(see paragraph before Eq. \ref{eq:RG 1H2E_1}).

\subsubsection{The Vertices}

The equations for the SC and SDW vertices are obtained in the same
way as before (see Fig. (\ref{fig:sc_sdw_1H2E})), but now
$\Delta_h$ is composed from fermions with $k$ and $-k$ belonging
to either of the two hole pockets. This leads to the equations for
the SDW vertex $\Delta_1$ and  SC vertices $\Delta_h$ and
$\Delta_e$ in the form \beq \Delta_1 = \Delta^o_1 \left(1 +
{\tilde u}_1 + {\tilde u}_3\right) \label{eq:SDW 2H2E} \eeq and
\bea \label{eq:SC 2H2E} \left(
\begin {array}{cc}
 1 -{\tilde u}_5 L & -{\tilde u}_3 L\\
-2{\tilde u}_3 L  & 1 -{\tilde u}_4 L\\
\end{array}
\right) \left(
\begin {array}{c}
 \Delta^o_h\\
 \Delta^o_e\\
\end{array}
\right) = \left(\begin {array}{c}
  \Delta_h\\
  \Delta_e\\
\end{array}\right) \eea
where $\tilde{u}_1=2u_1$, $\tilde{u}_3=2u_3$, $\tilde{u}_4=2u_4$ and
$\tilde{u}_5=4u_5$. Casting the results into $\Delta_i = \Delta_i^o (1 + \Gamma_i L)$, we obtain
 \bea\label{eq:4_pocket_other limit}
\Gamma^{SDW}&=&\tilde{u}_1+\tilde{u}_3\nonumber\\
\Gamma^{s\pm}&=& \frac{-(\tilde{u}_4+\tilde{u}_5) +
\sqrt{(\tilde{u}_4-\tilde{u}_5)^2+8(\tilde{u}_3)^2}}{2} \eea

\subsubsection{Parquet RG equations}

 The RG equations are obtained in the same way as before and are
\bea \label{4_pocket_other limit2}
\dot{\tilde{u}}_5 &=& -\left[\tilde{u}_5^2 + 2 \tilde{u}_3^2\right] \nonumber\\
\dot{\tilde{u}}_4 &=& -\left[\tilde{u}_4^2 + 2\tilde{u}_3^2 \right]\nonumber\\
\dot{\tilde{u}}_1 &=& +\left[\tilde{u}_1^2 + \tilde{u}_3^2\right]\nonumber\\
\dot{\tilde{u}}_3
&=&+\left[4\tilde{u}_1\tilde{u}_3-\tilde{u}_5\tilde{u}_3-\tilde{u}_4
\tilde{u}_3\right] \eea
We drop $u_2$ for simplicity as it eventually becomes smaller than other $u_i$

Comparing these equations and the equations for the vertices with
those for one hole FS (Eqs. \ref{eq:gap_eq_sp2} and \ref{eq:RG
1H2E_1}) we see that they are identical up to renormalizations
$u_i \to {\tilde u}_i$. Accordingly, the flow of the couplings and
the vertices is the same as in the limit when only one hole FS is
present. In both limits, SC vertex is secondary to SDW vertex at
large energies, but has larger slope and crosses over SDW vertex
at some energy, before the system reaches a fixed point. At
smaller energies, SC vertex is larger, i.e. if parquet RG flow
extends beyond the scale where the two vertices cross, the system
first develops a SC order even at perfect nesting. This SC order
can be either with or without nodes in the gaps along the two
electron FSs (see Fig. \ref{fig:phase_schem}).  The only
difference to the effective 3-pocket model is that now  at the
fixed point we have $\Delta_e = - \sqrt{2} \Delta_h$ for $\alpha
=0$ as opposed to $\Delta_h = - \sqrt{2} \Delta_e$ for the earlier
case.

As we said, the equivalence of the system behavior in the two
limits strongly suggests that the same behavior holds also in the
intermediate case.

\subsection{Summary of the results for the 4-pocket model}

Collecting all the points we have discussed  --
 we have shown that under suitable extent
of renormalization of the Coulomb repulsion and pair-hopping
couplings one can have SDW, nodal $s\pm$, and nodeless $s\pm$
state even at perfect nesting. The angular dependence of the
interaction between holes and electrons tends to drive the system
towards a nodal SC phase. The SC order develops if the fixed point
is reached within parquet RG, but if the scale $E_F$ is reached
before that, the system develops either SDW or SC order (either
nodeless or nodal), depending on at what $L$ the flow crosses over
from parquet to ladder RG. That the SC $s\pm$ order can emerge
even at perfect nesting is specific feature of the 4-pocket model.
This feature was not present in the 2-pocket model, where the
fixed point had an O(6) symmetry. This symmetry is clearly broken
in the 4-pocket model, even when $\alpha=0$. The `crossing' of the
SDW and SC vertices can be unambiguously attributed to the
presence of the other electron pocket because its presence helps SC
but not SDW.

Fig. \ref{fig:phase_schem} summarizes the implication of our
results towards the actual phase diagrams of Fe-pnictides. In the
SDW dominated region (red), SC  emerges after doping reduces SDW
order. In the other part where SC dominates, SC order prevails
already at perfect nesting. The $s\pm$ SC gap can be nodeless or
have nodes along electron FSs depending on  how strong is the
angular dependence of the interaction between electrons and holes.

\begin{figure}[htp]
\includegraphics[width = 2in]{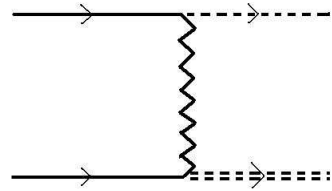}
\caption{\label{fig:forbidden} The scattering which takes two
fermions from the region near $k=0$ and scatters them to fermions
with momenta near $(\pi,0)$ and $(0,\pi)$. This process is not
allowed in the unfolded BZ because of momentum non-conservation, but
it is allowed as an umklapp process in the folded BZ, which knows
about $As$. }
\end{figure}

\emph{A final remark: }In the analysis above we considered only
the interactions which obey momentum conservation in the unfolded
BZ. These are direct interactions between fermionic states
obtained by the hybridization of 5 $Fe$ orbitals. There exists,
however, additional interactions which involve pnictide orbitals
as intermediate states. These additional interactions obey
momentum conservation in the folded BZ, but they do not always
obey momentum conservation in the unfolded, $Fe-$only BZ. An
example of such process is shown in  Fig. \ref{fig:forbidden}: two
fermions from the hole band near $k=0$ scatter into two fermions
at two {\it different} electron pockets. In the unfolded zone,
this process doesn't conserve momentum, and we didn't include it
into our consideration. In the folded zone, both electron FSs are
at $(\pi,\pi)$, and this process is an umklapp process. The
difference is indeed due to the fact that in reality such process
involves intermediate states on As.

Neither our RG procedure nor fRG calculations include such terms.
How important are they is not known.  On general grounds, such
interactions  tend to enhance the SDW vertex and might potentially
alter the picture that we presented. They also may alter the
ordering momentum of the SDW state.  This remains an open issue.

\section{5-pocket model}\label{section:4Bmodel}

We now extend the analysis from previous two sections to a
5-pocket model in which we include into consideration the
additional hole pocket appearing at $(\pi,\pi)$ point in the
unfolded BZ. We show below that the behavior of $5-$pocket model
is  similar to that for 2-pocket model in the sense that SDW
vertex exceeds SC vertex along the whole RG trajectory, and SDW
and SC vertices tend to the same value if the fixed point is
reached within parquet RG.

As in the previous section, we restrict our consideration to the
two limits, one when the two hole FSs centered at $(0,0)$ are
identical, and the other when one of these two hole FSs is
relatively weakly coupled to electronic states and can be
neglected. In the latter case, 5-pocket model reduces to an
effective 4-pocket model consisting of one hole FS at $(0,0)$, one
hole FS at $(\pi,\pi)$, and the two electron FSs at $(0,\pi)$ and
$(\pi,0)$. We show that the system behavior is again identical in
the two limits.

\begin{figure}[htp]
\includegraphics[width = 2.5in]{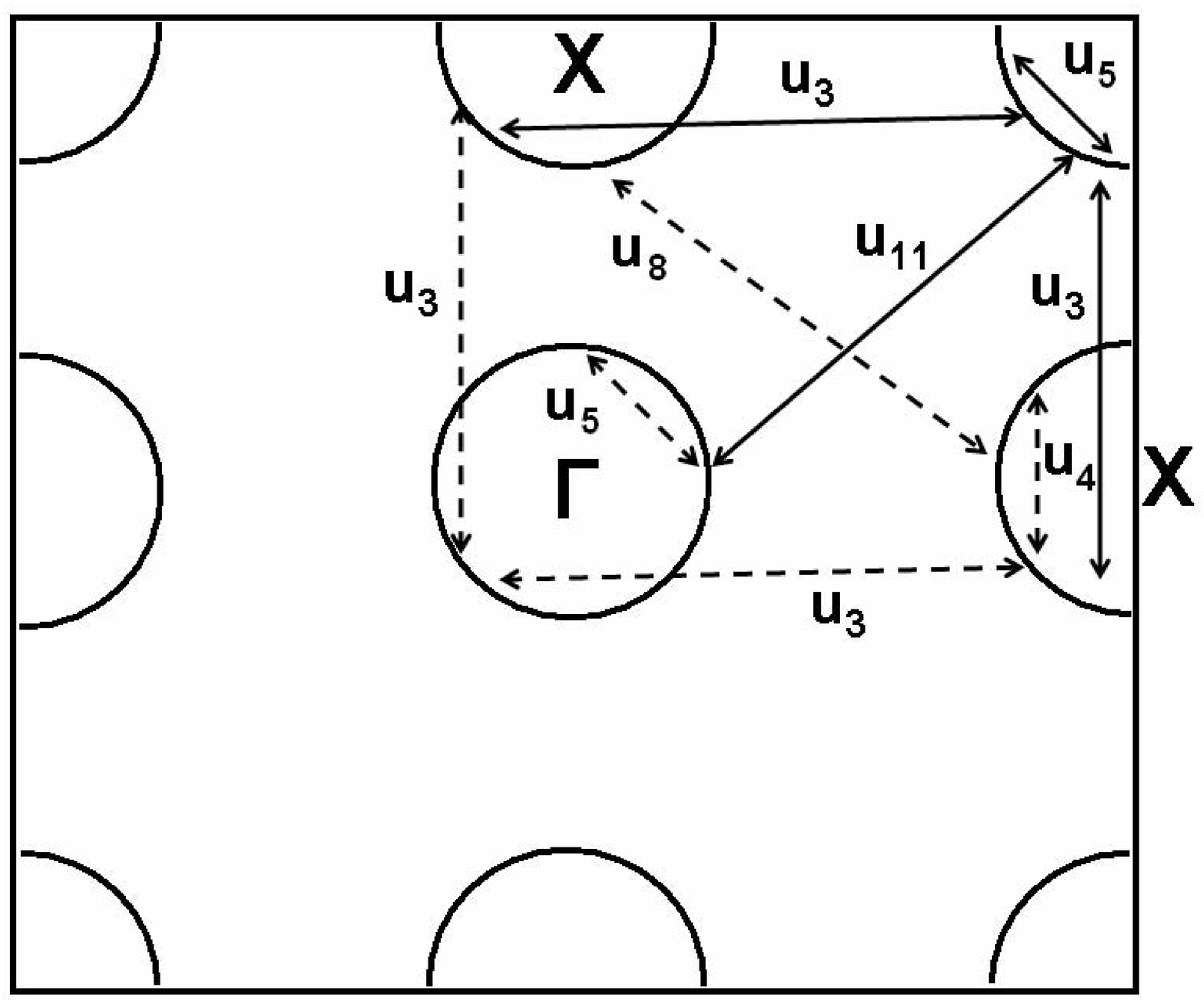}
\caption{\label{fig:UBZ_4band} The FSs and interactions in the
5-pocket model. Dashed lines mark the interactions already present
in 4-pocket model, solid lines mark the new pairing interactions
specific to 5-pocket model. As before, we only present
interactions which contribute to the pairing vertices. There are
other density-density and exchange interactions between electrons
belonging to different pockets.}
\end{figure}

\subsection{Effective model with one hole pocket at $(0,0)$}

The FS geometry and interactions contributing  to the SC vertex
for the effective 4-pocket model with hole pockets at $(0,0)$ and
$(\pi,\pi)$ are presented in Fig. \ref{fig:UBZ_4band}

\begin{figure}[htp]
\includegraphics[width = 3in]{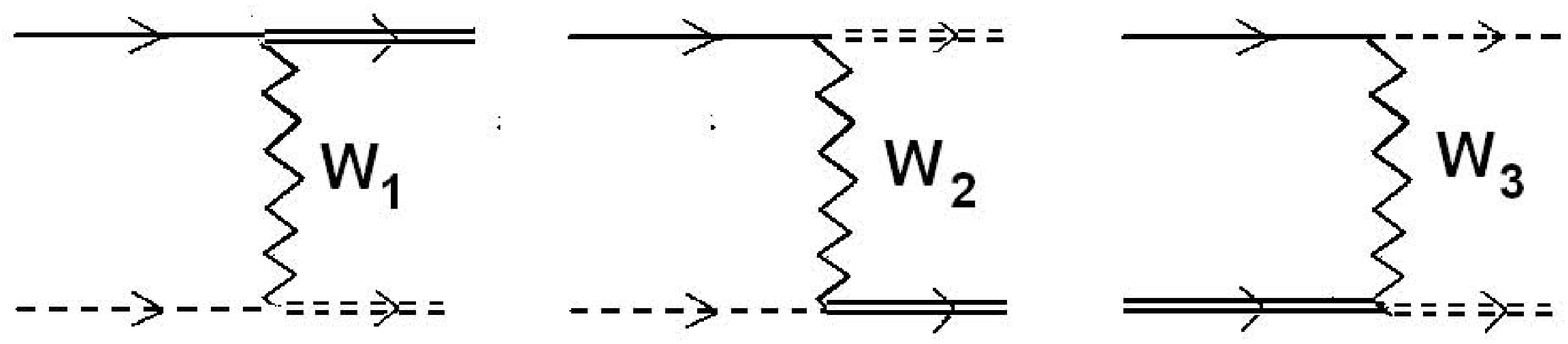}
\caption{\label{fig:W_vertex} The new interaction vertices for the
5-pocket model. Single and double solid lines denote fermions from
the two hole pockets, single and double dashed lines denote
fermions from the two electron pockets.}
\end{figure}

The Hamiltonian now contains three new terms $u_9$, $u_{10}$, and
$u_{11}$, which are density-density, exchange, and pair-hopping
interaction between fermions belonging to two different hole
pockets. In addition, we have three new vertices shown in Fig.
\ref{fig:W_vertex}. These include fermions from two different hole
and two different electron FSs. We call them $w_i$ vertices ($i$
runs from 1 to 3).

The Hamiltonian now has the form

\begin{widetext}
\bea \label{eq:H_int 2H2E} \frac{m}{2\pi} H_{int}&=&
        \sum u_1^{(1)}\;c^\dag_{1 p_1 s}f^\dag_{1 p_2 s'} f_{1 p_4 s'}c_{1 p_3
        s} +
        \sum u_2^{(1)}\;c^\dag_{1 p_1 s}f^\dag_{1 p_2 s'} c_{1 p_4 s'}f_{1 p_3
        s} +
        \sum \frac{u_3^{(1)}}{2}
        \left(
        c^\dag_{1 p_1 s}c^\dag_{1 p_2 s'} f_{1 p_4 s'}f_{1 p_3
        s}\;+\;h.c.
        \right)\nonumber\\
        &+& f_1 \leftrightarrow f_2\;(with~c_1~unchanged)~and\;u_i^{(1)} \leftrightarrow u_i^{(2)}
        ~+~ c_1 \leftrightarrow c_2\;(with~f_1~unchanged)~and\;u_i^{(1)} \leftrightarrow u_i^{(3)}\nonumber\\
        &+& c_1 \leftrightarrow c_2\;(with~f_2~unchanged)~and\;u_i^{(2)} \leftrightarrow u_i^{(4)}\nonumber\\
        &+& \sum \frac{u_4^{(1)}}{2}   f^\dag_{1 p_1 s}f^\dag_{1 p_2 s'}
        f_{1 p_4 s'}f_{1 p_3s} +
        \sum \frac{u_4^{(2)}}{2}  f^\dag_{2 p_1 s}f^\dag_{2 p_2 s'} f_{2 p_4 s'}f_{2 p_3s}\nonumber\\
        &+& \sum \frac{u^{(1)}_5}{2} c^\dag_{1 p_1 s}c^\dag_{1 p_2 s'}c_{1 p_4 s'}c_{1 p_3
        s} +
        \sum \frac{u^{(2)}_5}{2} c^\dag_{2 p_1 s}c^\dag_{2 p_2 s'}c_{2 p_4 s'}c_{2 p_3
        s}\nonumber\\
        &+& \sum u_6 \; f^\dag_{1 p_1 s}f^\dag_{2 p_2 s'} f_{2 p_4 s'}f_{1 p_3
        s} +\sum u_7\;  f^\dag_{1 p_1 s}f^\dag_{2 p_2 s'} f_{1 p_4 s'}f_{2 p_3
        s} +  \sum \frac{u_8}{2}
        \left(
        f^\dag_{1 p_1 s}f^\dag_{1 p_2 s'} f_{2 p_4 s'}f_{2 p_3
        s}\;+\;h.c. \right)\nonumber\\
        &+&c\leftrightarrow
        f~and~(u_6,u_7,u_8)\leftrightarrow(u_9,u_{10},u_{11})\nonumber\\
        &+& \sum w_1 \; c^\dag_{1 p_1 s}f^\dag_{1 p_2 s'}f_{2 p_4 s'}c_{2 p_3 s}\; +\;
        \sum w_2 \; c^\dag_{1 p_1 s}f^\dag_{1 p_2 s'}c_{2 p_4 s'}f_{2 p_3
        s}+
        \sum \frac{w_3}{2} \; c^\dag_{1 p_1 s}c^\dag_{2 p_2 s'}f_{1 p_4 s'}f_{2 p_3
        s}+(1\leftrightarrow2)...
\eea
\end{widetext}

The analysis of the 5-pocket model parallels that of the 4-pocket
model, so we will be rather brief and present only the results. We
verified that  RG equations for  $u_4, u_5, u_8$ and $u_{11}$ are
identical, and these four couplings tend to the same value at the
fixed point. To make the presentation compact, we set $u_4 = u_5 =
u_8 =u_{11}$ from the start and call all of them $u_4$. Similarly
we set $u_6 = u_7 = u_9 =u_{10}$ calling it $u_6$. It is
convenient to introduce ${\tilde u}_i = u_i + w_i$ ($i=1-3$), and
$\dtildeu_i = u_i - w_i$. We will use these variables below.

\subsubsection{The Vertices}

We first consider the case when the interactions  are angle-independent ($\alpha =0$) and then discuss system behavior at a nonzero $\alpha$.

The SC and SDW vertices are obtained in the same way as before,
but there are additional terms for the SDW term due to $w_i$
vertices (see Fig. \ref{fig:sdw_2H2E}).  Combining this with the
equations for the SC vertices at $(0,0)$ and $(\pi,\pi)$
($\Delta_{h1}$ and $\Delta_{h2}$ respectively) we obtain \bea
\label{eq:SC5} \left(
\begin {array}{ccc}
 1-2u_4 L & -u_3 L &  - u_3 L\\
 -2 u_3 L& 1- u_4 L  & -u_4 L\\
 -2 u_3 L& -u_4 L & 1-u_4 L\\
\end{array}
\right) \left(
\begin {array}{c}
  \Delta^o_e\\
  \Delta^o_{h1}\\
  \Delta^o_{h2}\\
\end{array}
\right) = \left(\begin {array}{c}
  \Delta_e\\
  \Delta_{h1}\\
  \Delta_{h2}\\
\end{array}\right)\eea

and

\beq \label{eq:SDW5} \Delta_1 = \Delta^o_1 (1 + ({\tilde u}_1 +
{\tilde u}_3)L) \eeq

(we recall that we set $u_4=u_5$). Casting the equations for the
SC vertices  in the form $\Delta_i = \Delta^o_i (1 + \Gamma_i)$
and neglecting repulsive vertex for $s^{++}$ SC, we obtain \bea
&&\Gamma^{SDW}=\tilde{u}_1+\tilde{u}_3\nonumber\\
&&\Gamma^{s\pm} = - 2 u_4 + \tilde {u}_3 + \dtildeu_3
\label{g_8}
\eea

Note that SDW and SC vertices contain different terms involving
$u_3$. The gap structure for the SC vertex $\Gamma^{s\pm}$ is
$\Delta_{h1} = \Delta_{h2} =-\Delta_e$.

\subsubsection{RG flow between $\Lambda$ and $E_F$}

The set of parquet RG equations is obtained in the same way as
before. Collecting the equations for the other variables we obtain
\bea \label{eq:2H2E RG}
\dot{{\tilde u}}_1 &=& \tilde {u}_1^2 + {\tilde u}_3^2\nonumber\\
\dot{{\tilde u}}_2 &=& 2\tilde {u}_1{\tilde u}_2-2{\tilde u}_2^2\nonumber\\
\dot{{\tilde u}}_3 &=& 4\tilde {u}_1 {\tilde u}_3 - 2{\tilde u}_2
{\tilde u}_3
- 2 {\tilde u}_3 (u_4+u_6) - 2 \dtildeu_3 (u_4-u_6)\nonumber\\
2\dot{u}_6 &=& -(2u_6)^2 - (\tilde {u}_3 -
{\dtildeu}_3)^2\nonumber\\
2\dot{u}_4 &=& -(2u_4)^2 - (\tilde {u}_3 + {\dtildeu}_3)^2
\nonumber\\
\dot{{\dtildeu}}_3 &=& 4\dtildeu_1 {\dtildeu}_3 - 2{\dtildeu}_2
{\dtildeu}_3 - 2{\dtildeu}_3 (u_4+u_6)-2 {\tilde u}_3 (u_4-u_6)\nonumber\\
\dot{{\dtildeu}}_1 &=& \dtildeu_1^2 + {\dtildeu}_3^2\nonumber\\
\dot{{\dtildeu}}_2 &=& 2\dtildeu_1{\dtildeu}_2-2{\dtildeu}_2^2
 \eea

This set of equations almost decouples between the subsets for
${\tilde u}_i$ and ${\dtildeu}_i$, the only places where the two
subsets mix are the equations for the flow of $u_4$ and $u_6$
whose r.h.s. contains both ${\tilde u}_3$ and ${\dtildeu}_3$.
Re-writing this set as equations for the ratios of the couplings,
we found four fixed points. One corresponds to ${\dtildeu}_i$
vanishing compared to ${\tilde u}_i$, another to  ${\tilde u}_i$
vanishing compared to ${\dtildeu}_i$, the third corresponds to
${\dtildeu}_3 = {\tilde u}_3$, ${\dtildeu}_1 = {\tilde u}_1$, and
the fourth corresponds to ${\dtildeu}_3 = - {\tilde u}_3$,
${\dtildeu}_1 = {\tilde u}_1$.  The first two fixed points are
attractive, the other two are saddle points.  Which fixed point
the system will flow to depends on the initial conditions. In
our case all interactions are positive (repulsive), i.e. at the
bare level ${\tilde u}_i$ are all positive and ${\tilde u}_i >
{\dtildeu}_i$.  For these initial conditions, we verified that the
system is outside the base of attraction of the second fixed
point as it can be reached only if bare $w_i$ are negative (at
this fixed point ${\tilde u}_i$ vanishes compared to
${\dtildeu}_i$ i.e., $w_i/u_i$ tends to $-1$).

At the first attractive fixed point  ${\dtildeu}_i$ vanishes
compared to ${\tilde u}_i$ i.e., $w_i/u_i$ tends to $1$. This is
consistent with our initial conditions. Near this fixed point,  ${\dtildeu}_3$
can be neglected compared to ${\tilde u}_3$ in
the equations for ${\dot u}_4$ and ${\dot u}_6$, and the first five RG
equations form a closed set:

\bea \label{eq:2H2E RG_1}
\dot{{\tilde u}}_1 &=& {\tilde u}_1^2 + {\tilde u}_3^2\nonumber\\
\dot{{\tilde u}}_2 &=& 2{\tilde u}_1{\tilde u}_2-2{\tilde u}_2^2\nonumber\\
\dot{{\tilde u}}_3 &=& 4{\tilde u}_1 {\tilde u}_3 - 2{\tilde u}_2 {\tilde u}_3
 - 2 {\tilde u}_3 (u_4+u_6) \nonumber\\
2\dot{u}_4 &=& -(2u_4)^2 - {\tilde u}^2_3 \nonumber\\
2\dot{u}_6 &=& -(2u_6)^2 - {\tilde u}^2_3\eea

Within the same approximation \bea
\label{g_10}
&&\Gamma^{SDW}=\tilde{u}_1+\tilde{u}_3\nonumber\\
&&\Gamma^{s\pm} = - 2 u_4 + {\tilde u}_3 \eea

\begin{figure}[htp]
\includegraphics[width = 3in]{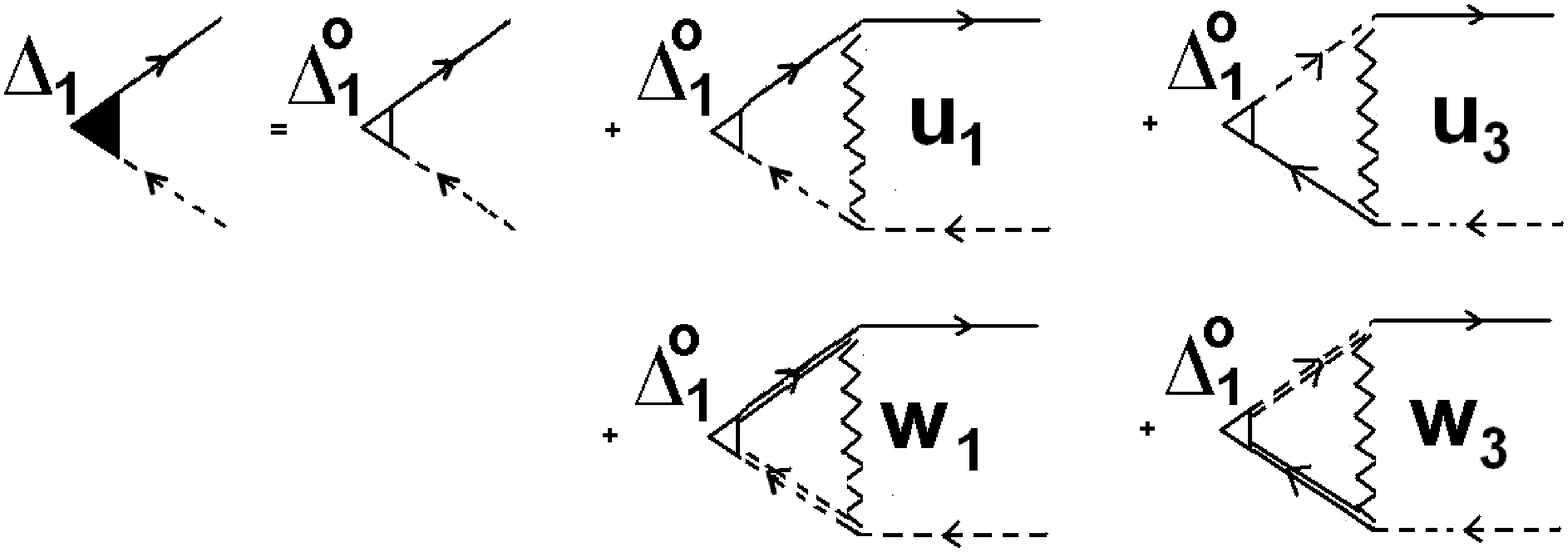}
\caption{\label{fig:sdw_2H2E} The diagrammatic equation for the
renormalized SDW vertex in the 5-pocket model. Comparing with the
corresponding  Fig. \protect\ref{fig:sc_sdw_1H2E}a for the
4-pocket case, there are extra diagrams which contribute to the
SDW vertex. This leads to effectively doubling $\Gamma^{SDW}$.}
\end{figure}

Comparing these equations with the ones we obtained for the
2-pocket model, Eqs. (\ref{eq:gamma}) and (\ref{eq:1H1E above E_f}),
we see that they are equivalent, up to overall renormalizations of
the couplings, if we identify $2u_6$ in Eq. (\ref{eq:2H2E RG_1}) with
$u_5$ in Eq. (\ref{eq:1H1E above E_f}). There is minor
 difference between the $\Gamma^{s\pm}$
in the two cases ($2u_4$ in (\ref{g_10}) vs $u_4+u_5$ in \ref{eq:gamma}),
but it vanishes at the fixed point.
Acordingly, the RG flow of the couplings and the vertices is
 the same as in the 2-pocket model, namely SDW vertex
remains dominant for all $L$ up to a fixed point, and SC vertex
changes sign at some $L$, becomes attractive at larger $L$ and
becomes equal to the SDW vertex at the fixed point if, indeed, the
fixed point is reached within parquet RG. This similarity with a
2-pocket model was first noted by K. Haule~\cite{haule} and can be
understood if we note that in the 5 pocket case SC pairing is the
same as in 4-pocket model (with extra combinatoric factor of 2
compared to 2-pocket case), but SDW pairing is now possible
between the two sets of electron pockets (see Fig.
\ref{fig:sdw_2H2E}), this adds the combinatoric factor of 2 also
to the renormalization of the SDW vertex.

We now need to understand what is the basis of attraction for this
fixed point. For this, we consider the two other fixed points for
which  ${\dtildeu}_3 = {\tilde u}_3$, ${\dtildeu}_1 = {\tilde
u}_1$, or ${\dtildeu}_3 = - {\tilde u}_3$, ${\dtildeu}_1 = {\tilde
u}_1$. We show that both are saddle points, and both are unstable
when the bare $u_i$ and $w_i$ are all positive.

Consider for example the fixed point at ${\dtildeu}_1 = {\tilde
u}_1$ and ${\dtildeu}_3 = {\tilde u}_3$. At this fixed point
$u_4/{\tilde u}_1 =-3$, ${\tilde u}_3/{\tilde u}_1 = \sqrt{15}$
and $u_6=0$. Expanding in $\delta = \tilde{u}_1 - {\dtildeu}_1$
and $\epsilon = {\tilde u}_3 - {\dtildeu}_3$, we obtain the set of
coupled linear differential equations

\bea
&&{\dot \delta} = 2 \tilde {u}_1 \left(\delta + \sqrt{15} \epsilon\right) \nonumber \\
&&{\dot \epsilon} = 4 \tilde {u}_1 \left(\sqrt{15}\delta + 4
\epsilon\right) \label{g_4} \eea

together with $\dot{\tilde{u}}_1 = \tilde{u}^2_1 + \tilde {u}^2_3
= 16 \tilde {u}^2_1$. The solution of the set is $\epsilon =
\epsilon_0 \left({\tilde u}_1\right)^\gamma,~\delta = \delta_0
\left({\tilde u}_1\right)^\gamma$. Substituting and solving for
the set of two linear equations for $\epsilon_0$ and $\delta_0$,
we obtain $\gamma_1 = 11/8$ and $\gamma_2 =-1/4$. For $\epsilon$
and $\delta$ which correspond to $\gamma = \gamma_1$, the fixed
point is unstable, for $\gamma = \gamma_2$, it is stable. A simple
analysis shows that $\gamma_1$ is the solution when
$\epsilon_0/\delta_0 >0$, while $\gamma_2$ is the solution when
$\epsilon_0/\delta_0 <0$. In our case, the bare values of
$\epsilon$ and $\delta$ are $2w_3$ and $2 w_1$, respectively, both
are positive. Hence this fixed point is unstable, and the RG flow
bring the system towards the stable fixed point at which
$\dtildeu_1$ and ${\dtildeu}_3$ are both small.  The stability
analysis of the fixed point at $\dtildeu_1 = \tilde u_1$, or
$\dtildeu_3 = - {\tilde u}_3$ yields the same results, leaving the
fixed point with $\dtildeu_i << {\tilde u}_i$ as the only stable
fixed point.

\begin{figure}[htp]
\includegraphics[width = 3in]{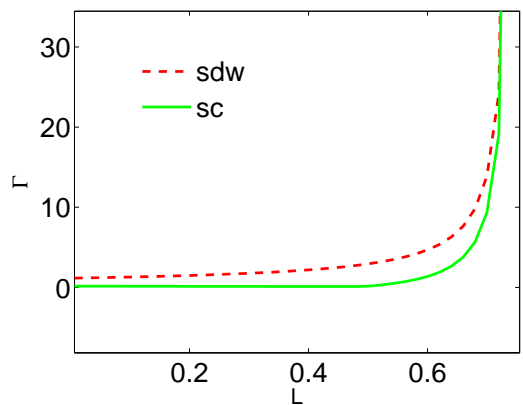}
\caption{\label{fig:H2E2_eff_vrtx} Parquet RG flow of SC and SDW
vertices for the 5-pocket model at  $\alpha =0.3$. The SDW vertex
remains the largest over the whole flow, and the ratio of the SDW
and SC vertices  approaches one at the fixed point of parquet RG.
This is very similar to the 2-pocket case except that here
$\Gamma^{SC}$ is  attractive for all $L$.}
\end{figure}

We next consider how the results are modified due to angular
dependence of the vertices. We found two effects. First, one of SC
vertices $\Gamma^{SC}_1$ can become attractive already at small
$L$ in the same way as in the  3-pocket model studies in the
previous section. Namely, the system adjusts $\cos 2 \phi$ and
angle-independent components of the gaps along the two electron
FSs to minimize the effect of intra-pocket Coulomb repulsion. Just
as for 3-pocket model, $\Gamma^{SC}_1$ is attractive and scales as
$\alpha^2$ if we only include angular dependence of the
pair-hopping $u_3$ and $w_3$ terms.   Second, SC and SDW vertices
do not become identical at the fixed point if $\alpha$ is nonzero.
If we only include angular dependence of $u_3, w_3$ and $u_1$ and
$w_1$ (and set them equal), we find that SC vertex becomes larger
than SDW vertex very near fixed point. However, the effect is
numerically very weak, even when $\alpha \sim 1$. In Fig.
\ref{fig:H2E2_eff_vrtx} we show the flow of SC and SDW vertices
for $\alpha =0.3$. SC vertex eventually becomes larger, but this
is truly weak effect.

The flow of SDW and SC vertices towards almost the same value in
the 5-pocket model has been found numerically  by Thomale et al
within fRG study~\cite{Thomale}. Our analytical RG results for
this case again agree with their fRG, what, in our opinion, is
another confirmation that the ``topology'' of the RG flow is
chiefly determined by combinatoric effects.

\subsubsection{RG flow below the scale of $E_F$}

The system behavior for the case when the fixed point of the
functional RG is not reached at $E > E_F$ is quite similar to the
2-pocket model with the only difference that now SC instability
always occurs when SDW order is destroyed by doping. Namely, at
perfect nesting the system develops SDW order. At finite doping,
the RG flow of the SDW vertex levels off, and SC vertex eventually
becomes larger. The SC gap has $s\pm$ structure either without or
with nodes along the electron FS, depending on the values of
$\alpha$ and of $\log \Lambda/E_F$. The phase diagram is similar
to that in Fig. \ref{fig:phase_schem}, but only has the ``SDW''
region in that figure.

\begin{figure}[htp]
\includegraphics[width = 3in]{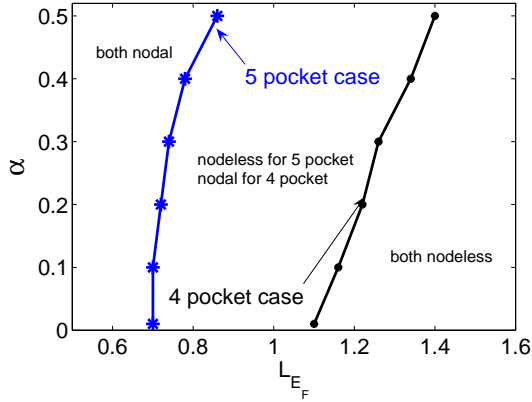}
\caption{\label{fig:comp_3_4_pocks} Comparison of the SC phases
for 4 and 5 pocket models.  The initial conditions are identical for the common parameters. The figure shows that there exists a parameter range
 where the SC gap is nodeless in the 5-pocket model but is nodal in the
 4-pocket model.}
\end{figure}

\subsection{5-pocket model with two equivalent hole FS at $(0,0)$}

We now consider the opposite limit of 5-pocket model when the
two hole pockets centered at $(0,0)$ are equivalent. Our goal is to verify whether the system behavior remains the same as in the limit when we keep only
one of these two hole pockets.

The computations in the case of two equivalent hole pockets at
$(0,0)$ are quite involved and we only present the results for
$\alpha =0$. Because the pockets at $(0,0)$ and $(\pi,\pi)$ are
now non-equivalent (in the sense that there are two pockets at
$(0,0)$ and only one at $(\pi,\pi)$),
 the interactions involving these pockets do not need to flow to the same value under RG, e.g, $u_i$ and $w_i$ need not to fly to the same value, and also
electron-hole and hole-hole interactions involving fermions from
near $(0,0)$ and $(\pi,\pi)$ need not to be the same. Finally, SC
gaps on the hole FSs at $(0,0)$ and $(\pi,\pi)$ also do not have
to be equal.

\subsubsection{The Vertices}

\begin{figure}[htp]
\includegraphics[width = 2.5in]{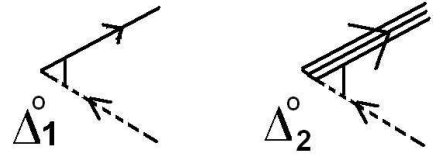}
\caption{\label{fig:sdw_new} The two non-equivalent SDW vertices
in the 5-pocket model with two identical hole FSs at $(0,0)$. The
triple solid line stands for the fermion (hole) at $(\pi,\pi)$.
These two vertices are present also for the case when there is
only one hole FSs at $(0,0)$, but in that case we set $\Delta^o_1
= \Delta^o_2$ and verified that the equivalence also holds for
renormalized $\Delta_{1,2}$.}
\end{figure}

\begin{figure}[htp]
\includegraphics[width = 2.5in]{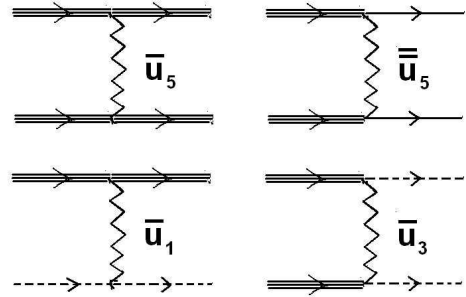}
\caption{\label{fig:5p_new_vtx} The interactions involving
fermions near the hole FS at $(\pi,\pi)$ (the triple solid line).
The dashed line stands for a fermion from an electronic pocket.
For the case of only one hole FSs at $(0,0)$ we set ${\bar u}_5 =
{\bar {\bar u}}_5 = u_5$, ${\bar u}_1 = u_1$, and ${\bar u}_3
=u_3$ from the beginning and verified that these relations hold
for running couplings. For the case of two identical hole pockets
at $(0,0)$, the fixed point values of the couplings involving
fermions near $(\pi,\pi)$ are different from those involving
fermions near $(0,0)$.}
\end{figure}

Keeping all this in mind and applying the same analysis as before
we obtain the equations for the SDW and SC vertices.  For the SDW,
we introduce two vertices $\Delta_1$ and $\Delta_2$, shown in Fig.
\ref{fig:sdw_new} and write the set of $2\times 2$ coupled
equations as \bea \label{eq:SDW 3H2E} \left(
\begin {array}{cc}
 1+2(u_1+u_3) L & (w_1+w_3) L\\
 2(w_1+w_3) L  & 1 + ({\bar u}_1 + {\bar u}_3) L\\
\end{array}
\right) \left(
\begin {array}{c}
 \Delta_1^o\\
 \Delta_2^o\\
\end{array}
\right) = \left(\begin {array}{c}
  \Delta_1\\
  \Delta_2\\
\end{array}\right) \eea
where the vertices ${\bar u}_1$ and ${\bar u}_3$ are shown in Fig.
\ref{fig:5p_new_vtx}

For the SC vertex we introduce, as before, $\Delta_e$,
$\Delta_{h1} = \Delta  (k=(0,0))$ and $\Delta_{h2} = \Delta
(k=(\pi,\pi))$ and obtain \bea \label{eq:SC 3H2E} \left(
\begin {array}{ccc}
 1-2u_4 L & -4u_3 L &  -{\bar u}_3 L\\
 -2u_3 L& 1-4 u_5 L  & -{\bar u}_5 L\\
 -2{\bar u}_3 L& -4 {\bar u}_5 L & 1- \bar {\bar u}_5 L\\
\end{array}
\right) \left(
\begin {array}{c}
  \Delta^o_e\\
  \Delta^o_{h_1}\\
  \Delta^o_{h_2}\\
\end{array}
\right) = \left(\begin {array}{c}
  \Delta_e\\
  \Delta_{h_1}\\
  \Delta_{h_2}\\
\end{array}\right)\eea
The vertices ${\bar u}_5$ and ${\bar {\bar u}}_5$ are shown in Fig
\ref{fig:5p_new_vtx}

\subsubsection{Parquet RG equations}

The total number of RG equations is quite large and we refrain
from writing all of them. Quite predictably, the combinatoric
factors associated with the existence of the two equivalent
pockets at $(0,0)$ give rise to relations  ${\bar u}_i = 2 u_i$
($i=1,3,5$), $u_4 = 4 u_5$, and $\bar {\bar u}_5 = 4 u_5$. Using
these relations, we obtain for the relevant couplings $u_1, w_1,
u_3, w_3$ and $u_4$ the set

\bea \label{g_6}
\dot{u}_1 &=& 2 (u_1^2 + u_3^2) + w^2_1 + w^2_3\nonumber\\
\dot{w}_1 &=&  4 w_1 u_1 + 4 w_3 u_3\nonumber\\
\dot{u}_3 &=&  4(2u_1u_3 + w_1w_3) -4 u_3 u_4  \nonumber\\
\dot{w}_3 &=& 8(u_1w_3+u_3w_1) -4u_6w_3\nonumber \\
2\dot{u}_4 &=& -(2u_4)^2 - 16u^2_3 \nonumber\\
2\dot{u}_6 &=& -(2u_6)^2 - 8w^2_3 \eea

Introducing now ${\tilde u}_1 = 2 u_1 + \sqrt{2} w_1,~\dtildeu_1 =
2 u_1 - \sqrt{2} w_1$, ${\tilde u}_3 = 2 u_3 + \sqrt{2}
w_3,~\dtildeu_3 = 2 u_3 - \sqrt{2} w_3$ and substituting into
(\ref{g_6}) we obtain

\bea \label{eq:3H2E RG}
\dot{{\tilde u}}_1 &=& {\tilde u}_1^2 + {\tilde u}_3^2\nonumber\\
\dot{{\tilde u}}_3 &=& 4{\tilde u}_1 {\tilde u}_3 - 2 {\tilde u}_3 (u_4+u_6) - 2 \dtildeu_3 (u_4-u_6)\nonumber\\
2\dot{u}_4 &=& -(2u_4)^2 - ({\tilde u}_3 + {\dtildeu}_3)^2
\nonumber\\
2\dot{u}_6 &=& -(2u_6)^2 - ({\tilde u}_3 -{\dtildeu}_3)^2
\nonumber\\
\dot{{\dtildeu}}_1 &=& {\dtildeu}_1^2 + {\dtildeu}_3^2\nonumber\\
\dot{{\dtildeu}}_3 &=& 4{\dtildeu}_1 {\dtildeu}_3  - 2{\dtildeu}_3
(u_4+u_6)- 2{\tilde u}_3 (u_4-u_6) \eea

This is exactly the same set as Eq. (\ref{eq:2H2E RG}) that we
obtained in the previous subsection (we skip the equation on $u_2$
which is irrelevant coupling anyway).

Under the same conditions ( ${\bar u}_i = 2 u_i$ ($i=1,3,5$), $u_4
= 4 u_5$, and $\bar {\bar u}_5 = 4 u_5$), the relevant SDW and SC
vertices become \bea
\Gamma^{SDW} &=& {\tilde u}_1 + {\tilde u}_3 \nonumber \\
\Gamma^{s\pm} &=& -2u_4 + {\tilde u}_3 + \dtildeu_3 \label{g_7}
\eea

These again are the same equations as Eqs (\ref{g_8}) for the case
of only one hole FS at $(0,0)$.  The only difference with the
other limit is that now the solutions corresponding to
$\Gamma^{SDW}$ and $\Gamma^{s\pm}$ from (\ref{g_7}) are $\Delta_2
= \sqrt{2} \Delta_1$ and $\Delta_{h2} = 2\Delta_{h1} = -\Delta_e$.

We see therefore that the system behavior in the two limits is
identical. Like in the 4-pocket case, this equivalence strongly
suggests that the same behavior holds also in the intermediate,
most generic 5-pocket model,  when the two hole pockets at $(0,0)$
are both present but are not identical.

\subsection{Summary of the results for 5-pocket model}

We see that the system behavior in a 5-pocket model is
``intermediate'' between 2-pocket and 4-pocket models. On one
hand, like in  a 4-pocket model, the largest SC vertex can be
positive already at the smallest $L$, even when intra-pocket
Coulomb repulsion is the dominant interaction.  If this is the
case, then there is no critical $L$ before which SC vertex is
repulsive, and  the system always becomes a SC when the competing
SDW instability is reduced.  The SC gap is either nodeless or with
nodes on electron FSs, depending on $\alpha$, much like in the
unshaded region in Fig. \ref{fig:new_nodal_pd}. On the other hand,
like in a 2-pocket model, SDW vertex remains larger than SC vertex
for all $L$ before the fixed point is reached, and the two
vertices flow to the same value at the fixed point of parquet RG
(Fig. \ref{fig:H2E2_eff_vrtx}). This last statement is exact when
$\alpha =0$ and remains numerically quite accurate even when
$\alpha \neq 0$ although strictly speaking, at a finite $\alpha$,
SC vertex eventually becomes larger than SDW vertex in the
immediate vicinity of the fixed point.

\subsection{Comparison of 4-pocket and 5-pocket models}

It is instructive to compare the structures of the SC gaps in
5-pocket and 4-pocket models for the same values of input
parameters (and using the same relations as above for extra
parameters of a 5-pocket model). This comparison is shown in  Fig.
\ref{fig:comp_3_4_pocks}. We see that there is quite wide
parameter range where in the 4-pocket model the gap has nodes
while in the 5 pocket model it is still  nodeless. Each point in
the phase diagram in Fig. \ref{fig:comp_3_4_pocks} corresponds to
some values of the couplings, hence this result implies that for a
certain range of input parameters 4-pocket model yields a gap with
nodes while 5-pocket model yields the gap without nodes. This
agrees with the number of RPA
studies~\cite{Kuroki_2,peter,peter_2} which found nodal gap for
5-pocket model and no-nodal gap for 4 pocket model. At the same
time, our results show that in both models there are regions of
parameters in which the SC gap is either no-nodal or has nodes.

\section{Conclusion}
We have done calculations addressing on the same footing the
issues of the interplay between intra and inter--pocket Coulomb
repulsion in the Fe-based superconductors, the competition between
SC and SDW orders, and the angular dependence of the SC gap.  We
considered 2-, 4-, and 5-pocket models for the pnictides and for
each model considered the flow of the couplings and of SDW and SC
vertices within analytical parquet RG scheme. We found that in all
models, fluctuations in the SDW and SC channels are coupled at
intermediate energies  $\Lambda > E  > E_F$ between the bandwidth
and the Fermi energy,  but decouple at energies below $E_F$.  The
system behavior below $E_F$ is governed by conventional ladder RG,
and each vertex flows according to $d \Gamma_i/dL = \Gamma^2_i$.

For the toy 2-pocket model, earlier results
showed~\cite{Chubukov,bib:Chu_physica} that  SDW instability is
the dominant one at perfect nesting. The SC vertex is repulsive at
large energies but changes sign under parquet RG and become
attractive above some RG scale.  The SDW and SC couplings flow to
the same value at the fixed point of RG equations, and the fixed
point of parquet RG has extended $O(6)$ symmetry.~\cite{podolsky}
If the scale of $E_F$ is reached before this fixed point, SDW
order prevails at zero doping but is reduced and eventually
destroyed at  finite doping.  Whether or not SC appears in place
of SDW order depends on whether SC vertex already changes sign and
becomes attractive at $E_F$. If superconductivity appears, the SC
gap has a simple plus-minus structure.

The main goal of this paper was to understand how this scenario is
modified in realistic 4-pocket and 5-pocket models. We considered
both models in the two limits: one when one of the two hole
pockets centered at $(0,0)$ is weakly coupled to other pockets and
can be neglected, and the other when the two hole pockets centered
at $(0,0)$ are equivalent. We found identical results in both
limit what gives us confidence that the system behavior in the
intermediate case of two non-equivalent hole pockets at $(0,0)$
remains the same as in the two limits.

Our main results are the following:

\begin{itemize}

\item{In both 4-pocket and 5-pocket models electron-hole and
electron-electron interactions are generally angle dependent. The
most relevant angle dependence is in the form $\cos 2 \phi$, where
$\phi$ is the angle along an electron FS.  Because of this angular
dependence, there are three different vertices in the SC channel.
One of these vertices turns out to be attractive,  in most cases,
beginning from the largest energies. The symmetry of the
attractive interaction is extended $s\pm$wave (as opposed to a
conventional $s++$). Other two SC vertices are repulsive at all
scales. }

\item{This attractive SC vertex favors the $s\pm$ state in which the gaps along
hole FSs are angle-independent (up to $\cos 4 \phi$ corrections),
while the gaps along the two electron FSs are in the form
$\Delta_e \pm \bar{\Delta}_{e} \cos 2 \phi$. The interplay between
$\Delta_e$ and $\bar{\Delta}_{e}$ depends on the strength of $\cos
2 \phi$ component of the interaction and also on the interplay
between intra-pocket and inter-pocket Coulomb repulsions.
Depending on the parameters, the electron gaps can be either
nodeless ($\Delta_e > \bar{\Delta}_{e}$), or have accidental nodes
($\Delta_e < \bar{\Delta}_{e}$).}

\item{In 5-pocket model at perfect nesting, the SDW vertex remains larger than this attractive SC vertex. The two flow up to the same values at the
fixed point, if this fixed point is at an energy larger than
$E_F$, and the fixed point has enlarged symmetry. This behavior is
exact when the vertices are angle-independent, but only very
weakly changes  due to  angular dependence of the vertices. If the
system flows down to $E_F$ without yet reaching the fixed point,
SDW order wins.  Away from perfect nesting SDW order is
suppressed, and the system eventually develops a SC instability.}

\item{In 4-pocket model, the situation is
similar at large $E$ (i.e., at small RG parameter $\log \Lambda/E$),
but before the fixed point of parquet RG is reached, SC
vertex becomes larger than SDW vertex. If this happens before the
scale of $E_F$ is reached, the system develops SC instability
already at perfect nesting, and SDW instability does not appear.
If SDW vertex remains the largest down to $E_F$, the system
develops SDW instability at and around perfect nesting, and SC
instability at larger dopings.}

\item{We found that the SC gap is more likely to have accidental nodes on electron FSs in 4 pocket model than in 5-pocket model. Namely, for the same input parameters, there is a parameter range where the gap is nodal in 4-pocket model
and no-nodal in 5-pocket model. This agrees with several RPA-type
studies based on spin fluctuations~\cite{Kuroki_2,peter,peter_2}.
Still, we found that in both 4-pocket and 5-pocket model the gap
can be either nodal or node-less, depending on parameters.}
\end{itemize}

Our analytical results are fully consistent with numerical fRG
study of 4-pocket and 5-pocket models by Thomale et
al~\cite{Thomale}.  We view this agreement as the evidence that
the differences between 4-pocket and 5-pocket models are
geometrical (different combinatorics in RG equations), and are
captured already within analytical one-loop RG. We note in this
regard that we found that the difference between 4-pocket and
5-pocket models is not caused by the angular dependence of the
interaction and holds even when interactions are
angle-independent.

The results for the 4-pocket case demonstrate that SDW order need
not be  pre-requisite to SC$\pm$ order, although for most part of
the phase diagram it does appear at perfect nesting, and SC only
appears upon doping. We also emphasize that the interplay between
SC and SDW is both `` mutual support'' and ``competition''.
Namely,  SC and SDW {\it fluctuations} tend to enhance each other,
what is relevant is the fact that in the applicability range of
parquet RG  (when SC and SDW fluctuations talk to each other),
both SC and SDW vertices diverge upon approaching the fixed point.
At the same time, SDW and SC {\it orders} compete with each
other~\cite{vor_vav_chub,joerg}, meaning that SC order only
emerges when SDW order is reduced enough by doping, and SDW order
does not emerge at all if SC order emerges first already  at
perfect nesting.

\section*{Acknowledgements}
We acknowledge helpful discussions with L. Benfatto, R. Fernandes,
W. Hanke, P. Hirschfeld, I. Eremin, Y. Matsuda, I. Mazin, R.
Prozorov, D. Scalapino, J. Schmalian, Z.~Tesanovic, R. Thomale, M.
Vavilov, and A. Vorontsov. We also thank I. Mazin for careful reading of the MS and useful comments. This work was supported by
NSF-DMR-0906953. Partial support from MPI PKS (Dresden) (S.M. and
A.V.C), Ruhr-Univerisity Bochum (S.M), and Humboldt foundation
(A.V.C)  is gratefully acknowledged.


\begin{thebibliography}{5}

\bibitem{review} For recent reviews see D.C. Johnston, Adv. Phys., {\bf 59}, 803 (2010); J-P Paglione and R.L. Greene, Nature Phys. {\bf 6}, 645 (2010).
\bibitem{bib:Hosono} Y. Kamihara, T. Watanabe, M. Hirano, H.
Hosono, J. Am. Chem. Soc. \textbf{130}, 3296(2008).
\bibitem{bib:X Chen} X. H. Chen, T. Wu, G. Wu,
R. H. Liu, H. Chen, D. F. Fang, Nature \textbf{453}, 761(2008).
\bibitem{bib:G Chen} G. F. Chen, Z. Li, D. Wu, G. Li, W. Z. Hu, J. Dong, P. Zheng, J.
L. Luo, and N. L. Wang, Phys. Rev. Lett. \textbf{100}, 247002
(2008).
\bibitem{bib:ZA Ren}Z.-A. Ren, \emph{et. al.}, Europhys. Lett. \textbf{83}, 17002(2008)

\bibitem{bib:Rotter} M. Rotter, M. Tegel, D. Johrendt,  Phys. Rev. Lett. \textbf{101}, 107006 (2008)
\bibitem{bib:Sasmal}K. Sasmal, B. Lv, B. Lorenz, A. M. Guloy, F. Chen, Y.-Y. Xue, and
C.-W. Chu, Phys. Rev. Lett. \textbf{101}, 107007 (2008),
\bibitem{bib:Ni}N. Ni, A. Thaler, J. Q. Yan, A. Kracher, E. Colombier, S.
L. Bud'ko, P. C. Canfield, Phys. Rev. B \textbf{82}, 024519
(2010).

\bibitem{bib:Wang}X. C. Wang, \emph{et. al.}, arXiv:0806.4688v3;
 S.V, Borisenko {\it et al} Phys. Rev. Lett. 105, 067002 (2010).

\bibitem{bib:Mizuguchi}Y. Mizuguchi, F. Tomioka, S. Tsuda,
T. Yamaguchi, and Y. Takano, Appl. Phys. Lett. \textbf{93}, 152505
(2008), F. C. Hsu et al., Proc. Natl. Acad. Sci. U.S.A.
\textbf{105}, 14 262 (2008), M. H. Fang et al., Phys. Rev. B
\textbf{78}, 224503 (2008), G. F. Chen et al., Phys. Rev. B
\textbf{79}, 140509(R) (2009), B. Zeng \emph{et.
al.},arXiv:1007.3597.


\bibitem{Cruz} C. de la Cruz \textbf{et. al.}, Nature, \textbf{453}, 899 (2008).
For the latest results on magnetic measaurements, see D. S.
Inosov, \emph{et. al.}, Nature Physics \textbf{6}, 178-181 (2010)
and references therein.

\bibitem{bib:Kamihara}Y. Kamihara et al., J. Am. Chem. Soc. \textbf{128}, 10012 (2006).

\bibitem{Li} M. Li \emph{et. al.} Phys. Rev. B \textbf{80}, 024515
(2009)

\bibitem{bib:Suchitra}S.E. Sebastian \emph{et. al.},  J. Phys.: Condens. Matter \textbf{20} 422203(2008).

\bibitem{bib:first}
D.J. Singh and M.-H. Du, Phys. Rev. Lett. \textbf{100}, 237003
(2008); M.J. Calderon, B. Valenzuela, and  E. Bascones, Phys. Rev.
B 80, 094531 (2009).
\bibitem{Boeri} L. Boeri, O. V. Dolgov, and A. A. Golubov, Phys. Rev. Lett. \textbf{101}, 026403 (2008)
\bibitem{Mazin} I. I. Mazin, D. J. Singh, M. D. Johannes, and M.
H. Du, Phys. Rev. Lett. \textbf{101}, 057003 (2008)

\bibitem{jap} Seiichiro Onari and  Hiroshi Kontani,  arXiv:1009.3882.

\bibitem{Kuroki}K. Kuroki \emph{et. al.}, Phys. Rev. Lett. \textbf{101}, 087004 (2008)
\bibitem{Kuroki_2}K. Kuroki, H. Usui, S. Onari, R. Arita, and H. Aoki, Phys. Rev. B
79, 224511 (2009).
\bibitem{Seo}K. Seo, B. A. Bernevig, and J. Hu, Phys. Rev. Lett. \textbf{101}, 206404 (2008).
\bibitem{Chubukov} A. V. Chubukov, D. V. Efremov, and I. Eremin, Phys. Rev. B \textbf{78}, 134512 (2008).
\bibitem{Stanev}V. Stanev, J. Kang, Z. Tesanovic, Phys. Rev. B \textbf{78}, 184509 (2008); V. Stanev, B. S. Alexandrov, P. Nokoli\'{c}, Z. Te\v{s}anovi\'{c}, arXiv:1006.0447.
\bibitem{Graser}S. Graser, T. A. Maier, P. J. Hirshfeld, D. J. Scalapino, New
J. Phys. \textbf{11}, 025016 (2009).
\bibitem{FWang}Fa Wang \emph{et. al.}, Phys. Rev. Lett.
\textbf{102}, 047005 (2009).
\bibitem{Rthomale} C. Platt, C.  Honerkamp, and Werner Hanke,
New J. Phys. \textbf{11}, 055058 (2009). ; R. Thomale \emph{et.
al.}, Phys. Rev. B \textbf{80}, 180505 (2009).
\bibitem{Thomale}R. Thomale, C. Platt, W. Hanke, B. A. Bernevig,
arXiv:1002.3599v1
\bibitem{bib:Chu_Vav_vor} A. V. Chubukov, M. G. Vavilov, A. B. Vorontsov,
Phys. Rev. B \textbf{80}, 140515(R)(2009).

\bibitem{peter} A. F. Kemper, T. A. Maier, S. Graser, H.-P. Cheng, P. J. Hirschfeld, and D. J. Scalapino, New J. Phys. {\bf 12},  073030 (2010).

\bibitem{peter_2} S. Graser, A. F. Kemper, T. A. Maier, H.-P. Cheng, P. J. Hirschfeld, and D. J. Scalapino, Phys. Rev. B 81, 214503 (2010).


\bibitem{bib:Nakai}Y. Nakai, \emph{et. al.}, J. Phys. Soc. Jpn. \textbf{77},
073701 (2008)
\bibitem{bib:Grafe}H-J. Grafe, \emph{et. al.}, Phys. Rev. Lett. \textbf{101},
047003 (2008)

\bibitem{bib:YWang}Y. Wang, \emph{et. al.},Supercond. Sci. Technol.
\textbf{22} 015018(2009).
\bibitem{bib:Millo}O. Millo, \emph{et. al.}Phys. Rev. B \textbf{78}, 092505 (2008)


\bibitem{bib:Fletcher}J. D. Fletcher \emph{et. al.}, Phys. Rev.
Lett., \textbf{102}, 147001(2009)
\bibitem{bib:Hicks}C. W. Hicks \emph{et. al.},Phys. Rev.
Lett., \textbf{103}, 177003(2009)
\bibitem{bib:Hashimoto1} K. Hashimoto \emph{et. al.}, Phys. Rev. B \textbf{81},
220501(R) (2010)
\bibitem{reid} J.-Ph. Reid \emph{et. al.}
Phys. Rev. B 82, 064501 (2010).
\bibitem{nakai_2}
Y. Nakai {\it et al}
Phys. Rev. Lett. 105, 107003 (2010).
\bibitem{gordon}
%
R. T. Gordon \emph{et. al.}
Phys. Rev. B 82, 054507 (2010).
\bibitem{hashimoto_3}
 K. Hashimoto \emph{et. al.},
Phys. Rev. B 82, 014526 (2010).

\bibitem{bib:Hashimoto2} K. Hashimoto \emph{et. al.}, Phys. Rev. Lett. \textbf{102}, 017002 (2009)
\bibitem{bib:Malone}L. Malone \emph{et. al.}, Phys. Rev. B \textbf{79}, 140501(R) (2009)

\bibitem{bib:Ding}
K. Nakayama  \emph{et. al.},
arXiv:1009.4236 and references therein.
\bibitem{borisenko}
D. S. Inosov  \emph{et. al.},
Phys. Rev. Lett. 104, 187001 (2010).
\bibitem{kaminski}
Chang Liu \emph{et. al.}, arXiv:1011.0980 and references therein.
\bibitem{kont} Seiichiro Onari and  Hiroshi
Kontani, Phys. Rev. Lett. 103, 177001 (2009).
\bibitem{ding_2}
Z.-H. Liu \emph{et. al.}, arXiv:1008.3265.

\bibitem{mcmillan} W. L. McMillan, Phys. Rev. 167, 331 (1968); N.N. Bogolubov,
V.V. Tolmachev, and D.V. Shirkov, Consultants Bureau, 1959.


\bibitem{IMazin}I. Mazin and J. Schmalian, Physica C,
\textbf{469}, 614 (2009)

\bibitem{bib:Chu_physica} A. V. Chubukov
Physica C \textbf{469}, 640(2009).

\bibitem{podolsky} Daniel Podolsky, Hae-Young Kee, and Yong Baek Kim,
Europhysics Letters 88, 17004 (2009).

\bibitem{Singh}D. J. Singh, Phys. Rev. B, \textbf{78}, 094511(2008).

\bibitem{new} I. Aleiner et al, in preparation

\bibitem{golden} J.F. Annett, N. Goldenfeld, and A.J.  Leggett.
JLTP {\bf 105}, 473 (1996).

\bibitem{pines} P. Monthoux and D. Pines, Phys. Rev. B {\bf 49}, 4261 (1994);
D.J. Scalapino, Phys. Rep. {\bf 250}, 329 (1995); Ar. Abanov, A. V. Chubukov, and M. R. Norman, Phys. Rev. B 78, 220507 (2008).

\bibitem{Eremin} I. Eremin and A.V. Chubukov, Phys. Rev. B 81, 024511 (2010).
\bibitem{joerg} R. M. Fernandes and J. Schmalian, Phys. Rev. B 82, 014521 (2010)
\bibitem{vor_vav_chub} A.B.Vorontsov, M.G.Vavilov, and A.V.Chubukov, Phys. Rev. B 81, 174538 (2010);  M. G. Vavilov, A. V. Chubukov, and A. B. Vorontsov, Supercond. Sci. Technol. 23, 054011 (2010).
\bibitem{haule} K. Haule, unpublished.
\bibitem{bib:unp2} Saurabh Maiti {\it et al}, unpublished.

\end{thebibliography}
\end{document}